\documentclass[12pt]{article} 
\pdfoutput=1

\usepackage[utf8]{inputenc}
\usepackage[T1]{fontenc} 
\usepackage{lmodern}
\usepackage[english]{babel}           
\usepackage{cancel}
\usepackage[pdftex]{graphicx}
\usepackage{epsfig}
\usepackage{graphicx}
\usepackage{comment}
\usepackage{latexsym}
\usepackage{hyperref}
\usepackage{amsmath}
\usepackage{mathrsfs}
\usepackage[usenames, dvipsnames]{color}
\usepackage{amsbsy}
\usepackage{amssymb}
\usepackage{amsthm}
\usepackage{amsfonts}
\usepackage{cite}
\usepackage{enumitem}
\usepackage{xcolor}
\usepackage{color}
\usepackage{mathtools}
\usepackage{caption} 
\usepackage{subcaption} 
\usepackage{euscript} 
\usepackage{float}
\usepackage{diagbox}
\usepackage[title]{appendix}
\usepackage{tabularx}

\newcommand{\Vol}{{\rm Vol}}
\newcommand{\Diff}{{\rm Diff}}
\newcommand{\Id}{{\rm Id}}
\renewcommand\d{{\rm d}}
\renewcommand{\b}{\boldsymbol{\rm b}}
\renewcommand{\c}{\boldsymbol{\rm c}}
\newcommand{\bell}{\bar \ell}
\newcommand{\ba}{\bar a}
\newcommand{\bq}{\bar q}
\newcommand{\bS}{\boldsymbol{{\rm S}}}
\newcommand{\bP}{\boldsymbol{{\rm P}}}
\newcommand{\deltaa}{\boldsymbol{\delta \rm a}}
\newcommand{\deltaq}{\boldsymbol{\delta \rm q}}
\newcommand{\deltatq}{\boldsymbol{\delta \rm  \tilde q}}
\newcommand{\bM}{\boldsymbol{\rm M}}

\newcommand{\M}{{\cal M}}
\newcommand{\N}{{\cal N}}
\newcommand{\D}{{\cal D}}
\newcommand{\I}{{\cal I}}
\newcommand{\K}{{\cal K}}
\newcommand{\C}{{\cal C}}
\newcommand{\E}{{\cal E}}
\newcommand{\T}{{\rm T}}
\renewcommand{\S}{{\cal S}}
\newcommand{\A}{{\cal A}}
\newcommand{\cR}{{\cal R}}
\newcommand{\Q}{{\cal Q}}
\renewcommand{\L}{{\cal L}}
\newcommand{\J}{{\cal J}}
\newcommand{\wt}{\widetilde}

\newcommand{\Pc}{\mathscr{P}}
\newcommand{\Nc}{\mathscr{N}}
\newcommand{\Cc}{\mathscr{C}}
\newcommand{\Hc}{\mathscr{H}}
\newcommand{\Fc}{\mathscr{F}}

\DeclareMathOperator\sign{sign}

\DeclareMathOperator\diag{diag}
\DeclareMathOperator\arccosh{arccosh}

\newtheorem{theorem}{Theorem}

\newcommand{\be}{\begin{equation}}
\newcommand{\ee}{\end{equation}}
\newcommand{\dis}{\displaystyle}
\renewcommand{\thefootnote}{\fnsymbol{footnote}}
\newcommand{\eq}[1]{\eqref{#1}}
\newcommand{\Eq}[1]{Eq.~\eqref{#1}}
\newcommand{\Eqs}[1]{Eqs.~\eqref{#1}}
\newcommand{\Refe}[1]{Ref.~\cite{#1}}

\newcommand{\Sect}[1]{Sect.~\ref{#1}}
\newcommand{\Sects}[1]{Sects.~\ref{#1}}

\newcommand{\R}{\mathbb{R}}
\newcommand{\Z}{\mathbb{Z}}
\renewcommand{\natural}{\mathbb{N}}
\newcommand{\complex}{\mathbb{C}}

\renewcommand{\O}{{\cal O}}
\newcommand{\cst}{{\rm cst.}}

\newcommand{\ie}{{\em i.e.} }
\newcommand{\eg}{{\em e.g.} }
\newcommand{\via}{{\it via} }
\newcommand{\apriori}{{\it a priori} }
\newcommand{\where}{\mbox{where}}
\newcommand{\with}{\mbox{with}}
\newcommand{\when}{\mbox{when}}
\renewcommand{\and}{\mbox{and}}

 \newcommand{\WDW}{Wheeler--DeWitt }

\newcommand{\esps}{\phantom{\!\!\!\overset{|}{a}}}
\newcommand{\esp}{\phantom{\!\!\overset{\displaystyle |}{|}}}

\newcommand{\bm}{\boldmath} 

\catcode`\@=11
\def\marginnote#1{}
\newcount\hour
\newcount\minute
\newtoks\amorpm
\hour=\time\divide\hour by60 \minute=\time{\multiply\hour by60
\global\advance\minute by-\hour}
\edef\standardtime{{\ifnum\hour<12 \global\amorpm={am}%
        \else\global\amorpm={pm}\advance\hour by-12 \fi
        \ifnum\hour=0 \hour=12 \fi
        \number\hour:\ifnum\minute<10 0\fi\number\minute\the\amorpm}}
\edef\militarytime{\number\hour:\ifnum\minute<10 0\fi\number\minute}
\def\draftlabel#1{{\@bsphack\if@filesw {\let\thepage\relax
   \xdef\@gtempa{\write\@auxout{\string
      \newlabel{#1}{{\@currentlabel}{\thepage}}}}}\@gtempa
   \if@nobreak \ifvmode\nobreak\fi\fi\fi\@esphack}
        \gdef\@eqnlabel{#1}}
\def\@eqnlabel{}
\def\@vacuum{}
\def\draftmarginnote#1{\marginpar{\raggedright\scriptsize\tt#1}}
\def\draft{\oddsidemargin -.2truein
        \def\@oddfoot{\sl preliminary draft \hfil
        \rm\thepage\hfil\sl\today\quad\militarytime}
        \let\@evenfoot\@oddfoot \overfullrule 3pt
        \let\label=\draftlabel
        \let\marginnote=\draftmarginnote
   \def\@eqnnum{(\theequation)\rlap{\kern\marginparsep\tt\@eqnlabel}%
\global\let\@eqnlabel\@vacuum}  }
\def\thebibliography#1{
\vskip 0.5cm \centerline{\bf \Large References}
\list{
[\arabic{enumi}]}{\settowidth\labelwidth{[#1]}
\leftmargin\labelwidth
\advance\leftmargin\labelsep
\usecounter{enumi}}
\def\newblock{\hskip .11em plus .33em minus .07em}
\sloppy\clubpenalty4000\widowpenalty4000
\sfcode`\.=1000\relax}

\renewcommand{\theequation}{\arabic{section}.\arabic{equation}}
\renewcommand{\section}{\setcounter{equation}{0}\@startsection
{section}{1}{0mm}{-\baselineskip}{0.5\baselineskip} {\normalfont\Large\bfseries}}
\renewcommand{\subsection}{\@startsection
{subsection}{2}{0mm}{-\baselineskip}{0.5\baselineskip} {\normalfont\large\bfseries}}
\renewcommand{\subsubsection}{\@startsection
{subsubsection}{3}{0mm}{-\baselineskip}{0.5\baselineskip}
{\normalfont\normalsize\slshape}}

\begin{document}


\begin{titlepage}
\begin{flushright}
CPHT-RR027.032021, March 2021
\vspace{0.0cm}
\end{flushright}
\begin{centering}
{\bm\bf \Large Wavefunction of the universe: \\ Reparametrization invariance and field redefinitions \\ of the minisuperspace path integral\\ }

\vspace{6mm}

 {\bf Herv\'e Partouche,$^1$\footnote{herve.partouche@polytechnique.edu} Nicolaos Toumbas$^2$\footnote{nick@ucy.ac.cy} and Balthazar de Vaulchier$^1$\footnote{balthazar.devaulchier@polytechnique.edu}}

 \vspace{3mm}

$^1$  {\em CPHT, CNRS, Ecole polytechnique, IP Paris, \\F-91128 Palaiseau, France}

$^2$ {\em Department of Physics, University of Cyprus, \\Nicosia 1678, Cyprus}

\end{centering}
\vspace{0.5cm}
$~$\\
\centerline{\bf\Large Abstract}\\
\vspace{-0.6cm}

\begin{quote}

We consider the Hartle--Hawking wavefunction of the universe defined as a Euclidean path integral that satisfies the ``no-boundary proposal.'' We focus on the  simplest minisuperspace model that comprises a single scale factor degree of freedom and a positive cosmological constant. The model can be seen as a non-linear $\sigma$-model with a line-segment base. We reduce the path integral over the lapse function to an integral over the proper length of the base and use diffeomorphism-invariant measures for the ghosts and the scale factor. As a result, the gauge-fixed path integral is independent of the gauge. However, we point out that all field redefinitions of the scale factor degree of freedom yield different choices of gauge-invariant path-integral measures. For each prescription, we compute the wavefunction at the semi-classical level and find a different result. We resolve in each case the ambiguity in the form of the Wheeler--DeWitt equation at this level of approximation. By imposing that the Hamiltonians associated with these possibly distinct quantum theories are Hermitian, we determine the inner products of the corresponding Hilbert spaces and find that they lead to a universal norm, at least semi-classically. Quantum predictions are thus independent of the prescription at this level of approximation. Finally, all wavefunctions of the Hilbert spaces of the minisuperspace model we consider turn out to be non-normalizable, including the no-boundary states. 

\end{quote}

\end{titlepage}
\newpage
\setcounter{footnote}{0}
\renewcommand{\thefootnote}{\arabic{footnote}}
 \setlength{\baselineskip}{.7cm} \setlength{\parskip}{.2cm}

\setcounter{section}{0}


\section{Introduction}

According to the inflationary paradigm, a tiny Planckian region of space underwent a period of rapid accelerated expansion, and grew large enough to encompass the entire observable universe. Quantum effects are crucial to understand the state of the universe at the initial stages of this inflationary era. Indeed, it is widely believed that quantum fluctuations seed the primordial density perturbations, which lead eventually to the large scale structure of the universe today. Hence, it would be desirable to obtain the wavefunction of the universe, and to uncover a statistical interpretation favouring initial conditions amenable for inflation.

 A step  toward this direction was  initiated by Hartle and Hawking~\cite{HH} in the  context of Einstein's theory of gravity for closed universes, in the presence of a positive cosmological constant $\Lambda >0$. In particular, they  proposed a definition  for the ``ground-state wavefunction,'' which Vilenkin  has interpreted as the probability amplitude for creating   from ``nothing'' a three-dimensional universe  with metric $h_{ij}$~\cite{Vilenkin1, Vilenkin2,Vilenkin3,Vilenkin4}. In practice, this wavefunction is computed \via  a Euclidean path integral  according to the ``no-boundary proposal.'' This  path integral involves a sum over compact four-geometries that end on a particular spatial slice with induced metric $h_{ij}$. Notice that the denomination of ``ground state'' is somehow misleading, since  in quantum gravity all states associated with a closed universe (including  the matter content) are degenerate, with vanishing energy. Moreover, as we will see, such wavefunctions are not necessarily normalizable. 

In  this work, we address and clarify four issues  related to this path integral approach to quantum gravity:

$(i)$ To begin with, since general relativity is invariant under diffeomorphisms, special attention must be paid to the gauge fixing of this symmetry. This problem can be analyzed in the simpler framework of minisuperspace models, where the universe is assumed to be homogeneous, with a finite number of degrees of freedom depending only on time.  In the literature, this gauge fixing of  time-reparametrizations has not always been implemented appropriately  for wavefunctions of the universe defined as path integrals, since the results depend on the chosen gauges~\cite{Halli}. 

In the present work, we consider the simplest  such minisuperspace model,  corresponding to a homogenous and isotropic universe  with a single dynamical degree of freedom, namely the scale factor $a$. The model can be  interpreted as a non-linear $\sigma$-model, where Euclidean time parametrizes a base manifold which is a line segment. Thus the scale factor is a coordinate in a one-dimensional target space. After gauge fixing of the Euclidean-time reparametrizations, the path integral over the lapse function reduces to an integral over the moduli space of the base manifold, which is parametrized by the proper length $\ell$ of the line segment. Moreover, we use diffeomorphism-invariant path-integral measures for the Faddeev--Popov ghosts and the scale factor. The outcome is a gauge-fixed path integral consistently independent of the choice of gauge.

$(ii)$ Field redefinitions of the scale factor, $a=A(q)$, 
leave the classical action invariant. However, at the quantum level, the path-integral measures $\D a$ and $\D q$ are not equivalent, as they are related to each other by a Jacobian. Since the fields $a$ and $q$ correspond to different coordinate systems of the $\sigma$-model target space, there is no preferred choice among these measures. As a result, infinitely many acceptable definitions of wavefunctions exist, which are associated with all possible field redefinitions of the scale factor degree of freedom.

For each choice of path-integral measure $\D q$, we compute the ground-state wavefunction using the steepest-descent method. This amounts to deriving all instanton solutions $\bar q$ and the corresponding values of the modulus $\bell$ of the base segment, and to computing the quantum fluctuations and modulus fluctuations around these solutions at quadratic order. This is  achieved by evaluating functional determinants by applying and refining the methods presented in \Refe{Coleman}.  

$(iii)$ Given a choice of measure $\D q$, the set of wavefunctions describing the allowed states of the universe are the solutions of a \WDW equation~\cite{DeWitt}, which is the quantum  analogue of the vanishing of the classical Hamiltonian on shell. However, in the process of canonical quantization of the Hamiltonian, the question of the ordering of the canonically conjugate variables $q$ and $\pi_q$ gives rise to an ambiguity in the exact form of the \WDW equation.\footnote{Since the path integrals computed in \Refe{Halli} depend on the choice of gauge for time-reparametrizations, the attempt described in this reference to use these wavefunctions to lift the ambiguity in the \WDW equation is not strictly valid.}  

Thanks to our derivation of the ground-state wavefunctions \via the steepest-descent method, we are able to compare the results with the solutions of the \WDW equation found by applying the WKB approximation. This allows us to resolve the ambiguity of the equation at the semi-classical level for every choice of measure $\D q$ in the path integrals. 

$(iv)$ A question related to points $(ii)$ and $(iii)$ arises then. Do the different choices of measures $\D q$ in the prescription of the wavefunctions, as well as the distinct associated Wheeler--DeWitt equations, lead to different quantum gravity theories with the same classical limits? We show that the answer to this question is negative, at least at the semi-classical level. 

In fact, by imposing that the Hamiltonians of these possibly distinct quantum theories are Hermitian, we find that the  inner products of the corresponding Hilbert spaces take different forms in each case. However, the point is that the norms of the wavefunctions in the WKB approximation turn out to be universal, \ie independent of the prescription. A non-trivial crosscheck of our results is provided by the particular choice of measure $\D q$ associated with a field $q$ 
 with a quadratic kinetic term. In that case we find that the \WDW equation reduces to the time-independent Schr\"odinger equation for vanishing energy,  and that the 
 inner product of the Hilbert space takes the standard form encountered in quantum mechanics. 

Moreover, in the minisuperspace model that comprises a single scale factor degree of freedom, 
it turns out that all wavefunctions of the two-dimensional Hilbert space are non-normalizable, including that of the ``ground-state.'' To construct normalizable states, extra degrees of freedom  must be added,  rendering the dimension of the Hilbert space  infinite. In some cases, it could then be possible to construct square-integrable wavefunctions by superimposing infinitely many solutions of the \WDW equation \cite{DeWitt}. 

 Other seminal work on the wavefunction of the universe \via the no-boundary proposal includes Refs.~\cite{Linde, HawkingL=0, HalliHawk, Schleich:1986db, Vilenkin5, Halli2, Turok1, Halli-Hartle, Quevedo, Davidson1, Davidson2}.

In \Sect{sGS}, we review the definition of the ground-state wavefunction in the minisuperspace model we consider. The implementation of the gauge fixing of the Euclidean-time diffeomorphisms introduces a Faddeev--Popov Jacobian. The latter is expressed in \Sect{FPdet} as a path integral over  anticommuting ghosts, which is evaluated by using gauge-invariant measures. The outcome of the computation is that the Faddeev--Popov determinant is a trivial (and irrelevant) constant. The following two sections are devoted to the evaluation of the gauged-fixed path integral by applying the steepest-descent method. This is done in \Sect{path-a} by using the measure $\D a$ of the scale factor, while in \Sect{field redef} the result is generalized to any choice of gauge-invariant measure $\D q$ associated with a field $q$ obtained by redefining the scale factor degree of freedom. We also 
 clarify the extent to which our calculations differ  with previous works such as Refs.~\cite{Turok1, Halli-Hartle,Halli, Halli2}, leading to different results. In \Sect{WDW}, we lift at the semi-classical level the ambiguities arising in the \WDW equations corresponding to the wavefunctions with  arbitrary prescriptions $\D q$ for the path-integral measure. The equivalence at this level of approximation of all Hilbert spaces is demonstrated in \Sect{Qequi}. Our conclusions and perspectives can be found in \Sect{conclu}. Various technical complements are reported in the Appendix.


\section{Ground state wavefunction}
\label{sGS}

In this section, we present the formal definition of the ground-state wavefunction of a closed, homogeneous and isotropic universe.
The wavefunction is expressed in terms of a Euclidean path integral over the lapse function and the scale factor. Our goal is to carry out carefully the gauge fixing of the Euclidean-time reparametrization group.  In the following sections we present the computations of the Faddeev--Popov determinant and the path integral over the scale factor.


\subsection{Minisuperspace of dimension one}

We are interested in Einstein's theory for spatially-closed universes in the presence of a positive cosmological constant $\Lambda>0$, formulated on Lorentzian four-manifolds $\M$ with space-like boundaries $\partial\M$.  The appropriate action reads
\begin{align}
S&=S_{\rm bulk}+S_{\rm boundary}\, ,\nonumber \\
\where \quad  S_{\rm bulk}&=\int_\M \d^4x \,\sqrt{-g}\, \Big[{R\over 2}-\Lambda\Big]\,,\quad S_{\rm boundary}=-\int_{\partial\M}\d^3x_{\rm b}\,\sqrt{h}\, K\, ,\esp\label{action}
\end{align}
where $R$ is the scalar curvature associated with the metric $g_{\mu\nu}(x)$ on $\M$, while $K$ is the trace of the extrinsic curvature on $\partial\M$ whose metric is denoted $h_{ij}(x_{\rm b})$.   

In this work, we consider the minisuperspace version 
where the degrees of freedom are reduced to a single scale factor depending only on time, $a(x^0)$. This amounts to restricting the manifolds $\M$ to be homogeneous and isotropic. The invariant infinitesimal length squared is given by
\be
\d s^2=-N(x^0)^2(\d x^0)^2+a(x^0)^2\, \d\Omega_3^2\, , 
\label{FRW}
\ee
where $N(x^0)\equiv \sqrt{g_{00}(x^0)}$ is the lapse function and $\d\Omega_3$ is the volume element of the unit 3-sphere. 
Moreover, $\partial\M$ is composed of 3-spheres at initial and final times $x^0_{\rm i}$ and $x^0_{\rm f}$, and the boundary action becomes  
\be
S_{\rm boundary}=3v_3\bigg[{a^2\over N}\, {\d a\over \d x^0}\bigg]_{x^0_{\rm i}}^{x^0_{\rm f}}\,\, , 
\ee
where $v_3=2\pi^2$ is the volume of the unit 3-sphere. 
This term cancels a similar boundary term generated upon integrating by parts the second derivative of the scale factor arising from the Ricci scalar. As a result, the full action is reduced to 
\be
S=3v_3\int_{x^0_{\rm i}}^{x^0_{\rm f}} \d x^0\, N\bigg[\!-\!{a\over N^2}\Big({\d a\over \d x^0}\Big)^2+a-\lambda^2a^3\bigg]\, ,\quad \where \quad \lambda=\sqrt{\Lambda\over 3}\, .
\label{action2}
\ee
Notice the negative sign of the kinetic energy compared to that of a conventional matter scalar field. In this form, the action is expressed in terms of a Lagrangian involving only the scale factor and its first derivative (along with $N$), which is suitable for canonical quantization, as well as  for deriving classical equations of motion, keeping the scale factor fixed on the boundaries.


\subsection{Ground-state wavefunction as a Euclidean  path integral}

The quantum state of the universe can be described by a wavefunction that depends on $h_{ij}$, defined as a path integral. 
 Hartle and Hawking have proposed to sum over all manifolds $\M$ with a space-like boundary, on which the induced metric is $h_{ij}$~\cite{HH}. Specifying other conditions on the class of paths summed over amounts to characterizing the state. For the ground state, the  Hartle--Hawking prescription is formulated in terms of the Euclidean action, while the four-manifolds should have no other boundary than that of metric $h_{ij}$. This is the ``no-boundary proposal'' for the ground-state wavefunction. It follows from this definition, that the wavefunction can be interpreted as the amplitude for creating the three-geometry from the empty set, \ie from ``nothing''~\cite{Vilenkin1, Vilenkin2,Vilenkin3,Vilenkin4}.\footnote{The ground state wavefunction of a closed universe cannot be defined as the state of lowest energy since the quantum Hamiltonian vanishes identically (see \Sect{WDW}).}

In minisuperspace, the boundary geometry is fully characterized by the scale factor $a_0$ on the boundary 3-sphere. As a result, a possible definition of the ground-state wavefunction is given by  
\be
\Psi(a_0)=\int {\D g_{00} \over \Vol(\Diff[g_{00}])} \int_{\textstyle \substack{\!\!a_{\rm i}=0 \\\, a_{\rm f}=a_0}}\!\!\D a \; e^{-{1\over \hbar}S_{\rm E}[g_{00},a]}\, , 
\label{Psi(a0)}
\ee
where we keep explicit the reduced Planck constant $\hbar$.  The following comments are in order:

$\bullet$ 
Two possible prescriptions for the continuation to imaginary time have been advocated in the literature, given by 
\be
x^0=s\, i\, x^0_{\rm E} \, , ~~\quad \where ~~\quad s \in\{1,- 1\}\, .
\label{LE}
\ee 
Hartle and Hawking~\cite{HH} take $s=-1$, while Vilenkin~\cite{Vilenkin1, Vilenkin2,Vilenkin3,Vilenkin4,Vilenkin5} and Linde~\cite{Linde} argue for $s=+1$. In various minisuperspace models, the predictions associated with the two choices can be drastically different. For instance, when the path integral is approximated by the steepest-decent method, with only one instanton solution taken into account, the wavefunction for $s=-1$ leads to the conclusion that the cosmological constant is 
likely to be null~\cite{HawkingL=0}. On the other hand, when $s=+1$ conditions amenable for inflation are favored~\cite{Linde,LindeBook}.  In the following, we 
 consider both options, so that the Euclidean action  $S_{\rm E}=-iS$
becomes 
\be
S_{\rm E}[g_{00},a]=3sv_3\int_{x^0_{\rm Ei}}^{x^0_{\rm Ef}} \d x^0_{\rm E}\,  \sqrt{g_{00}}\,\bigg[ a\, g^{00}\Big({\d a\over \d x^0_{\rm E}}\Big)^2+V(a)\bigg]\, ,
\label{action3}
\ee
 where we have defined
\be
V(a)=a-\lambda^2a^3\, .
\ee
Moreover, the paths  $a(x^0_{\rm E})$ must obey $a(x^0_{\rm Ef})\equiv a_{\rm f}=a_0$ and the ``no-boundary condition'' \mbox{$a(x^0_{\rm Ei})\equiv a_{\rm i}=0$}. Since both choices of $s$ yield actions $S_{\rm E}[g_{00},a]$ not bounded from below, special attention must be paid for defining convergent path integrals. We will come back to this issue in \Sect{Qfluc}. 

$\bullet$ The action $S_{\rm E}$ describes a non-linear $\sigma$-model where Euclidean time parametrizes a base manifold which is a segment of metric $g_{00}$, while the scale factor is a coordinate in a one-dimensional target space of metric
\be
G_{aa}=6v_3a\, . 
\ee
In other words, the system can be viewed as describing a worldline $a(x^0_{\rm E})$ in a dimension-one target minisuperspace, by analogy with the worldline trajectory of a particle in 
spacetime, or a worldsheet embedded in 
spacetime in string theory. 

$\bullet$ The action being invariant under Euclidean-time reparametrizations, 
all diffeomorphism-equivalent metrics $g_{00}$ 
yield an overcounting of physically equivalent configurations. 
As a result, the path-integral measure $\D g_{00}$ must be divided by the volume of this group, which is denoted by $\Diff[g_{00}]$.\footnote{We will show in Appendix~\ref{A3} that this group is actually independent of $g_{00}$. Therefore, we may ignore all arguments of the symbol $\Diff$ in the sequel.}

$\bullet$ For field configurations $a(x^0)$ and $a(\xi(x^0))$, where $\xi\in \Diff[g_{00}]$, to be truly equivalent, the path-integral measure $\D a$ must be invariant under this symmetry group. 

$\bullet$  However, diffeomorphism-invariant measures  are not unique. The definition of the wavefunction in \Eq{Psi(a0)} uses a particular choice of coordinate in the target space, namely the scale factor. As a result, $\Psi(a_0)$ is not equivalent to the wavefunction defined with the measure $\D q$, where $q=Q(a)$ is a field redefinition for an arbitrary function $Q$.  
This is despite the fact that such a transformation leaves the action invariant, as it corresponds to a change of coordinate in the target space. Since there is no 
preferred variable $q$ in the target space, there is no preferred measure $\D q$ in the definition of the wavefunction. All choices yield wavefunctions 
solving different Wheeler--DeWitt
equations. In \Sects{field redef} and~\ref{WDW}, we will determine the relations between all these avatars   and see in \Sect{Qequi} how they yield equivalent predictions, at least at the semi-classical level.


\subsection{Gauge fixing of the Euclidean-time reparametrizations }

In this subsection, 
we rewrite in a more practical way the path integral over the metrics $g_{00}$, which is weighted by the inverse of the volume of the diffeomorphism group. 

For any metric $g_{00}$ defined on the domain $[x^0_{\rm Ei},x^0_{\rm Ef}]$, let us denote the action of a change of coordinate as follows,
\be
\xi(x^0_{\rm E})=x^{\xi 0}_{\rm E}\, , ~~\quad g^\xi_{00}(x^{\xi0}_{\rm E}) = \Big({\d x^0_{\rm E}\over \d x^{\xi 0}_{\rm E}}\Big)^2\,g_{00}(x^0_{\rm E})\, , ~~\quad a^\xi(x^{\xi 0}_{\rm E})=a(x^0_{\rm E})\, .
\label{transfo}
\ee 
For an infinitesimal diffeomorphism  $\xi=\Id+\delta\xi$, this transformation rule yields 
\be
x^{\xi 0}_{\rm E}=x^{0}_{\rm E}+\delta x^{0}_{\rm E}\quad \Longrightarrow\quad \delta_\xi  g_{00} = -2\nabla_0 \delta x_{\rm E 0}+\O(\delta x_{\rm E 0}^2)\, ,
\label{xiloc}
\ee
where $\nabla_0$ is the covariant derivative associated with  $g_{00}$. 
Notice that not all metrics $g_{00}$  are equivalent up to diffeomorphisms since such transformations cannot change the proper length $\ell$ of the line segment,
\be
\ell = \int_{x_{\rm Ei}}^{x_{\rm Ef}} \d x^0_{\rm E} \,\sqrt{g_{00}}=\int_{\xi(x_{\rm Ei})}^{\xi(x_{\rm Ef})} \d x^{\xi0}_{\rm E} \,\sqrt{g^\xi_{00}}\, .
\label{di}
\ee  
As a result, the set of metrics can be divided in equivalence classes distinguished by the value of $\ell\in \R_+$. 
 In other words, a line segment admits a moduli space of real dimension one parametrized by $\ell$.\footnote{We review in Appendix~\ref{A3}  a formal proof of the fact that there is no other modulus than the length~$\ell$. 
 Since the proof uses ingredients from \Sect{gsect}, the reader can wait until then before reading it.} In practice, varying the modulus of a metric $g_{00}$ amounts to rescaling it while keeping fixed its domain of definition. 

Let us define reference metrics that will serve 
as gauge-fixed representatives of each equivalence class.
We begin by choosing in class $\ell=1$ an arbitrary fiducial metric defined on a domain $[\hat x^0_{\rm Ei}, \hat x^0_{\rm Ef}]$,  which we denote by $\hat g_{00}[1]$.   Then any class $\ell>0$ can be represented by the fiducial metric
\be
 \hat g_{00}[\ell]=\ell^2\,  \hat g_{00}[1] \quad \mbox{defined on} \quad [\hat x^0_{\rm Ei}, \hat x^0_{\rm Ef}]\, .
 \label{l2}
\ee
Any other metric $g_{00}$ in class $\ell$ can be 
 obtained by the action of a diffeomorphism $\xi$ on $\hat g_{00}[\ell]$ (see Appendix~\ref{A1}). However, such a coordinate transformation is not unique. For a base manifold of generic topology and dimension, there are continuous isometries, which by definition are diffeomorphisms that preserve the metric. In the case of a  line segment, the group of isometries, or Killing group, is of dimension 0.\footnote{In fact, an infinitesimal diffeomorphism that leaves the metric invariant must satisfy $\delta  g_{00}=0$. Hence, 
$$
0=-2\nabla_0 \delta x^{0}_{\rm E}=-2 \left(\partial_0-\partial_0\ln\sqrt{g_{00}}\right)\delta x^{0}_{\rm E}\, ,
$$
which implies $\delta x^{0}_{\rm E}=C \sqrt{g_{00}}$ for some constant $C$. However, the definition of the metric includes its domain of definition, which must also be left invariant by the diffeomorphism. Hence, $\delta x^{0}_{\rm E}(x^{0}_{\rm Ei})=\delta x^{0}_{\rm E}(x^{0}_{\rm Ef}) = 0$ which implies $C=0$. } It is nonetheless non-trivial and reduces to the $\Z_2$ generated by the discrete isometry that  reverses the orientation of the segment. Both of these well known 
 facts are  derived
in Appendix~\ref{A2}. As a result, the group of diffeomorphisms can be divided into two disconnected components,
\be
\Diff[g_{00}]=\Diff[g_{00}]_{\Id} \cup \big(\mbox{orientation reversal}\big)\!\cdot\!\Diff[g_{00}]_{\Id}\, , 
\label{decompo}
\ee
where $\Diff[g_{00}]_{\Id}$ is the subgroup connected to the identity.

 Next we replace the path integral over 
$g_{00}$ by two integrals: 
 An integral over the moduli space and 
 the other one over 
the 
orbit of each 
equivalence class, in order to cancel the volume of the diffeomorphism group. This operation
 yields a Jacobian that can be determined by applying the method of Faddeev and Popov, which is valid for any 
 local symmetry group. Let us define the quantity $\Delta_{\rm FP}[g_{00}]$ by
\be
{1\over \Delta_{\rm FP}[g_{00}]} = \int_0^{+\infty}\!\!\d\ell\int_{\Diff[\hat g_{00}[\ell]]}\!\!\D\xi\; \delta\big[g_{00}-\hat g^\xi_{00}[\ell]\big]\, ,
\label{fp}
\ee
where 
$\delta$  stands for a ``functional Dirac distribution,'' 
vanishing when $g_{00}\neq \hat g^\xi_{00}[\ell]$. In other words, it vanishes when the domains of definition $[x^0_{\rm Ei},x^0_{\rm Ef}]$ and $[\xi(\hat x^0_{\rm Ei}), \xi(\hat x^0_{\rm Ef})]$  differ,
or when $g_{00}$ and $\hat g^\xi_{00}[\ell]$ differ at any instance of time. 

Using \Eq{fp} in \Eq{Psi(a0)}, we obtain
\begin{align}
\Psi(a_0)&=\int_0^{+\infty}\!\!\d\ell\int_{\Diff[\hat g_{00}[\ell]]}\!\!\D\xi \int {\D g_{00} \over \Vol(\Diff[g_{00}])}\;\delta[g_{00}-\hat g^\xi_{00}[\ell]] \;\Delta_{\rm FP}[g_{00}] \int_{\textstyle \substack{\!\!a_{\rm i}=0 \\ \,a_{\rm f}=a_0}}\!\!\D a\nonumber \\
&\;\quad  \exp\Big\{\!-\!{1\over \hbar}\int_{x^0_{\rm E i}}^{x^0_{\rm Ef}}\d x^0_{\rm E}\,\sqrt{g_{00}}\; \L_{\rm E}(g_{00},a,\partial_{x^0_{\rm E}}a)\Big\}\,, \esp 
\end{align}
where $\L_{\rm E}$ is the Euclidean Lagrangian density associated with the action $S_{\rm E}$. 
Performing the path integral over $g_{00}$ and renaming 
$a$ into $a^\xi$
leads to 
\begin{align}
\Psi(a_0)&=\int_0^{+\infty}\!\!\d\ell\int_{\Diff[\hat g_{00}[\ell]]}\!\!\D\xi \; {1\over \Vol(\Diff[\hat g^\xi_{00}[\ell]])}\; \Delta_{\rm FP}[\hat g^\xi_{00}[\ell]] \int_{\textstyle \substack{\!\!a^\xi(\hat x^{\xi0}_{\rm E i})=0 \\ \,a^\xi(\hat x^{\xi0}_{\rm E f})=a_0}}\!\!\D a^\xi  \nonumber \\
&\;\quad  \exp\Big\{\!-\!{1\over \hbar}\int_{\hat x^{\xi0}_{\rm E i}}^{\hat x^{\xi0}_{\rm Ef}}\d \hat x^{\xi 0}_{\rm E}\,\sqrt{\hat g^\xi_{00}[\ell]}\; \L_{\rm E}\big(\hat g^{\xi}_{00}[\ell],a^\xi,\partial_{\hat x^{\xi 0}_{\rm E}}a^\xi\big)\Big\}\, . \esp
\end{align}
In the above expression, the action is diffeomorphism invariant. As 
 stressed in the previous subsection,  we also use a scalar-field measure satisfying this symmetry (see \Sect{Qfluc} for an explicit construction at the semi-classical level),
\be
\D a^\xi\equiv \D a^\xi(\hat x^{\xi 0}_{\rm {\rm E}})=\D a(\hat x^0_{\rm E})\equiv \D a\, , \quad \where\quad \hat x^{\xi 0}_{\rm E}=\xi(\hat x^0_{\rm E})\, .
\ee 
$\Delta_{\rm FP}[g_{00}]$ is also invariant under the action of the diffeomorphism group since
\begin{align}
{1\over \Delta_{\rm FP}[g^\xi_{00}]} &= \int_0^{+\infty}\!\!\d\ell\int_{\Diff[\hat g_{00}[\ell]]}\!\!\D\xi'\; \delta\big[g^\xi_{00}-\hat g^{\xi'}_{00}[\ell]\big]\nonumber \\
&= \int_0^{+\infty}\!\!\d\ell\int_{\Diff[\hat g_{00}[\ell]]}\!\!\D\xi'\; \delta\big[g_{00}-\hat g^{\xi^{-1}\circ\xi'}_{00}[\ell]\big]\nonumber \\
&=  \int_0^{+\infty}\!\!\d\ell\int_{\Diff[\hat g_{00}[\ell]]}\!\!\D\xi^{\prime\prime}\; \delta\big[g_{00}-\hat g^{\xi^{\prime\prime}}_{00}[\ell]\big]={1\over \Delta_{\rm FP}[g_{00}]}\, .
\end{align}
In the second line, we use the fact that the $\delta$-functional is diffeomorphism invariant, as will be seen explicitly in the next section. Moreover, the third line follows from the change of variable $\xi^{\prime\prime}=\xi^{-1}\circ \xi'$ and the invariance of the measure, $\D\xi'=\D\xi^{\prime\prime}$, which is the case because the symmetry of  reparametrizations is anomaly free.\footnote{In string theory, the local symmetry group contains the diffeomorphisms of the two-dimensional worldsheet, as well as Weyl transformations. It is only the Weyl symmetry that develops an anomaly at the quantum level, unless the latter vanishes by imposing the target-space dimension to be critical~\cite{Pol1}.}   Finally, the group $\Diff[\hat g^\xi_{00}[\ell]]$ is   by definition diffeomorphism invariant. As a result, the wavefunction simplifies to give
\begin{align}
\Psi(a_0)&=\int_0^{+\infty}\!\!\d\ell\int_{\Diff[\hat g_{00}[\ell]]}\!\!\D\xi \; {1\over \Vol(\Diff[\hat g_{00}[\ell]])}\; \Delta_{\rm FP}[\hat g_{00}[\ell]] \int_{\textstyle \substack{\!\!a(\hat x^{0}_{\rm E i})=0 \\ \,a(\hat x^{0}_{\rm E f})=a_0}}\!\!\D a  \;  e^{-{1\over \hbar}S_{\rm E}[\hat g_{00}[\ell],a]}\, , 
\end{align}
where the action is defined on the fixed domain of the fiducial metrics
\be
S_{\rm E}\big[\hat g_{00}[\ell],a\big]=\int_{\hat x^{0}_{\rm E i}}^{\hat x^{0}_{\rm Ef}}\d \hat x^0_{\rm E}\,\sqrt{\hat g_{00}[\ell]}\; \L_{\rm E}\big(\hat g_{00}[\ell],a,\partial_{\hat x^{0}_{\rm E}}a\big)\, .
\ee 
All dependency on $\xi$ being now trivial, the path integral over the diffeomorphisms can be carried out, 
 cancelling the volume factor. The ground-state wavefunction therefore takes the form of a gauged fixed path integral 
\be
\Psi(a_0)=\int_0^{+\infty}\!\!\d\ell \; \Delta_{\rm FP}[\hat g_{00}[\ell]] \int_{\textstyle \substack{\!\!a(\hat x^{0}_{\rm E i})=0 \\ \,a(\hat x^{0}_{\rm E f})=a_0}}\!\!\D a  \; e^{-{1\over \hbar}S_{\rm E}[\hat g_{00}[\ell],a]}\, .
\ee

In the  following two sections, 
we first compute the Jacobian $\Delta_{\rm FP}[\hat g_{00}[\ell]]$ and then evaluate the path integral over the scale factor.


\section{Faddeev--Popov Jacobian}
\label{FPdet}

 Actually, the Faddeev--Popov determinant $\Delta_{\rm FP}[g_{00}]$ 
 only depends on the equivalence class of its argument, {thanks to its invariance under the diffeomorphism group.} In what follows, however, we choose to 
 keep the notation $\Delta_{\rm FP}[\hat g_{00}[\ell]]$. Our goal is to compute this Jacobian by expressing it as a path integral over 
 ghost fields.


\subsection{Introducing ghost fields}
\label{gsect}

Since the reversal of the orientation of the segment is the only isometry, the path integral appearing in the expression
\be
{1\over \Delta_{\rm FP}[\hat g_{00}[\ell]]} = \int_0^{+\infty}\!\!\d\ell'\int_{\Diff[\hat g_{00}[\ell']]}\!\!\D\xi\; \delta\big[\hat g_{00}[\ell]-\hat g^\xi_{00}[\ell']\big]
\label{defD}
\ee
 is twice the contribution of the diffeomorphisms connected to the identity,~
\be
\int_{\Diff[\hat g_{00}[\ell']]}\!\!\D\xi ~~\longrightarrow ~~2\int_{\Diff[\hat g_{00}[\ell']]_{\Id}}\!\!\D\xi\, .
\ee
In that case, the $\delta$-functional enforces only $\ell'=\ell$ and $\xi=\Id$. In the vicinity of this point, the total variation of the fiducial metric is 
\begin{align}
\delta \hat g_{00}[\ell]&\equiv \hat g^{\Id+\delta\xi}_{00}[\ell+\delta\ell]-\hat g_{00}[\ell]=\delta_\xi \hat g_{00}[\ell]+\partial_\ell \hat g_{00}[\ell]\delta\ell+\cdots\nonumber \\
&=-2\hat \nabla_0\delta x_{\rm E0}+2\hat g_{00}[\ell]\,{\delta\ell\over \ell}+\cdots\, , \esp
\end{align}
where we have used \Eqs{xiloc} and~(\ref{l2}), while $\hat \nabla$ is the covariant derivative with respect to $\hat g_{00}[\ell]$. As a result, we obtain 
\be
{1\over \Delta_{\rm FP}[\hat g_{00}[\ell]]} = 2\int\d\delta\ell \int_{\textstyle \substack{\;\delta x^0_{\rm E}(\hat x^0_{\rm Ei})=0 \\ \; \delta x^0_{\rm E}(\hat x^0_{\rm Ef})=0}}   \D\delta x^0_{\rm E}\; \delta\Big[2\hat \nabla_0\delta x_{\rm E0}-2\hat g_{00}[\ell]{\delta\ell\over \ell}\Big]\,,
\ee
where the diffeomorphisms $\delta x^0_{\rm E}$ are required to vanish at the boundaries  $\hat x^0_{\rm Ei}$ and $\hat x^0_{\rm Ef}$ of the domain of $\hat g_{00}[\ell]$.  Otherwise the $\delta$-functional 
implies that they do not contribute at linear order. 

By writing the $\delta$-functional as a Fourier transform, the previous expression becomes
\begin{align}
{1\over2\, \Delta_{\rm FP}[\hat g_{00}[\ell]]} =& \int\d\delta\ell \int_{\textstyle \substack{\;\delta x^0_{\rm E}(\hat x^0_{\rm Ei})=0 \\ \; \delta x^0_{\rm E}(\hat x^0_{\rm Ef})=0}}   \D\delta x_{\rm E}\int\D \beta \nonumber \\
& \exp\Big\{2i\pi  \int_{\hat x^0_{\rm Ei}}^{\hat x^0_{\rm Ef}} \d \hat x^0_{\rm E}\, \sqrt{\hat g_{00}[\ell]}\; \beta^{00}\Big(2\hat \nabla_0\delta x_{\rm E0}-2\hat g_{00}[\ell]{\delta\ell\over \ell}\Big)\Big\}\,,\esp
\label{integ}
\end{align}
where the measure $\D \beta$ is assumed to be diffeomorphism invariant for the  $\delta$-functional to be invariant as well. Notice that we have not made any reference to the contravariant or covariant nature of the tensors in the measures $\D\delta x_{\rm E}$ and $\D\beta$  because the contractions in the argument of the exponential can be made in an arbitrary way. We justify  at the end of the section why this does not introduce ambiguities in the measures. 

At this stage, it is relevant to introduce a notation that will be used extensively in the following. Let $f^{0\cdots0}$ and  $h^{0\cdots0}$ be two tensors with $m\in\Z$ contravariant indices, where $m<0$ actually means that the tensors have $|m|$ covariant indices. We define the reparametrization-invariant quantity   
\be
(f, h)_\ell\equiv \int_{\hat x^0_{\rm Ei}}^{\hat x^0_{\rm Ef}} \d \hat x^0_{\rm E}\, \sqrt{\hat g_{00}[\ell]}\; f^{0\cdots0}\, h^{0\cdots0}\, \hat g_{00}[\ell]^m\, ,
\label{defin}
\ee
which we will especially use for $m=2$, $m=0$  and $m=-1$.
To compute explicitly $ \Delta_{\rm FP}[\hat g_{00}[\ell]]$ it is convenient to switch all integration variables into Grassmann ones, 
\be
\delta \ell  \longrightarrow \lambda\, ,\quad~~ \delta x_{\rm E0}(\hat x^0_{\rm E})  \longrightarrow c_0(\hat x^0_{\rm E})\, , \quad~~ \beta^{00}(\hat x^0_{\rm E})  \longrightarrow b^{00}(\hat x^0_{\rm E})\, , 
\ee
where $b^{00}$ and $c_0$ are ghost fields. This operation inverts the r.h.s. of \Eq{integ} up to an irrelevant numerical factor $\alpha$~\cite{Pol1}. Hence, we obtain  
\begin{align}
2\, \Delta_{\rm FP}[\hat g_{00}[\ell]] &= \alpha \int_{\textstyle \substack{\,c^0(\hat x^0_{\rm Ei})=0 \\\, c^0(\hat x^0_{\rm Ef})=0}}   \D c\int\D b\int\d\lambda \; \exp\Big\{2i\pi \Big(b,2\hat \nabla c-2\hat g[\ell]{\lambda\over \ell}\Big)_\ell\Big\}\nonumber \\
& = 4i\pi \alpha  \int_{\textstyle \substack{\,c^0(\hat x^0_{\rm Ei})=0 \\\, c^0(\hat x^0_{\rm Ef})=0}}   \D c\int\D b \, \Big(b,{\hat g[\ell]\over \ell}\Big)_\ell\,\exp\Big\{4i\pi \,(b,\hat \nabla c)_\ell\Big\}\, ,
\label{expre}
\end{align}
where the second equality follows by Berezin integration over $\lambda$. 


\subsection{Choice of fiducial metric and mode expansions}
\label{cfm}

Even though the content of this subsection can be  obtained
without specifying the fiducial metric, we find  it simpler to present it for a convenient 
 choice, remembering that $\Delta_{\rm FP}[\hat g_{00}[\ell]]$ is independent of this choice. 
 So let us choose a constant lapse function,
\be
\hat g_{00}[\ell](\tau)=\ell^2\quad  \mbox{defined on} \quad [\hat x^0_{\rm Ei} ,\hat x^0_{\rm Ef} ]= [0,1]\, ,
\label{lapseC}
\ee
where the variable $\hat x^0_{\rm E}$ is denoted $\tau$  for the sake of simplicity, while $\hat \nabla\equiv \partial_0$. Notice that $\tau$ 
is 
 proportional to the ``cosmological Euclidean time $t_{\rm E}$,'' which satisfies $\d t_{\rm E}=\ell\, \d\tau$. 

In \Eq{integ}, the paths  $\beta^{00}(\tau)$ are arbitrary  functions in $L^2([0,1])$, the vectorial space of functions  that are square-integrable on $[0,1]$. Therefore, they can be expanded  in the basis  $\big\{1/\sqrt{2}, \cos(k\pi \tau), k\in\natural^*\big\}$.  Likewise, the paths  $\delta x_{\rm E0}(\tau)$ are functions in $L^2([0,1])$ obeying the boundary conditions $\delta x_{\rm E0}(0)=0$ and $\delta x_{\rm E0}(1)=0$. Hence, they can be expanded in the basis $\big\{\sin(k\pi \tau), k\in\natural^*\big\}$.\footnote{The domain of definition of $\beta^{00}$ can be extended to $\R$ by imposing $\beta^{00}$ to be even on $[-1,1]$ and \mbox{2-periodic}. Similarly, the domain of definition of $\delta x_{\rm E0}$ can be extended to $\R$ by imposing $\delta x_{\rm E0}$ to be odd on $[-1,1]$ and  \mbox{2-periodic}. As a result, they can be expanded in the Fourier basis $\{1/\sqrt{2}, \cos(k\pi \tau), \sin(k\pi \tau), k\in\natural^*\}$ restricted to the even modes for $\beta^{00}$ and odd modes for $\delta x_{\rm E0}$. } 
In order to keep track of the tensorial natures of the paths, it is preferable to view the vectorial spaces as Hilbert spaces respectively equipped with the inner products
\be
(f, h)_\ell=\int_0^1 \d\tau \ell \, f^{00}(\tau)h^{00}(\tau)\, (\ell^2)^2\, , \qquad (f,h)_\ell=\int_0^1 \d\tau \ell \, f_{0}(\tau)h_{0}(\tau)\, \ell^{-2}\, .
\label{inner}
\ee
In that case, we can 
expand  $\beta^{00}$, $\delta x_{\rm E0}$ and the ghost fields  in orthonormal basis, 
\begin{align}
\beta^{00}(\tau) &= \sum_{k\ge 0} \boldsymbol{\beta}_k\,  \chi^{00}_k(\tau) \, , \qquad  \delta x_{\rm E0}(\tau) = \sum_{k\ge 1} \boldsymbol{\gamma}_k\,  \sigma_{0,k}(\tau) \,,\nonumber \\
b^{00}(\tau) &= \sum_{k\ge 0} \b_k\,  \chi^{00}_k(\tau) \,, \qquad ~~~\, c_{0}(\tau) = \sum_{k\ge 1} \c_k\,  \sigma_{0,k}(\tau) \,,
\end{align} 
where  $\boldsymbol{\beta}_k$, $\boldsymbol{\gamma}_k$ and anticommuting $\b_k$ and $\c_k$  are arbitrary constants, while   
\begin{align}
\chi_0^{00}(\tau)=\ell^{-2}\, {1\over \sqrt{\ell}}\, ,\quad   \chi_k^{00}(\tau)&=\ell^{-2}\sqrt{2\over \ell}\, \cos(k\pi \tau)\, , ~~ k\in \natural^*\, ,\nonumber \\
\sigma_{0,k}(\tau)&=\ell\,\sqrt{2\over \ell}\, \sin(k\pi \tau)\, , ~~~~\, k\in \natural^*\, ,
\end{align}
satisfy
\be
(\chi_k,\chi_{k'})=\delta_{kk'}\, , \qquad (\sigma_{k},\sigma_{k'})=\delta_{kk'}\, .
\ee

Using these conventions, we first obtain two pieces of \Eq{expre} which are
\be
 \Big(b,{\hat g[\ell]\over \ell}\Big)_\ell= {\b_0\over \sqrt{\ell}}\, , \qquad 4i\pi (b,\hat \nabla c)_\ell={4i\pi^2\over \ell} \sum_{k\ge 1}k\, \b_k\c_k\, .
 \label{pieces}
\ee
Next, we determine the correctly normalized path-integral measures of the ghosts. In the case of commuting tensors $\beta^{00}$ and $\delta x_{\rm E0}$, the norms associated with the inner products in \Eq{inner} 
yield
\begin{align}
||\beta||_\ell^2 &\equiv  (\beta,\beta)_\ell = \sum_{k\ge 0} \boldsymbol{\beta}_k^2 ~~~~~~\quad\Longrightarrow~~~\quad\D\beta = \bigwedge_{k\ge 0}\d \boldsymbol{\beta}_k\, , \nonumber \\
||\delta x_{\rm E}||_\ell^2 &\equiv (\delta x_{\rm E},\delta x_{\rm E})_\ell \,= \sum_{k\ge 1} \boldsymbol{\gamma}_k^2\quad\Longrightarrow\quad \D \delta x_{\rm E}=\bigwedge_{k\ge 1} \d\boldsymbol{\gamma}_k\, . 
\label{meacom}
\end{align} 
In the anticommuting case of interest, we thus adopt the analogous prescriptions,
 \be
\D b = \prod_{k\ge 0} \d\b_k\, , \qquad \D c = \prod_{k\ge 1}\d\c_k\, .
\label{meaanti}
\ee 
Taking into account these results, \Eq{expre} becomes, up to an ambiguous sign that depends on the precise ordering of the Grassmann integration variables, 
\begin{align}
\Delta_{\rm FP}[\ell^2] &=  2i\pi\alpha \int \d\b_0\, {\b_0\over \sqrt{\ell}}\; \prod_{k\ge 1}\int \d\c_k\, \d \b_k \; e^{{4i\pi^2\over \ell} k\,\b_k\c_k}\nonumber \\
&={2i\pi\alpha\over \sqrt{\ell}} \prod_{k\ge 1}\! \Big({4i\pi^2\over \ell}\, k\Big)\nonumber \\
& = \alpha\,  \sqrt{2i\pi}\, , 
\label{compu}
\end{align}
where the last equality follows by the zeta regularization formulas 
\be
\prod_{k\ge 1} z={1\over \sqrt{z}}\, , \qquad \prod_{k\ge 1} k = \sqrt{2\pi}\, .
\ee
Remembering that $\Delta_{\rm FP}[\ell^2]$ is actually a Jacobian,  $\alpha$ must be such that $\Delta_{\rm FP}[\ell^2]$ is a real non-negative number. 

The important outcome of the above calculation is that the Jacobian is a trivial constant, \ie it is {\it independent} of $\ell$. Note that this is not something that could be inferred at the outset, as this result applies to the case of a base manifold that is a segment in a specific way. Indeed, one can check that  in the case of a base manifold with the  topology of a circle, the Jacobian proves to be a constant times $1/\ell$, where $\ell$ is again the proper length of the 
circle~\cite{Lihui}.\footnote{This is the reason of the presence of a dressing $\int_0^{+\infty}\d\ell/\ell$  in one-loop computations performed in first-quantized formalism. To derive it,  one has to take into account the fact that $\Diff[\hat g_{00}[\ell]]$ contains a subgroup ${\rm CKG}[\hat g_{00}[\ell]]$, which is  the conformal Killing group generated by the translations of Euclidean time. } 

 To conclude this section, let us mention that since this is important, we make even clearer the independence of the Faddeev--Popov Jacobian on the choice of fiducial metric in Appendix~\ref{A4}. Moreover, we would like to justify the  claim below \Eq{integ}. The positions of the indices of $\beta^{00}$ and $\delta x_{\rm E0}$ appearing in the exponential in \Eq{integ} have been chosen arbitrarily, since in all cases we would have been able to contract them by using the metric $\hat g_{00}[\ell]$. Hence, the tensor structure of the $b$-ghost could have been defined as $b^{00}$, $b^0_0$ or $b_{00}$, and likewise for the $c$-ghost as $c^0$ or $c_0$. However, since the mode expansions and the norms are always defined by using the ``universal'' scalar product in \Eq{defin}, the relations given in  \Eqs{expre},~(\ref{pieces}),~(\ref{meacom}),~(\ref{meaanti}) are always valid and the final result of the Jacobian  is unchanged. Therefore, there is no ambiguity in denoting the measures and scalar products without  specifying
which tensor  structures are implicitly chosen.


\section{Scale-factor path integral and modulus integral} 
\label{path-a}

Having implemented the gauge fixing of Euclidean-time  reparametrizations, we now proceed to compute the path integral over the scale factor. We will work in the simplest gauge where the lapse function is constant, as displayed in \Eq{lapseC}. The expression of the wavefunction becomes
\be
\Psi(a_0)=\Delta_{\rm FP} \int_0^{+\infty}\!\!\d\ell   \int_{\textstyle \substack{\!\!a(0)=0 \\ \,a(1)=a_0}}\!\!\D a  \; e^{-{1\over \hbar}S_{\rm E}[\ell^2,a]}\, ,
\label{psa0}
\ee
where $\Delta_{\rm FP}$ is an irrelevant constant  and the action~(\ref{action3}) is 
\be
S_{\rm E}[\ell^2,a]=3sv_3\int_0^1\d \tau \,\bigg[ {a\over \ell} \Big({\d a\over \d \tau}\Big)^2+\ell\, V(a)\bigg]\, .
\label{SE}
\ee 
 Since the action is not quadratic, we will  make use of the method of  steepest-descent to approximate the evaluation of the wavefunction. In practice, one has to expand $S_{\rm E}[\ell^2,a]$ around its extrema and integrate over the fluctuations at quadratic order only. The validity of this approximation is guaranteed when $\hbar\to 0$, which corresponds to the semiclassical limit.   We would like to stress that our treatment to come differs to what is done in previous works~\cite{Halli,Halli2,Turok1,Halli-Hartle}, as will be explained in \Sect{cpw}.


\subsection{Instanton solutions}

In order to apply the method of steepest-descent, we extremize the action with respect to both $\ell$ and $a(\tau)$. Denoting such an extremum as $(\bell^2,\ba)$, we first have to solve
\be
0=\!\left.{\d S_{\rm E}\over \d \ell}\right|_{(\bell^2,\ba)}=3sv_3\int_0^1\d \tau \,\bigg[ \!-\!{\ba\over \bell^2} \Big({\d \ba\over \d \tau}\Big)^2+ V(\ba)\bigg]\, .
\label{e1}
\ee
Moreover, writing 
\begin{align}
a(\tau)&=\bar a(\tau)+\delta a(\tau)\, , \nonumber\\
\where \quad \ba(0)&=0\, , ~~ \ba(1)=a_0 \quad ~~\and ~~\quad \delta a(0)=0\, , ~~\delta a(1)=0\, , 
\label{BC}
\end{align}
the equation of motion for the scale factor reads
\be
0=\!\left.{\delta S_{\rm E}\over \delta a}\right|_{(\bell^2,\ba)}~\Longleftrightarrow~
-{3\over \ba ^2}\Big({\d \ba \over \d\tau}\Big)^2-2{\d\over \d\tau}\Big({1\over \ba}\, {\d \ba\over \d\tau}\Big)+{\bell^2\over \ba^2}\,V_a(\ba)=0\, ,
\label{e2}
\ee
where $V_a\equiv \d V/\d a$.

Particular solutions of \Eq{e1} can be found by imposing the integrand to vanish identically, which amounts to solving the Friedmann equation. However, one may wonder whether more general solutions  for which the integral vanishes while the integrand  is non-zero are possible. 
It turns out that this is not the case. This can be  inferred by noticing that the equation of motion for the scale factor is equivalent to
 \be
- {\ba\over \bell^2} \Big({\d \ba\over \d \tau}\Big)^2+ V(\ba) = {\E\over 3v_3}\, , 
\ee 
where $\E$ is an arbitrary integration constant, which can be related to the total energy of the universe.   Using this relation, \Eq{e1} gives $0=s\E$, showing that 
 it suffices to solve the Friedmann equation in order to extremize the action with respect to $\ell$ and $a(\tau)$. 

\begin{figure}
\captionsetup[subfigure]{position=t}
\begin{center}
\begin{subfigure}[h!]{0.48\textwidth}
\begin{center}
\includegraphics [scale=0.38]{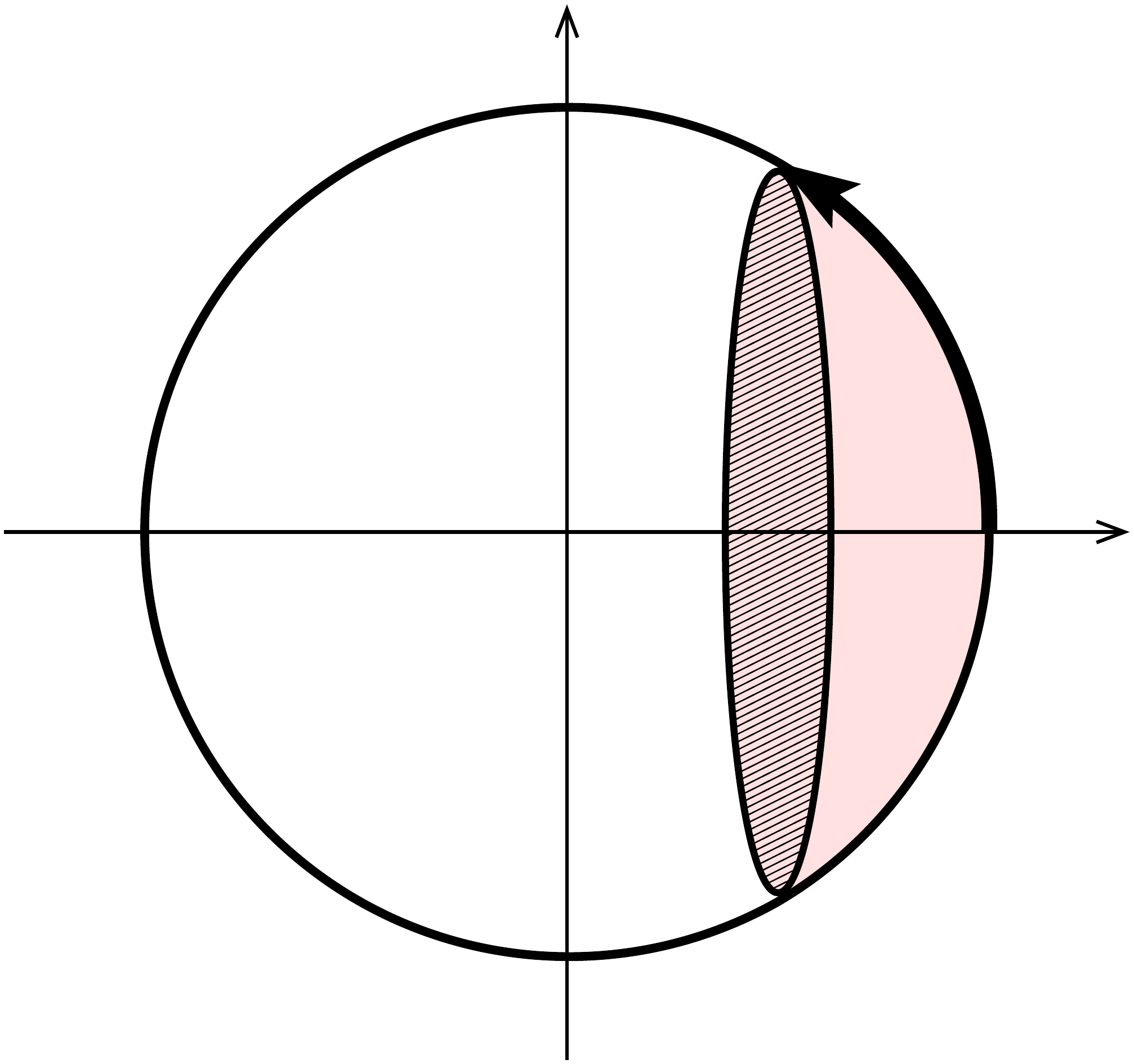}
\put(-32,143){$\lambda\bell_+\tau$}
\put(-24,90){${1\over\lambda}$}
\put(-126,172){$a_0-$}
\end{center}
\caption{\it \footnotesize Instanton solution $\ba_+$ corresponding to the ``small cap.'' }
\label{smaller}
\end{subfigure}
\quad
\begin{subfigure}[h!]{0.48\textwidth}
\begin{center}
\includegraphics [scale=0.38]{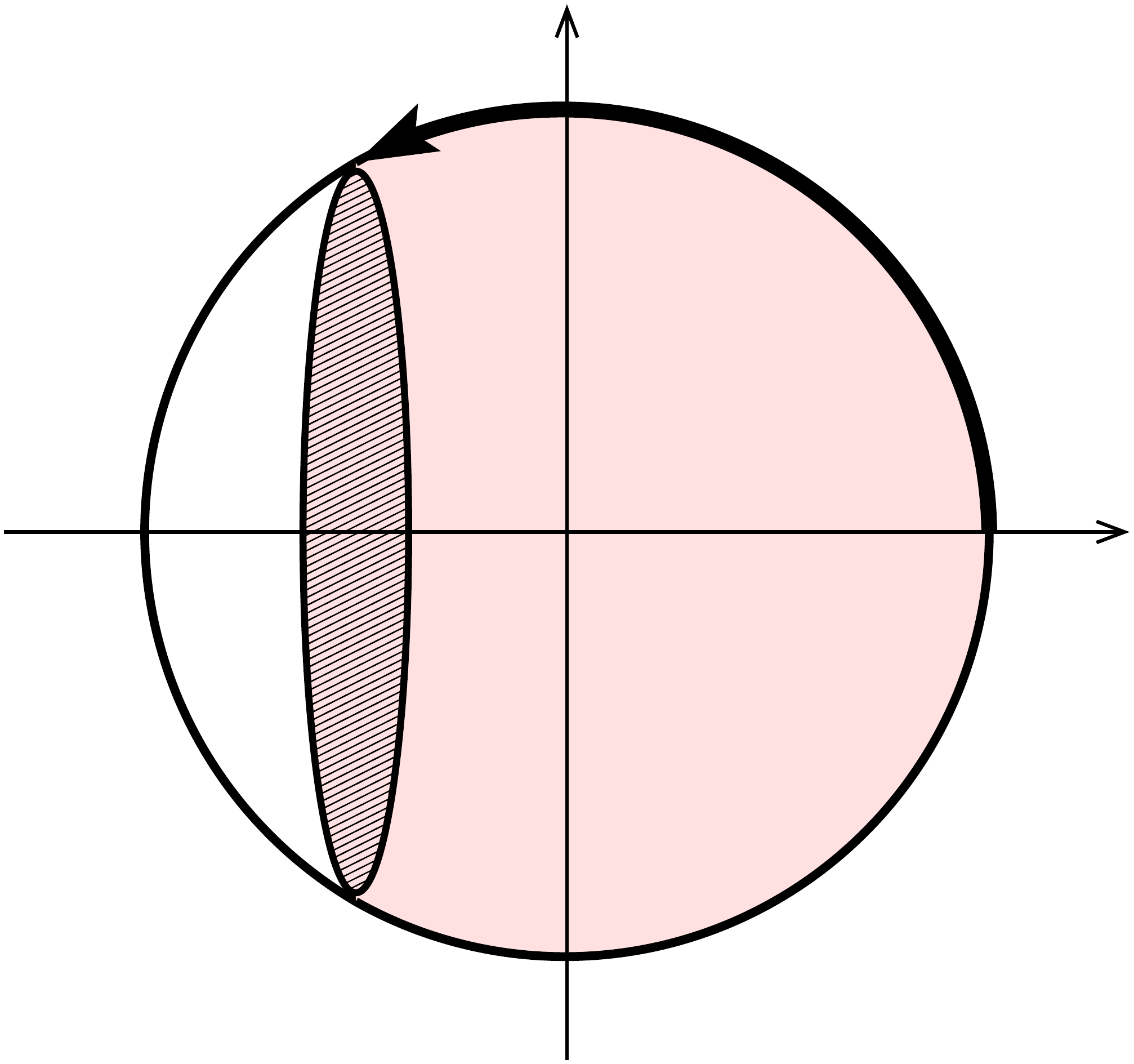}
\put(-54.5,170){$\lambda\bell_-\tau$}
\put(-24,90){${1\over\lambda}$}
\put(-126,172){$a_0-$}
\end{center}
\caption{\it \footnotesize Instanton solution $\ba_-$ corresponding to the `\mbox{`big cap.''}  }
\label{large}
\end{subfigure}
\caption{\it \footnotesize The instanton solutions are portions of the 4-sphere of radius $1/\lambda$.}
\label{4sphere}
\end{center}
\end{figure}
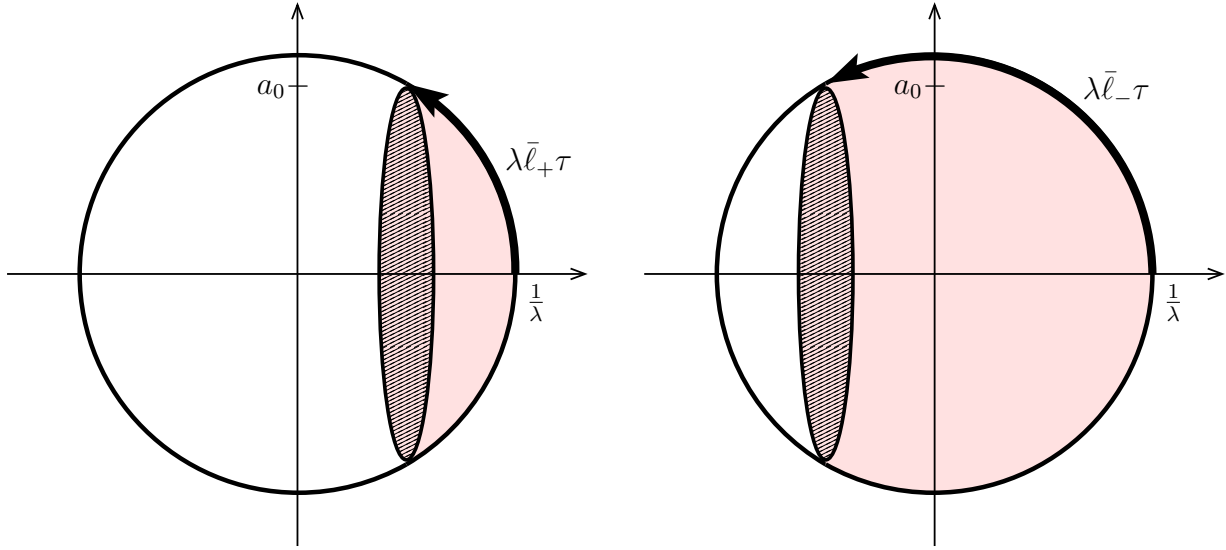

 The latter equation takes the form 
\be
\left({\d(\lambda \ba)\over \d(\lambda \bell \tau)}\right)^2+(\lambda \ba)^2=1\, , 
\label{eqE}
\ee
 with solution $\lambda \ba(\tau)=|\sin(\lambda \bell \tau+\cst)|$. Imposing $\ba(0)=0$ implies the constant to be zero, while the condition $\ba(1)=a_0$ fixes the modulus $\bell$ of the base segment. When
\be
\lambda a_0<1\, , 
\label{assume}
\ee 
$\bell$ can take discrete real  values. However only two yield smooth instanton solutions,
\begin{align}
\lambda \ba_\epsilon (\tau)&=\sin(\lambda  \bell_\epsilon\tau)\, , \quad \epsilon\in\{+1,-1\}\, , \nonumber\\
\where\quad~~ \lambda \bell_+ &= \arcsin(\lambda a_0)\, , \quad\lambda \bell_- = \pi- \arcsin(\lambda a_0)\, .\esp
\label{aepsilon}
\end{align}
The line elements of the associated  Euclidean  spacetimes are
\be
(\d \bar s_{\rm E}^\epsilon)^2={1\over \lambda^2}\Big[\d(\lambda \bell_\epsilon\tau)^2+\sin(\lambda \bell_\epsilon\tau)^2\, \d\Omega^2_3\Big]\,,\quad \tau\in[0,1]\, , 
\ee
 showing that the instantons describe portions of a 4-sphere of radius $1/\lambda$. Actually, $\epsilon=+1$ corresponds to a cap smaller than a hemisphere (see Fig.~\ref{smaller}), while $\epsilon=-1$ describes a cap bigger than a hemisphere (see Fig.~\ref{large}).
The actions associated with these solutions are\footnote{Other instanton solutions exist, which are continuous but not differentiable,
$$
\lambda \ba_{\epsilon,n}(\tau)=|\sin(\lambda \bell_{\epsilon,n} \tau)|\, ,  \quad \where \quad \bell_{\epsilon,n}=\bell_\epsilon+n\pi\, ,\quad n\in\natural^*\, . 
$$
The  associated four-dimensional Euclidean  spacetimes
look like  ``necklaces'' composed of $n$ beads (4-spheres).  The series terminates 
with a small or big cap. However, if one restricts the path integral sum to include
smooth four-manifolds only, these solutions should be excluded. It is nevertheless  interesting to relax this requirement,
as in \Refe{pearls}. 
 Then because the actions of these Euclidean solutions  are $$\bar S_{\rm E}^{\epsilon,n} = s\, {2v_3\over \lambda^2}\Big[2n+1-\epsilon \big(1-(\lambda a_0)^2\big)^{3\over 2}\Big]\, , $$ 
 the associated corrections to the wavefunction 
can be absorbed in $\O(\hbar)$ terms arising from the main instanton solutions  of \Eq{aepsilon}. }
\be
\bar S_{\rm E}^\epsilon = s\, {2v_3\over \lambda^2}\Big[1-\epsilon \big(1-(\lambda a_0)^2\big)^{3\over 2}\Big]\, .
\label{acInst}
\ee
In 
 this section, the semi-classical approximation of the wavefunction $\Psi(a_0)$ will be  carried out
by applying the steepest-descent method  to evaluate the path integral 
 when $0<\lambda a_0<1$. 
 The expression 
in the regime $\lambda a_0>1$ will be determined  to some extent in \Sect{WDW} by  analytically continuing
the solutions of the Wheeler--DeWitt equation.\footnote{For $\lambda a_0>1$, the instanton solutions are complex, since $\lambda \bell_\epsilon=\pi/2+i \epsilon  \arccosh(\lambda a_0)$. However, the contour of integration of $\ell$ can be deformed thanks to Cauchy's integral theorem in order to pass through these points when applying the steepest-descent method. } 


\subsection{Expansion  to quadratic order}

The action~(\ref{SE}) can be expanded around the  extremal
solutions $ (\bell_\epsilon^2, \ba_\epsilon)$. 
 Setting
\be
\ell=\bell_\epsilon+\delta \ell\, ,  ~~\quad a(\tau)=\ba_\epsilon (\tau)+\delta a(\tau)\, , 
\ee
where $\delta a(\tau)$ obeys the boundary conditions  of
\Eq{BC}, one obtains  to quadratic order
\begin{align}
S_{\rm E}[\ell^2,a]=\bar S_{\rm E}^\epsilon+3sv_3\int_0^1\d \tau \bell_\epsilon \Big[\delta a\, \Q_\epsilon \delta a + 2 \,\delta a\, V_a(\ba_\epsilon)\,{\delta\ell\over \bell_\epsilon}+ {\delta\ell\over \bell_\epsilon}\, V(\ba_\epsilon)\, {\delta\ell\over \bell_\epsilon}\Big]\!+\O(\delta^3)\, .
\label{qua}
\end{align}
 All linear terms vanish thanks to the equations of motion and $\O(\delta^3)$ stand for terms at least cubic in fluctuations. In the above equation, $\Q_\epsilon$ is  a quadratic differential operator,
\be
\Q_\epsilon = -{\ba_\epsilon\over \bell_\epsilon^2}\, {\d^2\over \d\tau^2}+{1\over \bell_\epsilon^2}\, {\d\ba_\epsilon\over \d \tau}\, {\d\over \d\tau}+{1\over 2}\, V_{aa}(\ba_\epsilon)\, .
\ee

Let us view the fluctuation $\delta a(\tau)$ as an 
element  in the Hilbert space of  $L^2([0,1])$ functions, obeying the boundary conditions $\delta a(0)=0$, $\delta a(1)=0$.  This space is equipped with the inner product (see \Eq{defin}) 
\be
(\delta a_1,\delta a_2)_{\bell_\epsilon}=\int_0^1\d\tau\bell_\epsilon\, \delta a_1\, \delta a_2\, .
\label{inprod}
\ee
The linear operator $\Q_\epsilon$ can be seen as an endomorphism of this Hilbert space.\footnote{This statement is not quite obvious. Indeed, if a function $f$ vanishes at 0 and 1, there is \apriori no reason for its derivatives to vanish at 0 and 1. Hence, the action of $\Q_\epsilon$ on the Hilbert space may  take
us outside of it. However, all vectors  in the Hilbert space can be expanded  in the (orthonormal) Fourier basis $\Big\{\sqrt{2/\bell_\epsilon}\sin(k\pi\tau), k\in\natural^*\Big\}$.  
 Requiring the derivatives of such Fourier series to vanish at 0 and 1 (this is at the price of 
possible discontinuities at 0 and 1), they can be expanded in the above basis and  the action of the operator can be viewed as internal.} Its adjoint, denoted $\Q_\epsilon^\T$, satisfies by definition 
\be
(\delta a_1,\Q_\epsilon\delta a_2)_{\bell_\epsilon}=(\Q_\epsilon^\T\delta a_1,\delta a_2)_{\bell_\epsilon}
\ee
and can be determined by integrating by parts the left hand side,
\be
\Q_\epsilon^\T = -{\ba_\epsilon\over \bell_\epsilon^2}\, {\d^2\over \d\tau^2}-{3\over \bell_\epsilon^2}\, {\d\ba_\epsilon\over \d \tau}\, {\d\over \d\tau}-{2\over \bell_\epsilon^2}\, {\d^2\ba_\epsilon\over \d \tau^2}+{1\over 2}\, V_{aa}(\ba_\epsilon)\, .
\ee
Let us then introduce  the ``symmetric'' and ``antisymmetric'' parts of 
$\Q_\epsilon$, 
\be
\S_\epsilon = {\Q_\epsilon+\Q^\T_\epsilon\over 2}\, , ~~\quad \A_\epsilon = {\Q_\epsilon-\Q^\T_\epsilon\over 2}\, .
\ee
They satisfy
\be
(\delta a_1,\S_\epsilon\delta a_2)_{\bell_\epsilon}=(\S_\epsilon\delta a_1,\delta a_2)_{\bell_\epsilon}\, ,~~ \quad (\delta a_1,\A_\epsilon\delta a_2)_{\bell_\epsilon}=-(\A_\epsilon\delta a_1,\delta a_2)_{\bell_\epsilon}\, ,
\ee
which means that $\S_\epsilon$ is self-dual while $\A_\epsilon$ satisfies $(\delta a,\A_\epsilon\delta a)_{\bell_\epsilon}=0$ for all $\delta a$. As a result, we  obtain 
\be
\delta a\, \Q_\epsilon \delta a=\delta a\, (\S_\epsilon+\A_\epsilon) \delta a = \delta a\, \S_\epsilon \delta a\, .
\label{QS}
\ee
Using the explicit expression of the instanton solution in \Eq{aepsilon}, the differential operator $\S_\epsilon$ can be  written in the following form,   
\be
\S_\epsilon = -{\ba_\epsilon\over \bell_\epsilon^2}\, {\d^2\over \d\tau^2}-{1\over \bell_\epsilon^2}\, {\d\ba_\epsilon\over \d \tau}\, {\d\over \d\tau}-2\lambda^2\ba_\epsilon\, .
\ee

In the next subsection, we will show that the operator $\S_\epsilon$ is invertible when\footnote{Actually, it is a regularized version of $\S_\epsilon$ that will satisfy this property. } 
\be
0<\lambda a_0<1\, .
\ee
Assuming for the moment this fact, we can diagonalize the integrand appearing in \Eq{qua} when $\lambda a_0$ is restricted  in the above range,
 \begin{align}
\delta a\, \S_\epsilon \delta a + 2 \,\delta a\, V_a(\ba_\epsilon)\,{\delta\ell\over \bell_\epsilon}+ {\delta\ell\over \bell_\epsilon}\, V(\ba_\epsilon)\, {\delta\ell\over \bell_\epsilon}&= \delta \check a\, \S_\epsilon \delta \check a +{\delta\ell\over \bell_\epsilon} \big[V(\ba_\epsilon)-V_a(\ba_\epsilon)\S_\epsilon^{-1} V_a(\ba_\epsilon)\big]{\delta\ell\over \bell_\epsilon}\, , \nonumber \\
\where \quad \delta \check a &= \delta a+{\delta\ell\over \bell_\epsilon}\, \S_\epsilon^{-1}V_a(\ba_\epsilon)\, .\esp 
\label{atilde}
\end{align}
The above expressions deserve a comment. {\em A priori}, $V_a(\ba_\epsilon)$ is a function which  satisfies  
\be
V_a(\ba_\epsilon) = 1-3(\lambda \ba_\epsilon)^2\quad \Longrightarrow\quad V_a(\ba_\epsilon(0))=1\, ,~~ V_a(\ba_\epsilon(1))=1-3(\lambda a_0)^2\, .
\label{pb}
\ee
From this point of view, it does not belong to the Hilbert space of functions vanishing at $0$ and $1$, on which the endomorphisms $\S_\epsilon$ and $\S^{-1}_\epsilon$ act. However, we may consider this function on the open set $(0,1)$,  and extend it so as to be odd on $(-1,0)\cup (0,1)$, and then 2-periodic on $\R\setminus \Z$. The  Fourier series of the resulting function involves only the modes $\big\{\sqrt{2/\bell_\epsilon}\sin(k\pi\tau), k\in\natural^*\big\}$ and vanishes at every $\tau\in\Z$, where the function is discontinuous. Hence, in \Eq{atilde}, $V_a(\ba_\epsilon)$ should be 
 thought of as its Fourier expansion in the above basis, which indeed belongs to the Hilbert space. 

Applying the steepest-decent method, the wavefunction in \Eq{psa0} reads\footnote{Since the actions $\bar S_{\rm E}^\epsilon$ approach each other when $\lambda a_0$ approaches 1, we do not absorb one of the two instanton contributions in the corrections $\O(\hbar)$.}
 \begin{align}
\Psi(a_0)=&\; \Delta_{\rm FP}\sum_{\epsilon=\pm1 } e^{-{1\over \hbar} \bar S_{\rm E}^\epsilon} \int \d\delta \ell \,\exp\Big\{\!-\!{3sv_3\over \hbar}\,   \K_\epsilon  \Big({\delta\ell\over \bell_\epsilon}\Big)^2\Big\}\nonumber \\
&\; \int_{\textstyle \substack{\;\delta a(0)=0 \\\; \delta a(1)=0}}\D \delta a  \, \exp\Big\{\!-\!{3sv_3\over \hbar}\,( \delta \check a, \S_\epsilon \delta \check a)_{\bell_\epsilon}\Big\}\,(1+\O(\hbar))\, ,
\end{align} 
where we have defined
\be
  \K_\epsilon = \int_0^1\d\tau \bell_\epsilon\, \big[V(\ba_\epsilon)-V_a(\ba_\epsilon)\S_\epsilon^{-1} V_a(\ba_\epsilon)\big]\, .
 \label{barI}
\ee
Since $\S_\epsilon^{-1}V_a(\ba_\epsilon)$ belongs to the Hilbert space,  so does $\delta\check a$.  As a result, we may perform a change of variable (actually field redefinition) in the path integral from $\delta a$ to $\delta \check a$, with identical vanishing boundary conditions at $\tau=0$ and $\tau=1$.  Furthermore, taking into account that the measures are related to each other by a trivial Jacobian
 \be
\D\delta a= \D\delta \check a \, \left|{\D\delta a\over \D\delta \check a}\right|= \D\delta \check a\, ,
\ee
we obtain that
\begin{align}
\Psi(a_0)&=\Delta_{\rm FP}\sum_{\epsilon=\pm 1} e^{-{1\over \hbar} \bar S_{\rm E}^\epsilon}\,Z_\epsilon(a_0) \int \d\delta \ell \,\exp\Big\{\!-\!{3sv_3\over \hbar}\,  \K_\epsilon  \Big({\delta\ell\over \bell_\epsilon}\Big)^2\Big\}\,(1+\O(\hbar))\nonumber \\
\where \quad Z_\epsilon(a_0)&= \int_{\textstyle \substack{\;\delta \check a(0)=0 \\ \;\delta \check a(1)=0}}\D \delta \check a  \, \exp\Big\{\!-\!{3sv_3\over \hbar}\,( \delta \check a, \S_\epsilon \delta \check a)_{\bell_\epsilon}\Big\}\, .\esp
\label{psii}
\end{align}


\subsection{Scale-factor quadratic fluctuations}
\label{Qfluc}

For the sake of simplicity we make the change of notation $ \delta\check a\to \delta a$ in the above expression of $Z_\epsilon(a_0)$. Our goal in the present subsection is to compute this contribution to the wavefunction. 


\subsubsection{Mode expansion}
\label{moexp}

Since $\S_\epsilon$ is a self-dual endomorphism of the Hilbert space, it is diagonalizable in an orthonormal basis. Denoting $\phi^\epsilon_k$ the eigenvector  with eigenvalue $\nu_k^\epsilon$, we have\footnote{We will show  later on that the eigenspaces are of dimension 1, so that an index $k$ is sufficient to label the eigenvectors unambiguously. Moreover, this label takes values in $\natural^*$ since $\Big\{\sqrt{2/\bell_\epsilon}\sin(k\pi\tau), k\in\natural^*\Big\}$ is another (orthonormal) basis. }
\be
\S_\epsilon\phi^\epsilon_k = \nu^\epsilon_k\phi^\epsilon_k\, , ~~ k\in\natural^*\, , \quad~~\where~~\quad  (\phi^\epsilon_k,\phi^\epsilon_{k'})_{\bell_\epsilon}=\delta_{kk'}\, ,~~ \nu^\epsilon_k\in\R\, .
\ee
Expanding the scale factor fluctuation as
\be
\delta a(\tau)=\sum_{k\ge 1}\deltaa_k\,  \phi^\epsilon_k(\tau)\, , 
\label{expanda}
\ee
the path-integral measure is derived from the norm
\be
||\delta a||_{\bell_\epsilon}^2 =  (\delta a,\delta a)_{\bell_\epsilon} = \sum_{k\ge 1} \deltaa_k^2 \quad\Longrightarrow\quad \D\delta a = \bigwedge_{k\ge 1}\d \deltaa_k\, .
\ee
Hence, we obtain 
\begin{align}
Z_\epsilon(a_0)&=\prod_{k\ge 1}\int \d\deltaa_k\; e^{-{3sv_3\over \hbar} \nu^\epsilon_k(\deltaa_k)^2}= \prod_{k\ge 1} \sqrt{\hbar\, \pi\over 3sv_3\, \nu^\epsilon_k}\nonumber\\
&= \left({3sv_3\over \hbar \, \pi}\right)^{1\over 4}{1\over \sqrt{\det \S_\epsilon}}\, , 
\label{Ze}
\end{align}
where the last equality follows by zeta regularization and $\det\S_\epsilon = \prod_{k\ge 1}\nu^\epsilon_k$. Notice that the prescription that we apply to the path integral is the following: Each mode coefficient $\deltaa_k$ is integrated from $-\infty$ to $+\infty$ when $s \nu_k^\epsilon>0$, and from $-i\infty$ to $+i\infty$ when $s \nu_k^\epsilon<0$.\footnote{The case $\nu_k^\epsilon=0$ is excluded since we assumed that $\S_\epsilon$ is invertible, as will be shown later on. } As a result,  all integrals are convergent. In fact, we will see at the end of the section   that $\det \S_+$ and $\det \S_-$ have opposite signs. 
Hence, whatever choice of sign $s$ is made in the definition of  Euclidean-time in \eq{LE}, imposing domains of integration along the real and imaginary axes is  necessary for both $Z_{+}(a_0)$ and $Z_{-}(a_0)$ to exist. In fact, rotating the domain of integrations for some mode coefficients can be seen as the result of an analytic continuation when the shape of the potential $V(a)$ is varied from cases where it is bounded from below to cases where it is not~\cite{Coleman,Callan}. 

Before we go any further, let us mention that an explicit check of the fact that $Z_\epsilon(a_0)$ is independent of the choice of fiducial metric is provided in Appendix~\ref{A4}.



\subsubsection{Computation of the determinants}
\label{compuDet}

To compute the determinant of $\S_\epsilon$ we apply the method presented in~\Refe{Coleman}.  Let us  consider the system
\be
\left\{\begin{array}{l}
\S_\epsilon \varphi^\epsilon_\nu = \nu \varphi^\epsilon_\nu \, , \\
\varphi^\epsilon_\nu(\tau_\epsilon)=0\, , \quad \dis {\d\varphi^\epsilon_\nu\over \d\tau}(\tau_\epsilon)=1\, ,\end{array}\right.
\label{difs}
\ee
where $\nu$ is real while $\tau_\epsilon\in(0,1)$ will serve as a regulator to be sent to 0 at the end of all computations.  Since this is a linear differential equation of order 2,
the initial conditions on $\varphi^\epsilon_\nu$ and its derivative select a unique solution for any $\nu$.  When such a solution satisfies $\varphi^\epsilon_\nu(1)=0$, we know that there is $k\in\natural^*$ such that $\nu=\nu^\epsilon_k$. Moreover, the dimension of any eigenspace of $\S_\epsilon$ cannot be of dimension $>1$ since otherwise there would be two or more solutions of the differential system for some $\nu=\nu^\epsilon_k$.\footnote{$\phi^\epsilon_k$ and  $\varphi_{\nu_k^\epsilon}^\epsilon/(\varphi_{\nu_k^\epsilon}^\epsilon,\varphi_{\nu_k^\epsilon}^\epsilon)_{\bell_\epsilon}$ are thus equal up to an arbitrary sign.}  We may  order them
such that $\nu_1^\epsilon<\nu_2^\epsilon<\nu_3^\epsilon<\cdots$. 

We can take a more general look to the same system for two operators of the Hilbert space that are identical to $\S_\epsilon$ up to  the terms involving no derivative. They can be written as 
\be
\S_\epsilon^{(i)} = -{\ba_\epsilon\over \bell_\epsilon^2}\, {\d^2\over \d\tau^2}-{1\over \bell_\epsilon^2}\, {\d\ba_\epsilon\over \d \tau}\, {\d\over \d\tau}+W^{(i)}_\epsilon\, ,\quad i\in\{1,2\}\, , 
\label{Wi}
\ee
where $W^{(i)}_\epsilon(\tau)$ are arbitrary  functions on $[\tau_\epsilon,1]$. Denoting $\varphi_\nu^{\epsilon(i)}$ the solutions of the associated differential systems and $\nu_k^{\epsilon(i)}$ the eigenvalues of $\S_\epsilon^{(i)}$, we have
\begin{align}
\varphi_\nu^{\epsilon(i)}(1)=0\quad &\Longleftrightarrow \quad \mbox{there exists $k\in\natural^*$ such that $\nu=\nu_k^{\epsilon(i)}$}\nonumber \\
&\Longleftrightarrow \quad  \det(\S^{(i)}_\epsilon-\nu)\equiv \prod_{k\ge 1}(\nu_k^{\epsilon(i)}-\nu)=0\, .
\end{align}
It then turns out that~\cite{Coleman}
\be
{\det(\S^{(1)}_\epsilon-\nu)\over \det(\S^{(2)}_\epsilon-\nu)}={\varphi_\nu^{\epsilon(1)}(1)\over \varphi_\nu^{\epsilon(2)}(1)}\, .
\ee
Indeed, because both sides of this equality are meromorphic functions of $\nu$ with identical simple zeros and poles, they are proportional. Moreover, thanks to the fact that $W^{(1)}_\epsilon$ and  $W^{(2)}_\epsilon$ are bounded on $[\tau_\epsilon,1]$, the two sides of the equality tend to 1 when $\nu\to \infty$ in the complex plane, which shows that they are equal for all $\nu\in\complex$.  Hence, there is a constant\footnote{ However it depends on the regulator $\tau_\epsilon$ and $a_0$. } $\N_\epsilon$, which  is independent of $W^{(i)}_\epsilon$, such that 
\be
\N_\epsilon\equiv {\det(\S^{(i)}_\epsilon-\nu)\over\varphi_\nu^{\epsilon(i)}(1)}\, , \quad i\in\{1,2\}\, . 
\ee
Taking $\nu=0$, we obtain a formal expression for the determinant of interest in terms of the universal constant $\N_\epsilon$, 
\be
\det \S^{(i)}_\epsilon=\N_\epsilon\, \varphi^{\epsilon(i)}_{0}(1)\, .
\label{detfor}
\ee

To apply this formula to $\det\S_\epsilon$, we have to compute $\varphi^\epsilon_{0}(1)$ by solving \Eq{difs} for $\nu=0$. Redefining\footnote{We choose the regulator $\theta_*$ to be independent of $\epsilon$ so that $\lambda \ba_+(\tau)=\lambda \ba_-(\tau)= \sin(\theta)$ describe portions of the 4-sphere with a common boundary at $\theta_*$. This  fact is required for instance for the regularized instantons to be identical in the limit $\lambda a_0\to 1$.}
\be
 \Phi_\epsilon(\theta)=\lambda \varphi^\epsilon_{0}(\tau)\, , \quad \where \quad \theta = \lambda \bell_\epsilon \tau \in[\theta_*,\lambda\bell_\epsilon]\, , \quad\theta_*=\lambda \bell_\epsilon\tau_\epsilon\, , 
 \label{redefi}
 \ee 
the system becomes
 \be
\left\{\begin{array}{l}
\dis -\sin\theta  \,{\d^2\Phi_\epsilon\over \d\theta^2}- \cos\theta \, {\d\Phi_\epsilon\over \d\theta}-2\sin\theta \,\Phi_\epsilon= 0\, , \\
\Phi_\epsilon(\theta_*)=0\, , \quad\dis  {\d \Phi_\epsilon\over \d\theta}(\theta_*)={1\over \bell_\epsilon}\, .\esp\end{array}\right.
\label{difs2}
\ee
The general solution of the differential equation can be written as
\be
\Phi_\epsilon(\theta)= A_\epsilon \cos \theta + B_\epsilon \Big[\!-\!1+{\cos \theta \over 2}\ln\!\Big({1+\cos \theta\over 1-\cos \theta}\Big)\Big]\, ,
\ee 
where $A_\epsilon$, $B_\epsilon$ are constants determined by the initial conditions.\footnote{It is at this stage that we see the relevance of introducing a cutoff $\tau_\epsilon$ \ie $\theta_*$. With $\theta_*=0$ we would have $A_\epsilon=B_\epsilon=0$ which is not allowed for an eigenvector.}  In the limit $\theta_*\to 0$, they satisfy
\be
A_\epsilon\underset{\theta_*\to 0}{\sim}{\theta_*\over \bell_\epsilon}\ln{1\over \theta_*}\, , \quad~~B_\epsilon\underset{\theta_*\to 0}{\sim}-{\theta_*\over \bell_\epsilon}\, .
\ee
Therefore, the mode proportional to $B_\epsilon$ is dominated by  the  mode  proportional to $A_\epsilon$, so that
\begin{align}
\det \S_\epsilon &= {\N_\epsilon \over \lambda}\, \Phi_\epsilon(\lambda \bell_\epsilon) \underset{\theta_*\to 0}{\sim} {\theta_*\over \lambda}\ln{1\over \theta_*}\,  {\N_\epsilon\over  \bell_\epsilon} \, \cos(\lambda \bell_\epsilon)\nonumber \\
&\!\:\!\!\!\underset{\theta_*\to 0}{\sim} {\theta_*\over \lambda}\ln{1\over \theta_*}\times {\N_\epsilon\over \bell_\epsilon} \, \epsilon\, \sqrt{1-(\lambda a_0)^2}\, .\esp
\label{determ}
\end{align}
 

\subsubsection{Computation of the normalization constants $\N_\epsilon$} 
\label{coN}

To find
the value of  $\N_\epsilon$, we need to compute the determinant of any operator $\S_\epsilon^{(1)}$ as the infinite product of its eigenvalues. Of course this is a  difficult task unless $W^{(1)}_\epsilon$ is chosen suitably. To this end  let us consider the system
\be
\left\{\begin{array}{l}
\S^{(1)}_\epsilon \psi^\epsilon_k = \nu_k^{\epsilon(1)}\,  \psi^\epsilon_k  \, , \\
\psi^\epsilon_k(\tau_\epsilon)=0\, , \quad  \psi^\epsilon_k(1)=0\, .\esps\end{array}\right.
\label{difs3}
\ee
The boundary condition at $\tau=\tau_\epsilon$ can be viewed as a linear relation between the two integration constants of the solutions, while  effectively the boundary condition at $\tau=1$ enforces $\nu_k^{\epsilon(1)}$ to be an eigenvalue of $\S_\epsilon^{(1)}$ acting on the Hilbert space of functions vanishing at $\tau_\epsilon$ and 1.\footnote{In fact, this is true unless the solutions of the system do not generate a dimension one vectorial space. This fact happens for some choices of $W^{(1)}_\epsilon$ and $\nu$ for which $\S^{(1)}_\epsilon \psi = \nu\,  \psi $, $\psi(\tau_\epsilon)=\psi(1)=0$ admits only the solution $\psi(\tau)\equiv 0$. Such values of $\nu$ are not eigenvalues.}

Let us first put the differential equation in canonical form. This can be done by  introducing a new Euclidean-time variable $\zeta$ and a new unknown function $\check \psi^\epsilon_k(\zeta)$ satisfying~ 
\be
\bell_\epsilon\,  \d\tau = \sqrt{\ba_\epsilon(\tau)}\, \d \zeta\, , \quad ~~\psi^\epsilon_k(\tau)=\ba_\epsilon(\tau)^{-{1\over 4}}\, \check\psi^\epsilon_k(\zeta)\, , 
\ee
in terms of which we obtain
\be
\left\{\begin{array}{l}
\dis -{\d^2\check\psi^\epsilon_k\over \d\zeta^2}+\!\big(W^{(1)}_\epsilon-{1\over 16} \,{1\over \ba_\epsilon} \, [1+3(\lambda \ba_\epsilon)^2]\big)\,\check \psi^\epsilon_k=\nu_k^{\epsilon(1)}\, \check \psi^\epsilon_k \, ,  \\
\check \psi^\epsilon_k(\zeta(\tau_\epsilon))=0\, , \quad  \check \psi^\epsilon_k(\zeta(1))=0\, .\esp\end{array}\right.
\label{difs4}
\ee
There is therefore a 
choice 
\be
W^{(1)}_\epsilon={1\over 16} \,{1\over \ba_\epsilon} \, [1+3(\lambda \ba_\epsilon)^2]
\ee
for which all eigenvectors and eigenvalues can be found trivially. They are given by 
\be
 \check \psi_{k}^{{\epsilon}}(\zeta)= C_k\sin\!\big[\sqrt{\nu_k^{\epsilon(1)}}\, (\zeta-\zeta(\tau_\epsilon))\big]\, ,\quad  \where\quad \sqrt{\nu_k^{\epsilon(1)}} = {k\pi\over \zeta(1)-\zeta(\tau_\epsilon)}\, , \quad k\in \natural^*\, ,
\ee
where $C_k$ are arbitrary constants. It is then straightforward to compute in this case the determinant by zeta regularization, 
\begin{align}
\det\S_\epsilon^{(1)} &= \prod_{k\ge 1} k^2\prod_{k'\ge 1}\Big({\pi\over \zeta(1)-\zeta(\tau_\epsilon)}\Big)^2\, , \nonumber \\
&=2\big[\zeta(1)-\zeta(\tau_\epsilon)\big]\, .\esps
\label{de1}
\end{align}

Next, we recompute this determinant by applying \Eq{detfor}. We thus consider the system
\be
\left\{\begin{array}{l}
\S^{(1)}_\epsilon \varphi^{\epsilon(1)}_0 = 0 \, , \\
\varphi^{\epsilon(1)}_0(\tau_\epsilon)=0\, , \quad \dis {\d\varphi^{\epsilon(1)}_0\over \d\tau}(\tau_\epsilon)=1\, ,\esps\end{array}\right.
\ee
on which we apply the change of variable and unknown function introduced before, 
\be
\varphi^{\epsilon(1)}_0(\tau) = \ba_\epsilon(\tau)^{-{1\over 4}}\,\check \varphi^{\epsilon(1)}_0(\zeta)\, ,
\ee
 yielding
\be
\left\{\begin{array}{l}
\dis -{\d^2\check \varphi_0^{\epsilon(1)}\over \d \zeta^2}= 0 \, , \\
\check \varphi^{\epsilon(1)}_0(\zeta(\tau_\epsilon))=0\, , \quad \dis {\d\check \varphi^{\epsilon(1)}_0\over \d\zeta}(\zeta(\tau_\epsilon))={\ba_\epsilon(\tau_\epsilon)^{3\over 4}\over \bell_\epsilon}\, .\esps\end{array}\right.
\ee
 Integrating, we find
\be
\check \varphi^{\epsilon(1)}_0(\zeta)={\ba_\epsilon(\tau_\epsilon)^{3\over 4}\over \bell_\epsilon} \, \big[\zeta-\zeta(\tau_\epsilon)\big]\quad \Longrightarrow\quad \varphi^{\epsilon(1)}_0(\tau) ={\ba_\epsilon(\tau_\epsilon)^{3\over 4}\over \bell_\epsilon} \,\ba_\epsilon(\tau)^{-{1\over 4}}\,  \big[\zeta(\tau)-\zeta(\tau_\epsilon)\big]\, , 
\ee
which we can use to get
\begin{align}
\det \S_\epsilon^{(1)}&=\N_\epsilon\, \varphi^{\epsilon(1)}_0(1) \nonumber \\
&= \N_\epsilon\, {\ba_\epsilon(\tau_\epsilon)^{3\over 4}\over \bell_\epsilon} \,a_0^{-{1\over 4}}\,  \big[\zeta(1)-\zeta(\tau_\epsilon)\big]\, .\esp
\label{de2s}
\end{align}

Identifying the above expression with \Eq{de1}, we obtain
\be
\N_\epsilon = 2\, \ba_\epsilon(\tau_\epsilon)^{-{3\over 4}}\, \bell_\epsilon \,a_0^{1\over 4} \underset{\theta_*\to 0}{\sim} 2\, \Big({\lambda\over \theta_*}\Big)^{3\over 4}\times \bell_\epsilon\, a_0^{1\over 4}\, , 
\ee
which can be inserted in \Eq{determ} to yield the final result
\be
\det \S_\epsilon \underset{\theta_*\to 0}{\sim} 2\, \Big({\theta_*\over \lambda}\Big)^{1\over 4}\ln{1\over \theta_*}\times \epsilon\, a_0^{1\over 4}\sqrt{1-(\lambda a_0)^2}\, .
\label{final_detS}
\ee
Notice that 
\be
\det\S_+>0\, ,\quad \det\S_-<0\, , \quad \when \quad 0<\lambda a_0<1\, ,
\ee
and so $\S_+$ and $\S_-$ are invertible, as announced before. 

\subsection{Quadratic fluctuations of the length}

The computation of the Gaussian integral over the fluctuation $\delta \ell$ of the modulus in the expression of the wavefunction in \Eq{psii} requires the evaluation of $ \K_\epsilon$ given in \Eq{barI}. Therefore, we determine $\S_\epsilon^{-1}V_a(\ba_\epsilon)$, which is the unique solution of the system 
\be 
\left\{\begin{array}{l}
\S_\epsilon f_\epsilon = V_a(\ba_\epsilon) \, , \\
f_\epsilon(\tau_\epsilon)=0\, , \quad f_\epsilon(1)=0\, ,\esps\end{array}\right.
\label{syss}
\ee
where $\tau_\epsilon$ is the regulator close to $0$ that was used in the derivation of $\det\S_\epsilon$ in the previous subsection. 
As explained below \Eq{pb}, $V_a(\ba_\epsilon)$ is understood in the above equation as the function $1-3(\lambda \ba_\epsilon)^2$ on the open set $(\tau_\epsilon, 1)$, vanishing at $\tau_\epsilon$ and $1$, where it is discontinuous. 

Redefining 
\be
 F_\epsilon(\theta)=\lambda f_\epsilon(\tau)\, 
 \ee
as in \Eq{redefi}, we have to solve
  \be
\left\{\begin{array}{l}
\dis -\sin\theta  \,{\d^2F_\epsilon\over \d\theta^2}- \cos\theta \, {\d F_\epsilon\over \d\theta}-2\sin\theta \,F_\epsilon= 1-3\sin^2\theta\, ,\quad \theta\in(\theta_*,\lambda \bell_\epsilon)\, ,  \\
F_\epsilon(\theta_*)=0\, , \quad F_\epsilon(\lambda \bell_\epsilon)=0\, .\esps\end{array}\right.
\ee
The general solution of the differential equation is 
\be
F_\epsilon(\theta)= -\theta \cos\theta+D_\epsilon \cos \theta + E_\epsilon \Big[\!-\!1+{\cos \theta \over 2}\ln\!\Big({1+\cos \theta\over 1-\cos \theta}\Big)\Big]\, ,
\ee
where $D_\epsilon$, $E_\epsilon$ are constants determined by the initial conditions.\footnote{Without introducing the cutoff $\tau_\epsilon$ \ie $\theta_*$, the condition $F_\epsilon(0)=0$ would yield $F_\epsilon(\theta)=-\theta\cos\theta$. However, we would have $F_\epsilon(\lambda \bell_\epsilon)=-\lambda \bell_\epsilon \epsilon\sqrt{1-(\lambda a_0)^2}\neq 0$ for $\lambda a_0<1$, and thus no solution.}  In the limit $\theta_*\to 0$, they satisfy
\begin{align}
D_\epsilon &= \lambda \bell_\epsilon\left\{1+{\epsilon\over \ln{1\over \theta_*}}\left[{-1\over \sqrt{1-(\lambda a_0)^2}}+{1\over 2}\ln\bigg({1+\sqrt{1-(\lambda a_0)^2}\over 1-\sqrt{1-(\lambda a_0)^2}}\bigg)\right] \!\Big(1+\O\Big({1\over \ln{1\over \theta_*}}\Big)\Big)\right\} ,\nonumber \\
E_\epsilon &=-{\lambda \bell_\epsilon\over \ln{1\over \theta_*}}\,\Big(1+\O\Big({1\over \ln{1\over \theta_*}}\Big)\Big)\, .
\end{align}
As a result, the function $f_\epsilon(\tau)$ is 
\begin{align}
\S_\epsilon^{-1}V_a(\ba_\epsilon) (\tau&)= \bell_\epsilon(1-\tau)\cos(\lambda\bell_\epsilon\tau)\nonumber \\
&+{\bell_\epsilon\over \ln{1\over \theta_*}}\,\Bigg\{\epsilon\left[{-1\over \sqrt{1-(\lambda a_0)^2}}+{1\over 2}\ln\bigg({1+\sqrt{1-(\lambda a_0)^2}\over 1-\sqrt{1-(\lambda a_0)^2}}\bigg)\right] \!\Big(1+\O\Big({1\over \ln{1\over \theta_*}}\Big)\Big)\cos(\lambda\bell_\epsilon\tau)\nonumber \esp\\
&~~~~~~~~~~~\,+ \Big(1+\O\Big({1\over \ln{1\over \theta_*}}\Big)\Big)\Big[1-{\cos (\lambda\bell_\epsilon\tau) \over 2}\ln\!\Big({1+\cos (\lambda\bell_\epsilon\tau)\over 1-\cos (\lambda\bell_\epsilon\tau)}\Big)\Big]\Bigg\}\,.\esp
\end{align}

Even if the above expression is quite involved, it turns out that the value of $\K_\epsilon$ is extremely simple. Indeed, the contribution to the integrand in \Eq{barI} that is independent of the cutoff $\theta_*$ vanishes after integration over $\tau$. As a result,  $\K_\epsilon$ is of order $1/\ln\theta_*$ and it simplifies to give
\be
\K_\epsilon \underset{\theta_*\to 0}{\sim} {\bell_\epsilon^2\over \ln{1\over \theta_*}}\, .
\ee
The quadratic fluctuations of the length around the extremal value $\bell_\epsilon$  then yield a  contribution
\be
\int \d\delta \ell \,\exp\Big\{\!-\!{3sv_3\over \hbar}\,  \K_\epsilon  \Big({\delta\ell\over \bell_\epsilon}\Big)^2\Big\}\underset{\theta_*\to 0}{\sim}\sqrt{{\pi\hbar\over 3sv_3}\, \ln{1\over \theta_*}}\, ,
\label{fluI}
\ee
provided the domain of integration is  from $-\infty$ to $+\infty$ when $s>0$, and from $-i\infty$ to $+i\infty$ when $s<0$. 

We are now ready to insert in the expression of the wavefunction given in \Eq{psii} our results for the Faddeev--Popov Jacobian $\Delta_{\rm FP}$ (\Eq{compu}), the instanton actions $\bar S_{\rm E}^\epsilon$ (\Eq{acInst}), the quadratic fluctuations  $Z_\epsilon(a_0)$ of the scale factor  (\Eqs{Ze},~(\ref{final_detS}))  and those of the modulus (\Eq{fluI}), to obtain the final result 
\be
\Psi(a_0) = \Cc_s(\theta_*)\sum_{\epsilon=\pm 1} {1\over \sqrt{\epsilon}}\, {\exp\!\Big[\epsilon s\, \dis{2v_3\over \hbar\lambda^2}\,\big(1-(\lambda a_0)^2\big)^{3\over 2}\Big]\over a_0^{1\over 8}\,\big(1-(\lambda a_0)^2\big)^{1\over 4}}\, (1+\O(\hbar))\, ,\quad 0<\lambda a_0<1\, .
\label{psia00}
\ee
In this expression, $\Cc_s(\theta_*)$ is a regulator-dependant coefficient 
\be
\Cc_s(\theta_*)=\alpha\,\sqrt{i\pi} \, \Big({\pi \hbar\over 3sv_3}\Big)^{1\over 4}\exp\!\Big[\!-\!s\, {2v_3\over \hbar \lambda^2}\Big]\Big({\lambda \over \theta_*}\Big)^{1\over 8}\, ,
\ee
which is an irrelevant multiplicative factor. Notice that even though the wavefunction defined in \Eq{Psi(a0)} seems to be real, it turns out 
to be complex due to the factor $1/\sqrt{\epsilon}$ arising from the opposite signs of $\det\S_+$ and $\det \S_-$. Notice that the Hartle-Hawking prescription, $s=-1$, and the Linde-Vilenkin prescription, $s=1$, lead to distinct wavefunctions with different physical behavior, see \eg Ref.~\cite{He:2020wzj}.


\section{Field redefinitions}
\label{field redef}

The classical action in \Eq{SE} is invariant under Euclidean-time reparametrizations and field redefinitions $a=A(q)$, where $A$ is an arbitrary function. However, at the quantum level, the formalism we have used to evaluate the wavefunction is only invariant under diffeomorphisms, taking into account the (trivial) Fadeev--Popov Jacobian arising  after gauge fixing and using reparametrization-invariant path integral measures. In this section, we want to  check how the choice of field $q$, which parametrizes the target minisuperspace, affects the wavefunction.


\subsection{Distinct gauge-invariant field measures}

To better understand the issue, let us apply a field redefinition of the scale-factor degree of freedom,
\be
a=A(q) ~~ \Longleftrightarrow ~~ q=Q(a)\, , 
\ee
where $Q=A^{-1}$ is an invertible function defined for $a> 0$.\footnote{The function $Q(a)$ is allowed to be finite or infinite at $a=0$, as well as when  $a\to +\infty$.} In fact, $q(\tau)$ is a field defined on $[0,1]$ satisfying fixed boundary conditions at $\tau=1$ and $\tau=0$, 
\be
q(1)\equiv q_0=Q(a_0)\, ,  \quad~~ q(0)= Q(0)\, .
\ee 
Around the small- or big-cap instanton solution, the fluctuations satisfy 
\be
\delta a=A'(\bq_\epsilon)\delta q+\O((\delta q)^2)\, ,~~\quad\where\quad ~~\bq_\epsilon=Q(\ba_\epsilon)\, , 
\label{pertur}
\ee
and where ``primes'' denote derivatives.

Introducing the cutoff $\tau_\epsilon$ as in the previous section,  $\delta a(\tau)$, $\delta q(\tau)$ and $A'(\bq_\epsilon(\tau))\phi^\epsilon_k(\tau)$ are all elements of the Hilbert space of functions in $L^2([\tau_\epsilon,1])$ that are vanishing at $\tau_\epsilon$ and 1.\footnote{$\{\phi^\epsilon_k, k\in\natural^*\}$ is a basis of this Hilbert space equipped with the  inner product defined as in \Eq{inprod} but with lower bound of the integral equal to $\tau_\epsilon$. } Hence, we may expand
\be
\delta a \equiv  \sum_{k'\ge 1}\deltaa_{k'}\, \phi^\epsilon_{k'}\, , ~~\quad  \delta q = \sum_{k\ge 1} \deltaq_k\, \phi^\epsilon_k\, ,\quad ~~A'(\bq_\epsilon)\phi^\epsilon_k= \sum_{k'\ge 1}\bM_{\epsilon,kk'}\, \phi^\epsilon_{k'}\, , 
\ee
where the coefficients $\deltaa_k$ were already defined in \Eq{expanda}, while  $\deltaq_k$ and $\bM_{\epsilon,kk'}$ are  real numbers. From \Eq{pertur}, we obtain 
\be
 \deltaa=\bM_\epsilon^\T\, \deltaq+\O((\deltaq)^2)\, ,
\ee
where the equality is to be understood as a relation between column matrices $ \deltaa$ and $\deltaq$. Using this matrix notation, the regularized path integral of the fluctuations $\delta a$ reads 
\begin{align}
Z_\epsilon(a_0)&= \int_{\textstyle \substack{\;\delta a(\tau_\epsilon)=0 \\ \;\,\:\delta a(1)=0}}\D \delta a  \, \exp\Big\{\!-\!{3sv_3\over \hbar}\,( \delta a, \S_\epsilon \delta a)_{\bell_\epsilon}\Big\}\nonumber \\
&= \int \bigwedge_{k\ge 1}\deltaa_k \,\exp\Big\{\!-\!{3sv_3\over \hbar}\,\deltaa^\T\,\bS_\epsilon\,\deltaa\Big\}\nonumber \\
&=\left| \det \bM_\epsilon^\T\right| \int \bigwedge_{k\ge 1}\deltaq_k \,\exp\Big\{\!-\!{3sv_3\over \hbar}\,\deltaq^\T\,(\bM_\epsilon\bS_\epsilon\bM_\epsilon^\T)\,\deltaq\Big\}\,(1+\O(\hbar))\, ,
\label{Zwedge}
\end{align}
where $\bS_\epsilon=\diag(\nu_k^\epsilon)$, while the factor $|\det \bM_\epsilon^\T|$ in the last equality arises from the complete antisymmetry of the wedge products in the measure.   Since the matrix $\bM_\epsilon\bS_\epsilon\bM_\epsilon^\T$ is real symmetric, it can be diagonalized with an orthogonal matrix $\bP_\epsilon$. Defining $\deltaq = \bP_\epsilon \deltatq$, we obtain
\be
Z_\epsilon(a_0)= |\det \bM^{\epsilon}|\int \bigwedge_{k\ge 1}\deltatq_k \, \exp\Big\{\!-\!{3sv_3\over \hbar}\,\deltatq^\T\,(\bP_\epsilon^{-1}\bM_\epsilon\bS_\epsilon\bM_\epsilon^\T\bP_\epsilon)\,\deltatq\Big\}\,(1+\O(\hbar))\, ,
\label{ZZ}
\ee
where $\bP_\epsilon^{-1}\bM_\epsilon\bS_\epsilon\bM_\epsilon^\T\bP_\epsilon$ is diagonal. Two possible outcomes to this calculation can be proposed. 

Firstly, we can compute the Gaussian integral as in \Eq{Ze} and check that it yields the same answer, up to $\O(\hbar)$ terms:
\begin{align}
Z_\epsilon(a_0)&=|\det \bM^{\epsilon}|\left({3sv_3\over \hbar \, \pi}\right)^{1\over 4}{1\over \sqrt{\det( \bP_\epsilon^{-1}\bM_\epsilon\bS_\epsilon\bM_\epsilon^\T\bP_\epsilon)}} \,(1+\O(\hbar))\nonumber \\
&= \left({3sv_3\over \hbar \, \pi}\right)^{1\over 4}{1\over \sqrt{\det\bS_\epsilon}} \,(1+\O(\hbar))\, .
\end{align}
This is a matrix proof of the fact that a ``change of field'' can be applied in the path integral  in analogy with a  ``change of variable''  in a conventional integral. In fact, $|\det \bM_\epsilon|$ is nothing but the Jacobian of this transformation. 

However, comparing \Eq{ZZ} with the two first lines of \Eq{Zwedge}, we can also write
\be
Z_\epsilon(a_0) = |\det \bM_\epsilon| \, \widetilde Z_\epsilon(q_0)\, (1+\O(\hbar))\, ,
\label{ZZs}
\ee
where $\widetilde Z_\epsilon(q_0)$ is a path integral based on the field $\delta q$, 
\be
\widetilde Z_\epsilon(q_0)= \int_{\textstyle \substack{\;\delta q(\tau_\epsilon)=0 \\ \;\,\: \delta q(1)=0}}\D \delta q  \, \exp\Big\{\!-\!{3sv_3\over \hbar}\,( \delta q, \widetilde \S_\epsilon \delta q)_{\bell_\epsilon}\Big\}\, .
\ee
In this expression, $\widetilde \S_\epsilon$ is the new operator satisfying 
\be
 \delta a\, \S_\epsilon \delta a \equiv \delta q\, \widetilde \S_\epsilon \delta q+\O((\delta q)^3)
\label{ShatS}
\ee
and whose eigenvalues are the diagonal elements of the diagonal matrix $ \bP_\epsilon^{-1}\bM_\epsilon\bS_\epsilon\bM_\epsilon^\T\bP_\epsilon$. Being represented by a real symmetric matrix, it is self-adjoint.

The conclusion to the above discussion is that for every field redefinition $a=A(q)$ in the classical action,
\be
S_{\rm E}[\ell^2,a]\equiv \widetilde S_{\rm E}[\ell^2,q] \, , 
\label{stilde}
\ee
it is legitimate to define a quantum wavefunction as in \Eq{psa0},
\be
\widetilde\Psi(q_0)=\Delta_{\rm FP} \int_0^{+\infty}\!\!\d\ell   \int_{\textstyle \substack{\; q(0)=Q(0) \\  \!\! \!\!\!q(1)=q_0}}\D q  \; e^{-{1\over \hbar}\widetilde S_{\rm E}[\ell^2,q]}\, ,
\label{psa1}
\ee
which is characterized by its own path-integral measure $\D q$. In the following we compute $\widetilde\Psi(q_0)$ in the semi-classical limit.


\subsection{Steepest-descent approximation}

The quadratic expansion of the classical action in \Eq{qua} (and \Eq{QS}) can be expressed  in terms of $\delta q$ and $\delta\ell$,
\begin{align}
\widetilde S_{\rm E}[\ell^2,q]&=\bar S_{\rm E}^\epsilon+3sv_3\int_0^1\d \tau \bell_\epsilon \Big[\delta q\, \widetilde \S_\epsilon \delta q + 2 \,\delta q\, \widetilde V_q(\bq_\epsilon)\,{\delta\ell\over \bell_\epsilon}+ {\delta\ell\over \bell_\epsilon}\,\widetilde  V(\bq_\epsilon)\, {\delta\ell\over \bell_\epsilon}\Big]\!+\O(\delta^3)\, ,\nonumber\\
 \where \quad \wt V(q)&\equiv V(a)\, .
\end{align}
The explicit form of the self-dual operator $\wt \S_\epsilon$ is obtained from its definition   in \Eq{ShatS},  which yields
\be
\wt\S_\epsilon = -A'\bigg[{AA'\over \bell_\epsilon^2}\, {\d^2\over \d\tau^2}+{2AA''+A'^2\over \bell_\epsilon^2}\, {\d \bq_\epsilon\over \d\tau}\, {\d\over \d\tau}+{AA''\over \bell_\epsilon^2}\, {\d^2\bq_\epsilon\over \d\tau^2}+{AA'''+A'A''\over \bell_\epsilon^2}\, \Big({\d\bq_\epsilon\over \d\tau}\Big)^2+2\lambda^2AA'\bigg]\, ,
\ee
where all functions $A$ and their derivatives are evaluated at $\bq_\epsilon$. Proceeding as we did for $\Psi(a_0)$ between \Eqs{atilde} and~(\ref{psii}), the steepest-descent approximation leads to 
\begin{align}
\wt\Psi(q_0)&=\Delta_{\rm FP}\sum_{\epsilon=\pm 1} e^{-{1\over \hbar} \bar S_{\rm E}^\epsilon}\,\wt Z_\epsilon(q_0) \int \d\delta \ell \,\exp\Big\{\!-\!{3sv_3\over \hbar}\,  \wt\K_\epsilon  \Big({\delta\ell\over \bell_\epsilon}\Big)^2\Big\}\,(1+\O(\hbar))\, , \nonumber \\
\where \quad\wt \K_\epsilon &= \int_0^1\d\tau \bell_\epsilon\, \big[\wt V(\bq_\epsilon)-\wt V_q(\bq_\epsilon)\wt\S_\epsilon^{-1} \wt V_q(\bq_\epsilon)\big]\, .\esp
\label{psihat}
\end{align}
Notice that we could have reached this result by simply substituting $Z_\epsilon(a_0)$ with $\wt Z_\epsilon(q_0)$ (related to each other by \Eq{ZZs})  in \Eq{psii}. In that case, $\wt\K_\epsilon$ would have not replaced $\K_\epsilon$. Consistently, we will find that $\wt\K_\epsilon=\K_\epsilon$.


\subsubsection{Determinants of the fluctuations of $q$}

As in \Eq{Ze}, the contributions to the wavefunction $\wt\Psi(q_0)$ arising from the  fluctuations $\delta q$ at quadratic order can be written in terms of the determinants of the self-dual operators~$\wt S_\epsilon$,
\be
\wt Z_\epsilon(q_0)= \left({3sv_3\over \hbar \, \pi}\right)^{1\over 4}{1\over \sqrt{\det \wt\S_\epsilon}}\, .
\label{ztilde}
\ee
To compute them we can work
exactly as we did in \Sect{compuDet}. This leads to the result 
\be
\det \wt\S_\epsilon=\wt\N_\epsilon\, \wt\varphi^{\epsilon}_{0}(1)\, ,
\label{d1}
\ee
where $\wt\N_\epsilon$ is a constant to be  determined and $\wt \varphi^{\epsilon}_{0}$ is the solution of the  system 
\be
\left\{\begin{array}{l}
\wt\S_\epsilon \wt\varphi^\epsilon_0 = 0 \, , \\
\wt\varphi^\epsilon_0(\tau_\epsilon)=0\, , \quad \dis {\d\wt\varphi^\epsilon_0\over \d\tau}(\tau_\epsilon)=1\, .\end{array}\right.
\label{d2}
\ee
However, the change of unknown function
\be
\wt\varphi^\epsilon_0(\tau) = K_\epsilon \, {\overset{\smile} \varphi_0{}^{\!\!\!\!\!\:\epsilon}\!\;(\tau) \over A'(\bq_\epsilon(\tau))}\, ,
\ee
where $K_\epsilon$ is a constant, transforms the differential equation back to a form involving the much simpler operator $\S_\epsilon$, 
\be
\left\{\begin{array}{l}
\S_\epsilon \overset{\smile} \varphi_0{}^{\!\!\!\!\!\:\epsilon} = 0 \, , \\
\overset{\smile} \varphi_0{}^{\!\!\!\!\!\:\epsilon}\!\;(\tau_\epsilon)=0\, , \quad \dis {\d\overset{\smile} \varphi_0{}^{\!\!\!\!\!\:\epsilon}\over \d\tau}(\tau_\epsilon)={A'(\bq_\epsilon(\tau_\epsilon))\over K_\epsilon}\, .\end{array}\right.
\ee
Hence, choosing $K_\epsilon  = A'(\bq_\epsilon(\tau_\epsilon))$, the functions $\overset{\smile} \varphi_0{}^{\!\!\!\!\!\:\epsilon}$ and $\varphi_0^\epsilon$ (encountered in the computation of $\det \S_\epsilon$) satisfy the same equation and initial conditions, as can be seen in \Eq{difs}. Therefore, they are equal, $\overset{\smile} \varphi_0{}^{\!\!\!\!\!\:\epsilon}=\varphi_0^\epsilon$, which leads to 
\begin{align}
\det \wt\S_\epsilon &= \wt\N_\epsilon \,  A'(\bq_\epsilon(\tau_\epsilon))\, {\varphi_0^\epsilon(1) \over A'(\bq_\epsilon(1))} \nonumber \\
&= \wt \N_\epsilon \, A'(\bq_\epsilon(\tau_\epsilon))\, {\Phi_\epsilon(\lambda \bell_\epsilon)\over  \lambda \,A'(q_0)} \, .
\end{align}
In this expression, it is important to note that 
\be
A'(\bq_\epsilon(\tau_\epsilon))={1\over Q'(\ba_\epsilon(\tau_\epsilon))} = {1\over Q'\big({\sin\theta_*\over \lambda}\big)}\, , 
\ee
which depends only on the cutoff $\theta_*$ (and not on $a_0$). Hence, we obtain 
\be
\det \wt\S_\epsilon \underset{\theta_*\to 0}{\sim} {1\over Q'\big({\sin\theta_*\over \lambda}\big)}\, {\theta_*\over \lambda}\ln{1\over \theta_*}\times  {\wt\N_\epsilon\over \bell_\epsilon} \, \epsilon\, Q'(a_0) \sqrt{1-(\lambda a_0)^2}\, .\esp
\label{detmi}
\ee


\subsubsection{Computation of the normalization constants $\wt\N_\epsilon$}

The computation of $\wt\N_\epsilon$ can be done by following steps similar to those presented in \Sect{coN}. Let us start by deriving explicitly the infinite product of the eigenvalues $\tilde \nu^{\epsilon(1)}_k$ of an operator 
\be
\wt\S_\epsilon^{(1)} = -A'\bigg[{AA'\over \bell_\epsilon^2}\, {\d^2\over \d\tau^2}+{2AA''+A'^2\over \bell_\epsilon^2}\, {\d \bq_\epsilon\over \d\tau}\, {\d\over \d\tau}\bigg]+\wt W^{(1)}_\epsilon\, ,
\ee
where $\wt W^{(1)}_\epsilon(\tau)$ is a function on $[\tau_\epsilon,1]$ to be chosen suitably. These eigenvalues are such that the system
\be
\left\{\begin{array}{l}
\wt\S^{(1)}_\epsilon \wt\psi^\epsilon_k = \tilde \nu_k^{\epsilon(1)}\, \wt \psi^\epsilon_k  \, , \\
\wt\psi^\epsilon_k(\tau_\epsilon)=0\, , \quad \wt \psi^\epsilon_k(1)=0\esps
\end{array}\right.
\label{difs31}
\ee
admits a dimension one vectorial space of solutions. By defining a new Euclidean-time variable $\tilde \zeta$ satisfying
\be
\bell_\epsilon\,  \d\tau = \sqrt{A(\bq_\epsilon(\tau))}\,A'(\bq_\epsilon(\tau))\,  \d \tilde \zeta 
\ee
and a new unknown function $\check{\wt\psi^\epsilon}_{\!\!\!k}(\tilde \zeta)$ by
\be
\wt\psi^\epsilon_k(\tau)=r(\tilde \zeta)\, \check{\wt\psi^\epsilon}_{\!\!\!k}(\tilde \zeta)\, , \quad \where\quad r(\tilde \zeta)=\ba_\epsilon(\tau)^{-{1\over 4}} \,|A'(\bq_\epsilon(\tau))|^{-{1\over 2}}\, , 
\ee
the differential equation takes a canonical form
\be
\left\{\begin{array}{l}
\dis -{\d^2 \check{\wt\psi^\epsilon}_{\!\!\!k}\over \d\tilde \zeta^2}+\!\Big[\wt W^{(1)}_\epsilon- {1\over r}\, {\d^2 r\over \d \tilde \zeta^2} +{2\over r^2}\,\Big( {\d r\over \d \tilde \zeta}\Big)^2\Big]\check{\wt\psi^\epsilon}_{\!\!\!k}=\tilde \nu_k^{\epsilon(1)}\, \check{\wt\psi^\epsilon}_{\!\!\!k} \, ,  \\
\check{\wt\psi^\epsilon}_{\!\!\!k}(\tilde \zeta(\tau_\epsilon))=0\, , \quad  \check{\wt\psi^\epsilon}_{\!\!\!k}(\tilde \zeta(1))=0\, .\esp\end{array}\right.
\ee
Choosing $\wt W^{(1)}$ such that the term in brackets vanishes identically, the system becomes identical to the one analyzed in \Eq{difs4}. As a result, we obtain that
\be
\det\wt \S_\epsilon^{(1)} =2\big[\tilde \zeta(1)-\tilde \zeta(\tau_\epsilon)\big]\, .
\label{de11}
\ee

The next step is to obtain an alternative form of the determinant in terms of the constant $\wt\N_\epsilon$. The latter being ``universal,''  it enters the formula analogous to \Eq{d1},
\be
\det \wt\S^{(1)}_\epsilon=\wt\N_\epsilon\, \wt\varphi^{\epsilon(1)}_{0}(1)\, ,
\ee
where $ \wt\varphi^{\epsilon(1)}_{0}(\tau)$ is the unique solution of 
\be
\left\{\begin{array}{l}
\wt\S^{(1)}_\epsilon \wt\varphi^{\epsilon(1)}_0 = 0 \, , \\
\wt\varphi^{\epsilon(1)}_0(\tau_\epsilon)=0\, , \quad \dis {\d\wt\varphi^{\epsilon(1)}_0\over \d\tau}(\tau_\epsilon)=1\, .\esps\end{array}\right.
\ee
Applying the change of unknown function  
\be
\wt\varphi^{\epsilon(1)}_0(\tau) = r(\tilde\zeta)\, \!\:\!\!{\!\!\!\!\check {\;~~\wt\varphi^{\epsilon(1)}}_{\!\!\;\!\!\!\!\!\!\!\!\!\!0}\;\;\;(\tilde\zeta)}\, , 
\ee
this system becomes 
\be
\left\{\begin{array}{l}
\dis -{\d^2 \!\:\!\!{\!\!\!\!\check {\;~~\wt\varphi^{\epsilon(1)}}_{\!\!\;\!\!\!\!\!\!\!\!\!\!0}}\phantom{si}\over \d \tilde \zeta^2}= 0 \, , \\
\!\:\!\!{\!\!\!\!\check {\;~~\wt\varphi^{\epsilon(1)}}_{\!\!\;\!\!\!\!\!\!\!\!\!\!0}\;\;\;(\tilde\zeta(\tau_\epsilon))}=0\, , \quad \dis {\d \!\:\!\!{\!\!\!\!\check {\;~~\wt\varphi^{\epsilon(1)}}_{\!\!\;\!\!\!\!\!\!\!\!\!\!0}}\phantom{si}\over \d \tilde \zeta}(\tilde\zeta(\tau_\epsilon))={\ba_\epsilon(\tau_\epsilon)^{3\over 4}\,\sqrt{|A'(\bq(\tau_\epsilon))|}\,A'(\bq(\tau_\epsilon))\over \bell_\epsilon}\, .\esps\end{array}\right.
\ee
Integrating and multiplying by $r(\tilde\zeta)$, one obtains
\begin{align}
 \wt\varphi^{\epsilon(1)}_0(\tau) &={\ba_\epsilon(\tau_\epsilon)^{3\over 4}\,\sqrt{|A'(\bq(\tau_\epsilon))|}\,A'(\bq(\tau_\epsilon))\over \bell_\epsilon}  \, \ba_\epsilon(\tau)^{-{1\over 4}}\,  |A'(\bq_\epsilon(\tau))|^{-{1\over 2}}\, \big[\tilde \zeta(\tau)-\tilde \zeta(\tau_\epsilon)\big]\, ,
\end{align}
which can be evaluated at $\tau=1$ to find 
\be
\det \wt\S_\epsilon^{(1)}= \wt\N_\epsilon\, {\ba_\epsilon(\tau_\epsilon)^{3\over 4}\,\sqrt{|A'(\bq(\tau_\epsilon))|}\,A'(\bq(\tau_\epsilon))\over \bell_\epsilon}  \, a_0^{-{1\over 4}}\,  |A'(q_0)|^{-{1\over 2}}\, \big[\tilde \zeta(1)-\tilde \zeta(\tau_\epsilon)\big]\, .
\label{de2s2}
\ee

The identification of the above expression with \Eq{de11} leads to 
\be
\wt \N_\epsilon \underset{\theta_*\to 0}{\sim} 2\, \Big({\lambda\over \theta_*}\Big)^{3\over 4}\, \sqrt{\Big|Q'\Big({\sin\theta_*\over \lambda}\Big)\Big|}\,Q'\Big({\sin\theta_*\over \lambda}\Big)\times \bell_\epsilon\, a_0^{1\over 4}\,  |Q'(a_0)|^{-{1\over 2}}\, , 
\ee
which can be inserted in \Eq{detmi} to yield the final result
\be
\det \wt\S_\epsilon \underset{\theta_*\to 0}{\sim}  2\, \Big({\theta_*\over \lambda}\Big)^{1\over 4}\ln{1\over \theta_*}\,  \sqrt{\Big|Q'\Big({\sin\theta_*\over \lambda}\Big)\Big|}\times \epsilon\,\sign(Q')\,a_0^{1\over 4}\,  \sqrt{|Q'(a_0)|}\,\sqrt{1-(\lambda a_0)^2}\, .
\label{final_detS2}
\ee
In the above expression, $\sign Q'$ is a sign independent of $a_0$ because the function $Q(a)$ is invertible and thus monotonic. Moreover, we also have $Q'(a)\neq 0$ for all $a>0$ since otherwise $A'(q)=1/Q'(a)$ would not be defined at some $ q\neq Q(0)$.\footnote{We may however have $ Q'(0)=1/A'(Q(0))\in\{0, +\infty, -\infty\}$.} Hence, we conclude that:

$\bullet$ The small- and big-cap instantons yield invertible operators $\wt\S_\epsilon$ when $0<a_0<1$.

$\bullet$ When $Q$ and $A$ are increasing, we have $\det\wt \S_+ >0$ and $\det\wt \S_- <0$. On the contrary,  when $Q$ and $A$ are decreasing, the converse is true \ie $\det\wt \S_+ <0$ and $\det\wt \S_- >0$.


\subsubsection{Quadratic fluctuations of the length}

What remains to be done to obtain the expression of $\wt\Psi(q_0)$ in the semi-classical approximation is to find the value of $\wt \K_\epsilon$ defined in \Eq{psihat}. To this end, we determine $\wt\S_\epsilon^{-1} \wt V_q(\bq_\epsilon)$, which satisfies
\be
\left\{\begin{array}{l}
\wt\S_\epsilon \wt f_\epsilon = \wt V_q(\bq_\epsilon) \, , \\
\wt f_\epsilon(\tau_\epsilon)=0\, , \quad \wt f_\epsilon(1)=0\, .\esps\end{array}\right.
\ee
Defining
\be
\wt f_\epsilon(\tau) =  {\overset{\smile} {\!\!f}_{\!\!\!\epsilon}(\tau)\over A'(\bq_\epsilon(\tau))}\, ,
\ee
the above system can be re-written in terms of the operator $\S_\epsilon$,
 \be
\left\{\begin{array}{l}
\S_\epsilon\, \overset{\smile} {\!\!f}_{\!\!\!\epsilon} = V_a(\ba_\epsilon) \, , \\
\overset{\smile} {\!\!f}_{\!\!\!\epsilon}(\tau_\epsilon)=0\, , \quad \overset{\smile} {\!\!f}_{\!\!\!\epsilon}(1)=0\, .\esps\end{array}\right.
\ee
Since it  turns out to be identical to the one given in \Eq{syss}, we conclude that $\overset{\smile} {\!\!f}_{\!\!\!\epsilon} =f_\epsilon$. As a result, we obtain that 
\be
\left\{\begin{array}{l}
\wt\S_\epsilon^{-1} \wt V_q(\bq_\epsilon)= \dis  {1\over A'(\bq_\epsilon(\tau))}\, \S_\epsilon^{-1}V_a(\ba_\epsilon)\\
\wt V_q(\bq_\epsilon) = A'(\bq_\epsilon(\tau)) \, V_a(\ba_\epsilon)\esps
\end{array}\right.\; \Longrightarrow\quad \wt\K_\epsilon=\K_\epsilon\, ,
\ee 
which was anticipated before and shows that the result in \Eq{fluI} remains unmodified.\footnote{ Notice that we also obtain that $ \delta \check a \equiv \delta a+{\delta\ell\over \bell_\epsilon}\, \S_\epsilon^{-1}V_a(\ba_\epsilon) = A'(\bq_\epsilon(\tau))\! \left(\delta q +{\delta\ell\over \bell_\epsilon}\, \wt\S_\epsilon^{-1}\wt V_q(\bq_\epsilon) \right)\!+\O((\delta q)^2)\equiv A'(\bq_\epsilon(\tau))\,\delta\check q+\O((\delta q)^2$. Hence, it is equivalent to present the computation in \Eq{Zwedge} as a change of variable from $\delta a$ to $\delta q$ or from $\delta \check a$ to $\delta \check q$ (see \Eq{psii}).}

Using the expression of the Gaussian fluctuations $\wt Z_\epsilon(q_0)$ of $q$ given in  \Eqs{ztilde} and~(\ref{final_detS2}), the wavefunction $\wt\Psi(q_0)$ approximated semi-classically reads  
\be
\wt\Psi(q_0) = {\wt\Cc}_s(\theta_*)\sum_{\epsilon=\pm 1} {1\over \sqrt{\epsilon\,\sign(Q')}}\, {\exp\!\Big[\epsilon s\, \dis{2v_3\over \hbar\lambda^2}\,\big(1-(\lambda a_0)^2\big)^{3\over 2}\Big]\over a_0^{1\over 8}\;|Q'(a_0)|^{1\over 4}\,\big(1-(\lambda a_0)^2\big)^{1\over 4}}\, (1+\O(\hbar))\, ,\quad 0<\lambda a_0<1\, ,
\label{1st ex}
\ee
where $\wt\Cc_s(\theta_*)$ is the regulator-dependant coefficient 
\be
\wt\Cc_s(\theta_*)=\alpha\,\sqrt{i\pi} \, \Big({\pi \hbar\over 3sv_3}\Big)^{1\over 4}\exp\!\Big[\!-\!s\, {2v_3\over \hbar \lambda^2}\Big]\Big({\lambda \over \theta_*}\Big)^{1\over 8}\, \Big|Q'\Big({\sin\theta_*\over \lambda}\Big)\Big|^{-{1\over 4}}\, .
\ee
Alternatively, we may express the wavefunction explicitly in terms of $q_0$ as follows, 
\be
\wt\Psi(q_0) = {\wt\Cc}_s(\theta_*)\sum_{\epsilon=\pm 1} {1\over \sqrt{\epsilon\,\sign(A')}}\, {\exp\!\Big[\epsilon s\, \dis{2v_3\over \hbar\lambda^2}\,\big(1-(\lambda A(q_0))^2\big)^{3\over 2}\Big]\over A(q_0)^{1\over 8}\;|A'(q_0)|^{-{1\over 4}}\,\big(1-(\lambda A(q_0))^2\big)^{1\over 4}}\, (1+\O(\hbar))\, .
\label{psifinal}
\ee


\subsection{Difference with previous works}
\label{cpw}

Before concluding, we would like to stress that even if the analysis presented in our paper shows similarities with those developed in previous works, it is to a large extent different and leads to different results. 

For instance, the starting point in \Refe{Turok1} is \Eqs{psa0} and~(\ref{SE})  (considered in Lorentzian time), which describe the amplitude for the universe to evolve from $a=0$ to $a=a_0$ in a proper time $\ell$. Following Refs.~\cite{Halli, Halli2}, the authors then apply the following change of time and field redefinition 
\be
\ell \,\d \tau = {\ell\over a(\tau)}\,\d u\, , \quad ~~a=A(q)=\sqrt{q}\, .
\ee
Imposing without loss of generality $u(0)=0$, the action becomes (keeping a Euclidean signature) 
\be
S_{\rm E}=3sv_3\int_0^{u(1)}\d u \,\bigg[ {1\over 4\ell} \Big({\d q\over \d u}\Big)^2+\ell\, (1-\lambda q) \bigg]\, , \quad \where\quad u(1) = \int_0^1\d\tau\, \sqrt{q(\tau)}\, . 
\label{acu}
\ee
Notice that in general $u(1)\neq 1$, as stressed in Chapter~5.1 of \Refe{Pol1}. However, the authors of Refs. \cite{Turok1, Halli-Hartle,Halli, Halli2} take the upper bound of the action to be 1. As a result, $\ell$ (denoted by a constant $N$ in these works) {\it is not} any more the proper length of the line segment (or the proper time in Lorentzian signature). Hence, their integral over $\ell$ is not an integral over the inequivalent classes of metrics $g_{00}$ modulo diffeomorphisms, \ie not an integral over the moduli space of the base segment.

In fact, this very notion of integration over the moduli space of a base manifold is central for defining gauged-fixed quantum amplitudes. For instance, in closed bosonic string, the base manifold is a Riemann surface of genus $\mathfrak{g}$ and metric $g_{ab}$, $a,b\in\{0,1\}$, while the target space is a $26$-dimensional spacetime parametrized by $X^\mu$, with $\sigma$-model metric $G_{\mu\nu}(X)$. In complete analogy with our analysis, amplitudes in closed bosonic string theory are defined as integrals over the moduli space of the genus-$\mathfrak{g}$ base and paths integrals over the fields $X^\mu$. 

So, the difference between the analyses of the above mentioned references and ours is that we do keep $u(1)$ in the upper bound of the action in \Eq{acu}. In that case, the fact that the Lagrangian becomes quadratic for this specific choices of Euclidean time and field $q$ is of no use. Indeed, the principle of least action cannot be applied in these variables since the upper bound yields an unusual extra contribution when varying $q\to q+\delta q$. Hence, Gaussian integral formulas cannot be applied  under these conditions. 
In a similar way, using Euclidean conformal time $\eta_{\rm E}$ satisfying $a(\tau)\, \d\eta_{\rm E} = \ell\, \d \tau$ simplifies the expression of the Lagrangian of the scale factor but prevents from having fixed conformal-time boundaries for the action. In general, it is therefore important to implement the gauge fixing of the Euclidean-time reparametrizations by choosing any fiducial metric as shown in \Eq{l2}, \ie independently of the path of the scale factor.      


\section{Wheeler--DeWitt equation}
\label{WDW}

The Wheeler--DeWitt equation is the differential equation whose solutions are the wavefunctions associated with the states of the entire quantum gravity Hilbert space~\cite{DeWitt}. However, its derivation leads to an ambiguity that  arises
when applying 
canonical quantization on its classical counterpart. In this section, we lift this ambiguity for each given choice of measure $\D q$ in the definition of the  wavefunctions, at least at the semi-classical level. 


\subsection{Ambiguity in the Wheeler--DeWitt equation}

In this subsection, we derive the Wheeler--DeWitt equation 
 satisfied by the wavefunctions corresponding to the choice of measure $\D q$ in the path integrals. This will be done by working in Lorentzian 
 signature.


\subsubsection{Hamiltonian constraint}

The starting point is to note that the path integral of a total  functional derivative vanishes,
\be
0=\int {\D N \over \Vol(\Diff[N^2])}\, {\delta\over \delta N(x^0)}\, e^{i\wt S[N^2,q]}\, , \quad \mbox{for all $x^0$}\, . 
\ee
Indeed, the right-hand side contains a multiplicative factor that is a boundary term, 
\begin{align}
&\int \d N(x^0)\, {\d\over \d N(x^0)} \, e^{i\wt L\left(N(x^0),q(x^0),\dot q(x^0)\right)}\, , \nonumber\\
\where  \quad & \wt L(N,q,\dot q)= 3 v_3\Big(\!-\!{A(q)A'(q)^2\over N}\,\dot q^2+N \wt V(q)\Big)\, ,\quad \dot q\equiv {\d q\over \d x^0}\, , 
\end{align}
and which vanishes for a suitable contour of integration of $N(x^0)$,
\be
\left[\exp\!\left\{i3 v_3 \Big(\!-\!{A(q)A'(q)^2\over N}\,\dot q^2+N \wt V(q)\Big)\Big|_{x^0}\right\}\right]_{N(x^0)={i 0_+}}^{N(x^0)=i\infty \sign[\wt V(q(x^0))]}=0-0\, .
\ee
Using this result, we have 
\be
0=\int_\C {\D N \,\D q\over \Vol(\Diff[N^2])} \left.i{\partial \wt L\over \partial N}\right|_{x^0}\, e^{i\wt S[N^2,q]}\, , 
\label{dLdN}
\ee
where the path integral is over any class of paths denoted by $\C$. 

The  conjugate variables of the generalized coordinates $q$ and $N$ are given by
\be
\pi_q ={\partial\wt L\over \partial \dot q} = -6 v_3\,{AA'^2\over N}\,\dot q\, , ~~\quad \pi_N ={\partial\wt L\over \partial \dot N} =0\, , 
\ee 
and the classical Hamiltonian obtained by  a Legendre transformation reads
\be
\wt H = \pi_q\dot q+\pi_N\dot N-\wt L = N\Big(\!- \!{1\over 12v_3}\, {\pi_q^2\over AA'^2}-3v_3\wt V\Big)\, . 
\label{H/N}
\ee
As a result
\be
{\partial \wt H\over \partial N} =  { \wt H\over  N} = -3 v_3\Big({AA'^2\over N^2}\,\dot q^2+ \wt V(q)\Big) = -{\partial \wt L\over \partial N}\, ,
\ee
which can be used in \Eq{dLdN} to obtain
\be
0=-i \int_\C {\D N \,\D q\over \Vol(\Diff[N^2])} \left.{ \wt H\over N}\right|_{x^0}\, e^{i\wt S[N^2,q]}\, .
\ee

The above result, which is valid for any class of paths $\C$, is equivalent in  the Hilbert-space formalism to saying that all matrix elements of the quantum Hamiltonian divided by the lapse function are vanishing. This quantum operator is thus annihilating all kets, 
\be
{\wt \Hc\over \Nc}| \wt\Psi_\C\rangle = 0\, , \quad\mbox{for all $| \wt\Psi_\C\rangle$}\, .
\ee
In terms of wavefunctions $\wt\Psi_\C$, the canonical quantization of the classical expression of $\wt H/N$ given in \Eq{H/N}  is obtained by replacing
\be
q\longrightarrow  q_0\, ,  ~~\quad \pi_q\longrightarrow  -i\hbar\, {\d\over \d q_0}\, . 
\ee
Indeed, this prescription provides a representation of the 
 commutator relation $[q,\pi_q]=i\hbar$ on the Hilbert space of wavefunctions.
However, because classically 
\be
{\pi_q^2\over AA'^2} = {1\over AA'^2\, \rho_1\,\rho_2}\, \pi_q \,\rho_1\, \pi_q\,\rho_2 \quad \mbox{for any functions $\rho_1(q_0)$, $\rho_2(q_0)$}\, , 
\ee
the process of quantization yields an ambiguity in the form of the Wheeler--DeWitt equation,~
\be
{\wt H\over N}\,\wt\Psi_\C\equiv {\hbar^2\over 12v_3}\, {1\over AA'^2 \rho_1\rho_2}\, {\d\over \d q_0}\Big[\rho_1\, {\d\over \d q_0}\big(\rho_2\wt \Psi_\C\big)\Big]- 3v_3\wt V\wt \Psi_\C=0\, .
\ee
 Defining 
\be
\wt \rho= \rho_1\rho_2^2\, , \quad ~~\wt \omega={(\rho_1\rho_2')'\over \rho_1\rho_2}\, , 
\ee
this simplifies to 
\be
{\wt H\over N}\,\wt\Psi_\C\equiv {\hbar^2\over 12v_3}\,  {1\over AA'^2} \bigg[{1\over \wt \rho}\, {\d\over \d q_0}\Big(\wt \rho\, {\d\wt \Psi_\C\over \d q_0}\Big)+\wt \omega\wt\Psi_\C\bigg]- 3v_3\wt V\wt \Psi_\C=0\, ,
\label{WDWtil}
\ee
where $A$, $A'$ along with $\wt\rho$, $\wt\omega$ depend on $q_0$. 


\subsubsection{Alternative formulation}

In some cases, it is convenient to formulate  the dependance of the wavefunctions on boundary data in terms of the scale factor $a_0\in[0,+\infty)$ instead of $q_0=Q(a_0)\in [Q(0),Q(+\infty))$. 
This corresponds to applying the change of variable
\be
\Psi_{A\C}(a_0)\equiv \wt\Psi_\C(Q(a_0))\, .
\label{chvar}
\ee
Notice that even though the functions $\Psi_{A\C}$ depend on $a_0$, they still correspond to path integrals based on the measure $\D q$. For instance,  among all wavefunctions $\Psi_{A\C}(a_0)$, the one denoted $\Psi_A(a_0)$ corresponding to the ground-state  is to be distinguished from $\Psi(a_0)$ given in \Eq{psia00}, which corresponds actually to the particular case $A(q)=q$ so that $\D a\equiv \D q$.  

Applying the  change of variable in the Hamiltonian constraint in \Eq{WDWtil}, one obtains the equivalent equation 
\begin{align}
{\wt H\over N}\,\wt\Psi_\C\equiv{H_A\over N}\,\Psi_{A\C}&\equiv {\hbar^2\over 12v_3}\,  {1\over a_0} \bigg[{1\over \rho_A}\, {\d\over \d a_0}\Big(\rho_A\, {\d\Psi_{A\C}\over \d a_0}\Big)+ \omega_A \Psi_{A\C}\bigg]- 3v_3 V  \Psi_{A\C}=0\, ,\nonumber \\
\where\quad~~ \rho_A(a_0)&= {\wt\rho(Q(a_0))\over |Q'(a_0)|}\, , ~~\quad \omega_A(a_0) = \wt\omega(Q(a_0))\, Q'(a_0)^2\, .\esp
\label{WDWtil2}
\end{align}
In fact, the above result can be obtained by deriving the Hamiltonian constraint in terms of the conjugate variables $a$, $\pi_a$ and applying the canonical quantization prescription~ 
\be
a\longrightarrow  a_0\, ,  ~~\quad \pi_a\longrightarrow  -i\hbar\, {\d\over \d a_0}\, ,
\ee
while taking into account the ambiguity of the classical quantity  $a^{-1}\pi_a^2$. 


\subsection{Resolution of the ambiguity}
\label{reWD}

In \Sect{field redef}, we have determined at the semi-classical level the ground-state wavefunction $\wt \Psi(q_0)$, which must be a particular solution of the Wheeler--DeWitt equation. Hence,  we can lift the ambiguity of the equation by imposing that this is indeed the case.

To this end, we put \Eq{WDWtil} in the following form, 
\be
\wt\Psi_\C''+{\wt\rho'\over \wt \rho}\, \wt\Psi'_\C+\wt\omega\wt\Psi_\C = {\Pc\over \hbar^2}\, \wt \Psi_\C\, , \quad \where\quad \Pc(q_0) = 36 v_3^2 \,AA'^2\,\wt V\, ,
\ee
and apply the WKB method~\cite{wkb}. This is done by writing 
\be
\wt \Psi_\C = \exp\!\Big[ {i\over \hbar} \Big( \Fc_0 + {\hbar\over i}\Fc_1 + \O(\hbar^2) \Big) \Big] \, ,
\ee
where $\Fc_0$, $\Fc_1$ are functions of $q_0$, 
 leading to
\be 
\Big[-{1\over \hbar^2}\, \Fc_0'^2 + {i\over \hbar}\Big(\Fc_0''+2\Fc_0'\Fc_1'+{\wt\rho'\over \wt\rho}\, \Fc_0'\Big)+\O(1)\Big]\wt\Psi_\C = {\Pc\over \hbar^2}\, \wt \Psi_\C\, .
\ee
The unknown function $\wt\omega$ is absorbed in the $\O(1)$ terms and thus cannot be determined at the semi-classical level. On the contrary, $\Fc_0$ is found by  identifying the contributions  $\O(1/\hbar^2)$, while $\Fc_1$ is such that the term  $\O(1/\hbar)$ vanishes. Two cases can be considered:

$\bullet$ When $ \Pc(q_0)>0$ \ie $0<\lambda a_0<1$, we find two imaginary solutions for $\Fc_0$, 
\begin{align}
\Fc_0&=\epsilon s\sign(Q')\, i \int_{Q(1/\lambda)}^{q_0}\d q\,  \sqrt{\Pc(q)}+\cst= -\epsilon s\,i\, \dis{2v_3\over \lambda^2}\,\big(1-(\lambda a_0)^2\big)^{3\over 2}+\cst \, ,  \nonumber \\
\Fc_1&= -{1\over 4}\ln\!\big(  \Pc\, \wt \rho^2\big)+\cst\, , 
\end{align}
where $\epsilon s$ denotes either $+1$ or $-1$. Consistently, there  are two linearly independent modes which lead to 
\be
\wt\Psi_\C(q_0) = \sum_{\epsilon=\pm 1} N_{\C\epsilon}\, {\exp\!\Big[\epsilon s\, \dis{2v_3\over \hbar\lambda^2}\,\big(1-(\lambda A(q_0))^2\big)^{3\over 2}\Big]\over |\wt \rho(q_0)|^{1\over 2}\,A(q_0)^{1\over 2}\,|A'(q_0)|^{1\over 2}\, \big(1-(\lambda A(q_0))^2\big)^{1\over 4}}\, (1+\O(\hbar))\, ,\quad  0<\lambda A(q_0)<1\, ,
\ee
where $N_{\C\epsilon}$ are two integration constants. Comparing this result with \Eq{psifinal} we can identify  $\wt\rho$,
\be
\tilde \rho(q_0) = A(q_0)^{-{3\over 4}}\,  |A'(q_0)|^{-{3\over 2}}\, . 
\label{rhoo1}
\ee
Using the relation between $\wt \rho$ and $\rho_A$, we also obtain the alternative form of the wavefunctions
\begin{align}
\Psi_{A\C}(a_0) &= \sum_{\epsilon=\pm 1} N_{\C\epsilon}\, {\exp\!\Big[\epsilon s\, \dis{2v_3\over \hbar\lambda^2}\,\big(1-(\lambda a_0)^2\big)^{3\over 2}\Big]\over |\rho_A(a_0)|^{1\over 2}\,a_0^{1\over 2}\, \big(1-(\lambda a_0)^2\big)^{1\over 4}}\, (1+\O(\hbar))\, ,\quad 0<\lambda a_0<1\, ,\nonumber \\
\where \quad \rho_A(a_0) &= a_0^{-{3\over 4}}\,  |Q'(a_0)|^{1\over 2}\,, 
\label{rhoo}
\end{align}
which is consistent with \Eq{1st ex}.
Hence, we have $\wt \rho(q_0)>0$ and $\rho_A(a_0)>0$ for $0<\lambda a_0<1$  (thanks to the remark below \Eq{final_detS2}). 
Moreover, the values $N_\epsilon$ of the mode coefficients  $N_{\C\epsilon}$ (with arbitrary normalization) that select the ground state are given by
\be
N_\epsilon = {1\over \sqrt{\epsilon \sign(Q')}}\, .
\ee

$\bullet$ For completeness, let us mention that in the second case $ \Pc(q_0)<0$ \ie $\lambda a_0>1$, the two solutions for $\Fc_0$ are real, 
\begin{align}
\Fc_0&=\epsilon s\sign(Q')\int_{Q(1/\lambda)}^{q_0}\d q\, \sqrt{-\Pc(q)}+\cst= \epsilon s\, \dis{2v_3\over \lambda^2}\,\big((\lambda a_0)^2-1\big)^{3\over 2}+\cst \, ,  \nonumber \\
\Fc_1&= -{1\over 4} \ln\!\big(\!-\!\Pc\, \wt \rho^2\big)+\cst \, , 
\end{align}
which lead to 
\be
\wt\Psi_\C(q_0) = \sum_{\epsilon=\pm 1} M_{\C\epsilon}\, {\exp\!\Big[i\epsilon s\, \dis{2v_3\over \hbar\lambda^2}\,\big((\lambda A(q_0))^2-1\big)^{3\over 2}\Big]\over |\wt \rho(q_0)|^{1\over 2}\,A(q_0)^{1\over 2}\,|A'(q_0)|^{1\over 2}\, \big((\lambda A(q_0))^2-1\big)^{1\over 4}}\, (1+\O(\hbar))\, ,\quad \lambda a_0>1\, ,
\ee
where $M_{\C\epsilon}$ are integration constants. However, we do not have an explicit expression of a solution of the Wheeler--DeWitt equation derived by evaluating a path integral when $\lambda a_0>1$.\footnote{We hope to provide such a computation at the semi-classical level in a forthcoming work.} Hence, we cannot determine $\wt \rho(q_0)$ when $\lambda a_0>1$.\footnote{However, notice that  the constants $N_{\C\epsilon}$ and  $M_{\C\epsilon}$ can be related, whatever the functions $\wt\rho$ and $\wt\omega$ are. This can be done by solving the Wheeler--DeWitt equation in the neighbourhood of $\lambda a_0=1$, where the solutions are expressed in terms of Airy functions. It  is then possible to match the solutions and thus the  integration constants in the three regions $ 0<\lambda a_0<1$, $\lambda a_0\simeq 1$ and $\lambda a_0>1$.} Applying the change of variable from $q_0$ to $a_0$, we also have
\be
\Psi_{A\C}(a_0) = \sum_{\epsilon=\pm 1} M_{\C\epsilon}\, {\exp\!\Big[i\epsilon s\, \dis{2v_3\over \hbar\lambda^2}\,\big((\lambda a_0)^2-1\big)^{3\over 2}\Big]\over |\rho_A(a_0)|^{1\over 2}\,a_0^{1\over 2}\, \big((\lambda a_0)^2-1\big)^{1\over 4}}\, (1+\O(\hbar))\, ,\quad \lambda a_0>1\, .
\label{rhoo2}
\ee

It should be noted that the asymptotic form of the function $\rho_A(a_0)$, which governs the ambiguity of the Wheeler--DeWitt equation at the semiclassical level, can be fixed by imposing boundary conditions for the modulus of the wavefunction at $a_0=0$ and/or $a_0\to \infty$~\cite{He:2015wla}. A more complete theory of quantum gravity may also lead to a resolution of the ambiguities -- see \eg Ref.~\cite{Nelson:2008vz} for work towards this direction, in the context of loop quantum cosmology. 

However, our approach in this work is different. We do not prescribe specific boundary conditions to select $\rho_A(a_0)$ and solve for the wavefunction. Rather, we construct a large family of no-boundary wavefunctions, based on the various choices $\D q$ of gauge-invariant path-integral measures, and then proceed to determine the Wheeler--DeWitt equation each such wavefunction satisfies. Equivalently, for each positive function $\rho_A(a_0)$, we can construct, \via a suitable field redefinition, a path-integral wavefunction that implements the no-boundary proposal. In the next section, we will show that, at least at the semi-classical level,  all these prescriptions yield identical observable predictions, despite the different asymptotic behavior the corresponding no-boundary wavefunctions exhibit. In particular, the inner product measure that renders the Hamiltonian Hermitian is such that the quantum probability density is universal, irrespectively of the choice $\D q$. 

\subsection{Crosscheck with the Schr\"odinger equation}
\label{cross}

A non-trivial check  of  the results achieved so far is obtained by choosing a particular field redefinition $a=A(q)$ for which the exact form of the Wheeler--DeWitt equation is  familiar. 

In terms of Euclidean time $t_{\rm E}=\ell \tau$, the action $\wt S_{\rm E}$ defined in \eq{stilde} is  
\be
\wt S_{\rm E}[\ell^2,q]=3sv_3\int_0^\ell\d t_{\rm E} \,\bigg[ A(q) A'(q)^2\Big({\d q\over \d t_{\rm E}}\Big)^2+ \wt V(q)\bigg]\, .
\ee 
 A relevant choice of target-minisuperspace coordinate $q$ is such that the metric of the $\sigma$-model action is canonical. Indeed,  imposing 
\be
G_{qq}=3v_3AA'^2={1\over 2}\quad \Longrightarrow\quad A(q) = \Big({3\over 2}\, {1\over \sqrt{6v_3}}\Big)^{2\over 3}\, \big|q-Q(0)\big|^{2\over 3}\, ,
\label{qcano}
\ee
the function $\wt \rho$ in \Eq{rhoo} turns out to be a constant, 
\be
\wt\rho(q_0) = (6v_3)^{3\over 4}\, . 
\label{rQM}
\ee
Therefore, the Wheeler--DeWitt equation~(\ref{WDWtil}) reduces to 
\be
\bigg[{\hbar^2\over 2}\, {\d^2\ \over \d q_0^2} +{\hbar^2\over 2}\, \wt\omega- 3v_3\wt V\bigg]\wt \Psi_\C= 0\, .
\label{tiS}
\ee
Omitting the contribution of  $\O(\hbar^2)$ to the potential, we recognize the time-independent Schr\"o\-din\-ger equation for a stationary state of vanishing energy. This is the expected result for the following reasons.

In quantum mechanics, the Lorentzian action of a particle of unit mass moving on a line of coordinate $q$ and subject to a force derived from the potential $3v_3\wt V(q)$ is 
\be
S^{\rm part}=\int_0^\ell\d t  \,\Big[{1\over 2}\, \Big({\d q\over \d t }\Big)^2-3v_3 \wt V(q)\Big]\, .
\ee
The Lagrangian is {\em opposite} to that corresponding to the gravity system (see \Eq{action2} in terms of the field $a=A(q)$), which has a negative kinetic term and a positive sign in front of the potential $3v_3\wt V(q)$.  
The amplitude for the particle to travel from an initial position $q_{\rm i}$ to a final one $q_0$ in a real 
 time $\ell$ is given by 
\be
U(q_{\rm i},q_0;\ell) = \int_{\textstyle \substack{\,q(0)=q_{\rm i} \\\,\, q(\ell)=q_0}}\!\D q \; e^{{i\over \hbar} S^{\rm part}}\, .
\ee
By discretizing time and varying the final position, one derives using standard manipulations that this amplitude satisfies the Schr\"odinger equation~\cite{Peskin},
\be
\bigg[\!-\!{\hbar^2\over 2}\, {\partial^2\ \over \partial q_0^2} + 3v_3\wt V(q_0)\bigg]U(q_{\rm i},q_0;\ell)=i\hbar\, {\partial\over \partial\ell}\,U(q_{\rm i},q_0;\ell)\, .
\ee
The stationary states $\wt \Psi^{\rm part}_{q_{\rm i},E}(q_0)$ of energy $E$  satisfy
\be
 \bigg[\!-\!{\hbar^2\over 2}\, {\d^2\ \over \d q_0^2} + 3v_3\wt V\bigg]\wt \Psi^{\rm part}_{q_{\rm i},E}=E \,\wt \Psi^{\rm part}_{q_{\rm i},E}\, .
\ee
Hence, for the gravity problem for which the total (kinetic + potential) energy is  $-E=0$,
the exact Wheeler--DeWitt equation is \Eq{tiS}, with $\wt \omega=0$.


\section{Quantum equivalence at the semi-classical level}
\label{Qequi}

In the previous section, we have identified for each choice of measure $\D q$ in the path integrals the expression 
of the function $\rho_A$ (or $\wt \rho$) appearing in the \WDW equation. This expression for $\rho_A$ is valid for $0<\lambda a_0<1$. However, we will 
show in this section that all these prescriptions yield identical observable predictions, at least at the semi-classical level. In fact, the identification of $\rho_A$ turns out to be somehow superfluous at this level of approximation, provided it is assumed to be positive. Moreover, any such function can be obtained for a suitable choice of field redefinition $q=Q(a)$ and associated path integral measure $\D q$, at least when $0<\lambda a_0<1$, as can be seen by integrating \Eq{rhoo} which yields 
\be
|Q(a_0)|=|Q(0)|+\int_0^{a_0}\d a\, a^{3\over 2}\, \rho_A(a)^2 \, .
\label{power}
\ee
For instance, $\rho_A(a_0)=a_0^p$, where $p$ is a real constant, is often considered in the literature. Since multiplying $\rho_A$ by any constant leaves \Eq{WDWtil2} invariant, the choice $q=Q(a) = a^{2p+{5\over 2}}$ yields the \WDW equation with a monomial form of $\rho_A$.  
 

\subsection{Hermiticity of the quantum Hamiltonians}
\label{hqh}

To 
 define probability amplitudes, we need to specify for each choice of path integral measure $\D q$ an inner product of the corresponding Hilbert space of wavefunctions. To this end, let us set
\be
\langle \Psi_{A1},\Psi_{A2}\rangle_A =\int_0^{+\infty} \d a_0\,  \mu_A(a_0) \,  \Psi_{A1}(a_0)^*\,  \Psi_{A2}(a_0)\, ,
\label{pr1}
\ee
where $\Psi_{A1}$, $\Psi_{A2}$ are at this stage arbitrary complex functions. Moreover, $\mu_A$ is a real positive function that we would like to interpret as a ``measure'' on the space of functions. Alternatively, we may apply the change of variable shown in \Eq{chvar} and define
\be
\langle \wt \Psi_1,\wt \Psi_2\rangle =\sign(Q')\int_{Q(0)}^{Q(+\infty)} \d q_0\, \wt \mu(q_0) \, \wt \Psi_1(q_0)^*\, \wt \Psi_2(q_0)\, ,
\label{pr2}
\ee
where $\wt \mu$ is related to $\mu_A$ so that the two  Hermitian forms are equivalent,
\be
\mu_A(a_0)=|Q'(a_0)|\, \wt \mu(Q(a_0))\quad \Longrightarrow\quad \langle \Psi_{A1},\Psi_{A2}\rangle_A= \langle \wt \Psi_1,\wt \Psi_2\rangle \, .
\ee


\subsubsection{Determining $\mu_A$ and $\wt\mu$} 
\label{detmu}

In order to find $\mu_A$, we are going to impose that the Hamiltonian differential operator $H_A/N$ in \Eq{WDWtil2} is Hermitian on a suitable Hilbert space. To this end, we derive the following identity by integrating by parts,
\begin{align}
\big\langle \Psi_{A1},{H_A\over N}\,\Psi_{A2}\big\rangle = &\;\big\langle {H_A^\dag\over N}\,\Psi_{A1},\Psi_{A2}\big\rangle\nonumber \\
&\, + {\hbar^2\over 12 v_3}\left[\rho_A\!\left({\mu_A\over a_0\rho_A}\, \Psi^*_{A1} \, {\d \Psi_{A2}\over \d a_0}-{\d\over \d a_0}\Big({\mu_A\over a_0\rho_A}\, \Psi^*_{A1}\Big)\Psi_{A2}\right)\right]_0^{+\infty}\, , 
\label{con}
\end{align}
where we have defined
\be
{H^\dag_A\over N}\,\Psi_{A\C}\equiv {\hbar^2\over 12v_3}\,  {1\over a_0}\bigg[{a_0\over \mu_A}\, {\d\over \d a_0}\Big(\rho_A\, {\d\over \d a_0}\Big({\mu_A\over a_0\rho_A}\, \Psi_{A\C}\Big)+ \omega_A \Psi_{A\C}\bigg]- 3v_3 V  \Psi_{A\C}\, .
\label{hdag}
\ee
For the Hamiltonian to be Hermitian, two conditions must be met: 

$\bullet$ We must have 
\be
{H_A\over N}\, \Psi_{A\C}={H^\dag_A\over N}\, \Psi_{A\C}\, , 
\label{hh}
\ee
which can be seen to be equivalent to  
\be
\kappa'\, {\Psi'_{A\C}\over \Psi_{A\C}} = -{1\over 2} \,\Big(\kappa''+\frac{\rho_A'}{\rho_A}\,\kappa'\Big)\, ,\quad \where\quad  \kappa = {\mu_A\over a_0\rho_A}\, .
\ee
If $\kappa'$ is non-vanishing,
we obtain a  separated differential equation by dividing by $\kappa'$, which leads upon integration to an expression of $\Psi_{A\C}$ in terms of $\kappa'\rho_A$. However, this cannot be true since $\Psi_{A\C}$ is arbitrary. Hence, $\kappa'\equiv 0$ \ie $\kappa$ is a constant. Because the normalization of wavefunctions is irrelevant, we may take $\kappa=1$ so that 
\be
\mu_A (a_0)= a_0\, \rho_A(a_0)\, , \quad \lambda a_0\ge 0\, ,\quad \and\quad \wt \mu (q_0)= A(q_0)\,A'(q_0)^2\, \wt\rho(q_0)\, . 
\label{fin}
\ee
As a result, since  $\mu_A$ and $\wt \mu$  are positive, the functions $\rho_A(a_0)$ and $\wt \rho(q_0)$ appearing in the \WDW equations must also be positive for all $\lambda a_0\ge 0$, and not only in the range $0<\lambda a_0<1$ as shown in \Sect{reWD}. This is necessary for the whole picture to be consistent. 

$\bullet$ Moreover, the boundary term in \Eq{con} must vanish for all functions of the Hilbert space,
\be
0=\left[\rho_A\Big( \Psi^*_{A1} \, {\d \Psi_{A2}\over \d a_0}-{\d\Psi^*_{A1}\over \d a_0} \, \Psi_{A2}\Big)\right]_0^{+\infty}\quad i.e. \quad 0=\left[\wt \rho\,\Big( \wt \Psi^*_{1} \, {\d \wt \Psi_{2}\over \d q_0}-{\d\wt \Psi^*_{1}\over \d q_0} \, \wt \Psi_{2}\Big)\right]_{Q(0)}^{Q(+\infty)}\,.
\label{bterm}
\ee

In Appendix~\ref{A5}, we recover the result $\mu_A=a_0\rho_A$  and derive for any complex functions $\Psi_{A_1}$, $\Psi_{A2}$ the following identity valid for all $\lambda a_0\ge 0$,
\begin{align}
\mu_A\, \bigg\{\Big({H_A\over N}\,\Psi_{A1}\Big)^*\, \Psi_{A2}& - \Psi_{A1}^*\, {H_A\over N}\,\Psi_{A2}\bigg\}\nonumber \\
&\equiv -{\hbar^2\over 12v_3}\, {\d\over \d a_0}\left\{\rho_A\Big( \Psi^*_{A1} \, {\d \Psi_{A2}\over \d a_0}-{\d\Psi^*_{A1}\over \d a_0} \, \Psi_{A2}\Big)\right\}  .
\label{local}
\end{align}
Applying this result to wavefunctions annihilated by the Hamiltonian, the left hand side is identically zero
and we obtain 
\be
\rho_A\Big( \Psi^*_{A1} \, {\d \Psi_{A2}\over \d a_0}-{\d\Psi^*_{A1}\over \d a_0} \, \Psi_{A2}\Big) = \cst\, , \quad \lambda a_0\ge 0\, .
\ee
As a result,  the vanishing of the boundary term in \Eq{bterm} is trivially true for all solutions of the \WDW equations. To put it another way, the two dimensional vectorial space of solutions of the \WDW equation is  a Hilbert space equipped with the Hermitian product defined in \Eq{pr1} (or \Eq{pr2}), 
 under which the Hamiltonian operator is Hermitian. 


\subsubsection{Crosscheck with quantum mechanics} 
\label{ccheck}

We can check our conclusions, 
by applying them to  the case where 
the path integral
measure $\D q$ corresponds to
a field $q$ with canonical kinetic term. As seen in \Sect{cross}, this amounts
to choosing $q=A^{-1}(a)$ given in \Eq{qcano}, for which we know that $\wt \rho(q_0)=(6v_3)^{3\over 4}$ when $0<A(q_0)<1$. Indeed, using \Eq{fin}, we find that $\wt\mu(q_0)$ is a constant
\be
\wt\mu(q_0) = (6v_3)^{-{1\over 4}}\quad \when \quad 0<A(q_0)<1\, .
\ee
This is the correct answer, as it reproduces the well known prescription in quantum mechanics where the wavefunction norm 
 involves a constant measure. 


\subsection{Universality at the semi-classical level}
\label{univ}

The considerations of the previous subsection are valid for any choice of positive function $\rho_A(a_0)$, $\lambda a_0\ge 0$, in the WDW equation and corresponding measure $\mu_A$ in the Hilbert space of wavefunctions. An immediate consequence is that at the semi-classical level, the probability amplitudes $\sqrt{\mu_A}\,\Psi_{A\C}$ are universal, since \Eqs{rhoo} and~(\ref{rhoo2}) yield
\be
\sqrt{\mu_A(a_0)} \Psi_{A\C}(a_0) =  \left\{ \begin{array}{lr}
\dis \sum_{\epsilon=\pm 1} N_{\C\epsilon}\, {\exp\!\Big[\dis \epsilon s\, {2v_3\over \hbar\lambda^2}\,\big(1-(\lambda a_0)^2\big)^{3\over 2}\Big]\over \dis \big(1-(\lambda a_0)^2\big)^{1\over 4}}\, (1+\O(\hbar))\, ,& 0<\lambda a_0<1\, ,\\
\dis \sum_{\epsilon=\pm 1} M_{\C\epsilon}\, {\exp\!\Big[\dis i\epsilon s\, {2v_3\over \hbar\lambda^2}\,\big((\lambda a_0)^2-1\big)^{3\over 2}\Big]\over \dis \big((\lambda a_0)^2-1\big)^{1\over 4}}\, (1+\O(\hbar))\, ,& \lambda a_0>1\, .
\end{array}
\right.
\ee
In fact, the choices of functions $\rho_A$ and $\omega_A$ in the \WDW equation~(\ref{WDWtil2}) can only affect corrections in $\hbar$ to the  Hermitian product defined in \Eq{pr1} 
beyond the semi-classical level.\footnote{
For completeness, one must also show that this fact remains true at $\lambda a_0\simeq 1$ and $\lambda a_0\to 0$. This turns out to be the case, as can be demonstrated by applying the  WKB method to the \WDW equation with arbitrary $\rho_A$ and  $\omega_A$,  in neighbourhoods of $\lambda a_0= 1$ and $\lambda a_0= 0$. The solutions in these cases can be written in terms of Airy functions and parabolic cylinder functions, respectively.} 


\subsubsection{ Normalizability}
\label{nonor}

Another consequence is that none of the wavefunctions of  this Hilbert space is normalizable. Indeed, when $|M_{\C+}|\neq |M_{\C-}|$ we have 
\be
\mu_A(a_0)\, |\Psi_{A\C}|^2\underset{a_0\to+\infty}\sim {1\over a_0}\, \bigg| \sum_{\epsilon=\pm1}M_{\C\epsilon} \exp\!\Big[\dis i\epsilon s\, {2v_3\over \hbar\lambda^2}\,\big((\lambda a_0)^2-1\big)^{3\over 2}\Big]\bigg|^2>{1\over a_0}\, \big||M_{\C+}|-|M_{\C-}|\big|^2\, ,
\ee
while for $|M_{\C+}| = |M_{\C-}|\neq 0$, 
\be
\mu_A(a_0)\, |\Psi_{A\C}|^2\underset{a_0\to+\infty}\sim {2|M_{\C+}|^2\over a_0}\Big(1+\cos\!\big[2v_3\lambda \,a_0^3\big]\Big)\, .
\ee
In these cases the ``norms'' are infinite due to the logarithmic divergences of the integrals at $a_0\to +\infty$, while for $M_{\C+}=M_{\C-}=0$ the norm is of course vanishing.  Hence, in the simplest minisuperspace model we are considering, there are no normalizable  states in the Hilbert space, including 
the  no-boundary ``ground state.'' So at best we can use these wavefunctions to define relative probabilities, in terms of ratios of the probability densities evaluated at different points of minisuperspace.  Interesting work on defining probabilities and constructing observables in quantum cosmology includes Refs.~\cite{Gibbons,Vilenkin5, HHH1, HHH2, HHH3}.

In more realistic  cases, the minisuperspace models comprise extra matter degrees of freedom, besides the scale factor $a$. DeWitt considers at least one matter field associated with a dust filling universe~\cite{DeWitt}, while Hartle and Hawking include a conformally coupled scalar field~\cite{HH}. One may also consider including an inflaton field with appropriate potential, for applications to inflationary cosmology. As a result, the Hilbert spaces of solutions of the corresponding \WDW equations become infinite dimensional. As discussed in \Refe{DeWitt}, one may choose a suitable basis labelled by quantum numbers associated  with the matter Hamiltonian, and define a proper Hermitian product. Presumably, in the more involved general cases, one can construct normalizable solutions of the \WDW equation, in terms of superpositions of the basis states. Moreover, there should be superpositions exhibiting classical behavior, which can be interpreted as wave-packets ``moving'' in minisuperspace~\cite{DeWitt}. It would be desirable to construct normalizable solutions in terms of path integrals, implementing the no-boundary proposal. Notice however that for the conformally coupled scalar field of \cite{HH}, the \WDW equation can be cast into a separable form, and the large $a_0$ asymptotic behavior of the solutions becomes independent of the matter Hamiltonian quantum number. So in this case, the wavefunction factorizes in the large $a_0$ limit, exhibiting similar asymptotic behavior to the one described in this work. As a result, the solutions in this conformally coupled scalar field case continue to be non-normalizable.


\subsubsection{Comparison with the literature} 
\label{complit}

We can compare our analysis with that of \Refe{DeWitt}. In this work, the minisuperspace model does not implement a cosmological term but takes into account, as outlined above, matter particles of dust characterised by their own Hamiltonian. The author makes a choice of  a  \WDW equation associated with the coupled system ``scale factor + dust'' corresponding to 
\be
\rho_A(a_0)=a_0^p \quad \with \quad p=-1\, , \quad  \and \quad \omega_A(a_0)={7\over 16}\, {1\over a_0^2}\, .
\ee    
From our discussion below \Eq{power}, this choice of $\rho_A$ is valid if one chooses to define the wavefunctions with the measure $\D q$ associated with the field 
\be
q=Q(a)=\sqrt{a}\, .
\ee
However, we do not know what the correct expression of the function $\omega_A(a_0)$ should be in this case. Finally, the Hermitian product used in \Refe{DeWitt} uses a trivial measure $\mu_A$, which is consistent with our analysis since 
\be 
\mu_A(a_0) = 1 = a_0\, \rho_A(a_0)\, .
\ee

It is also interesting to mention the discrepancy between our results and some conclusions  reported in~\Refe{HH}, where the discussion of the measure $\mu_A$ is presented in the case of the minisuperspace model we study.  They consider the \WDW equation in a form where $\rho_A(a_0)=a_0^p$ and  $\omega_A(a_0)=0$.  From our analysis, we conclude that the Hilbert space measure must be $\mu_A(a_0)=a_0^{p+1}$. However, the authors of \Refe{HH} take $\mu_A(a_0)=a_0^{p}$, which leads to the conclusion that all solutions of the \WDW equation are normalizable. The origin of the disagreement is that instead of imposing  hermiticity of the Hamiltonian $H_A/N$, they  impose hermiticity of what they call the ``\WDW operator,'' which is $a_0H_A/N$. In practice, this amounts to changing $\mu_A\to \mu_A a_0$ everywhere in our derivation of \Sect{hqh}, and thus to replacing \Eq{fin} with $\mu_A a_0 = a_0 \rho_A$ \ie $\mu_A=\rho_A$.


\section{Conclusion}
\label{conclu}

In this work, we have considered the minisuperspace model describing a homogeneous and isotropic universe with positive cosmological constant. More specifically, we focused on the wavefunction defined as a Euclidean path integral satisfying the ``no-boundary proposal.'' 

The invariance under redefinitions of the lapse function imposes a gauge fixing of the reparametrization symmetry of Euclidean time. To do so, we have applied a  procedure 
 from first-quantized string theory, where quantum amplitudes are defined as path integrals of  
two-dimensional field theories. In both cases, the gauged-fixed path integrals involve integrals over the moduli spaces of metrics, and the results do not depend on  the gauge choice. On the contrary,  the redefinitions of the scale factor lead to inequivalent choices of diffeomorphism-invariant path-integral measures.
All prescriptions yield different forms of the \WDW equation and thus {\em a priori} distinct quantum theories. However, the quantum Hilbert spaces of wavefunctions lead to universal predictions, at least at the semi-classical level.  It would be highly interesting to  determine whether this remains true beyond this  level of approximation, or even exactly.

In our analysis, the ground-state wave functions are evaluated using the steepest-descent method when the scale factor on the boundary 3-sphere satisfies $0<\lambda a_0 <1$. We hope to extend  this result in the near future to the case $\lambda a_0>1$, for which the instanton solutions are complex. Computing the path integrals for $\lambda a_0=0$ is also challenging since there  exists a one-parameter family of instanton solutions in this case, implying that a  functional determinant possesses a vanishing eigenvalue. 

The techniques we have used to fix the gauge of the diffeomorphisms as well as to compute the quantum fluctuations of the scale factor 
can be implemented in richer setups. In particular, it would be very natural to consider minisuperspace models involving scalar fields with potentials possessing non-negative (local) minima, and possibly to account for the statistically natural emergence of  a long enough period of inflation. Taking into account extra degrees of freedom also  allows for the existence of wave-packets ``moving'' in minisuperspace~\cite{DeWitt}. Hence, it would be very interesting to see whether normalizable states constructed as path integrals obeying the no-boundary proposal  exist. 
In this work, we have considered the amplitude describing the transition from ``nothing'', the ``space'' reduced to the empty set,  to a 3-sphere. This serves as the definition of the ``ground state'' wavefunction of Hartle and Hawking. It would be worth investigating whether more general transitions may be related to ``excited states.''  This may require insertions of operators in the path integrals, in the spirit of insertions of vertex operators in string amplitudes, which  describe excited states of the string.


\section*{Acknowledgements}

The authors would like to thank Lihui Liu for useful inputs during the realization of this work. H.P. would like to thank the University of Cyprus for hospitality.  


\begin{appendices}
\makeatletter
\DeclareRobustCommand{\@seccntformat}[1]{%
  \def\temp@@a{#1}%
  \def\temp@@b{section}%
  \ifx\temp@@a\temp@@b
  \appendixname
  \else
  \csname the#1\endcsname\quad%
  \fi
} 
\makeatother

\section{}
\label{A0}
\renewcommand{\theequation}{A.\arabic{equation}}

For completeness, we display in this Appendix proofs of simple theorems and technical details used in the core of the paper.   


\subsection{On the fiducial metric}
\label{A1}

\begin{theorem}  Any metric $g_{00}$ in class $\ell$, defined in an arbitrary domain $[x^0_{\rm Ei},x^0_{\rm Ef}]$, can be written as $g_{00}=\hat g^\xi_{00}[\ell]$ for some $\xi\in\Diff[\hat g_{00}[\ell]]$. 
\end{theorem}

\noindent In fact, any diffeomorphism $\xi$ defined on the domain $[\hat x^0_{\rm Ei},\hat x^0_{\rm Ef}]$ and satisfying the separable differential equation
\be 
{\d\xi\over \d \hat x^0_{\rm E}}=\sqrt{\hat g_{00}[\ell](\hat x_{\rm E}^{0})\over g_{00}(\xi(\hat x_{\rm E}^{0}))}
\ee
yields $\hat g_{00}^\xi[\ell](\xi(\hat x^0_{\rm E}) ) =g_{00}(\xi(\hat x^0_{\rm E}) )$ 
thanks to \Eq{transfo}. Choosing 
\be 
\xi(\hat x^0_{\rm E})=\int_{\hat x^0_{\rm Ei}}^{\hat x^0_{\rm E}} \d \hat x_{\rm E}^{0\prime}\, \sqrt{\hat g_{00}[\ell](\hat x_{\rm E}^{0\prime})\over g_{00}(\xi(\hat x_{\rm E}^{0\prime}))}+x^0_{\rm Ei} 
\ee
also leads to $\xi(\hat x^0_{\rm Ei})=x^0_{\rm Ei}$. Applying \Eq{di} for $\hat g_{00}[\ell]$ and using the above results,  we have finally 
\be
\ell= \int_{\xi(\hat x^0_{\rm Ei})}^{\xi(\hat x^0_{\rm Ef})}\d \hat x^{\xi 0}_{\rm E}\, \sqrt{\hat g_{00}^\xi[\ell](\hat x^{\xi0}_{\rm E})}= \int_{x_{\rm Ei}^0}^{\xi(\hat x^0_{\rm Ef})}\d \hat x^{\xi 0}_{\rm E}\, \sqrt{g_{00}(\hat x^{\xi0}_{\rm E})}\, , 
\ee
which implies  that $\xi(x^0_{\rm Ef})=x^0_{\rm Ef}$ since $g_{00}$ is in  class $\ell$. 


\subsection{Isometries of the line segment}
\label{A2}

\begin{theorem}  
The isometry group of  a line segment with metric $g_{00}$ is $\Z_2$ generated by orientation reversal.
\end{theorem}

\noindent Any isometry $\I$ satisfies $g^\I_{00}=g_{00}$. Since $g_{00}$ in class $\ell$ can be obtained
by acting on  $\hat g_{00}[\ell]$ with a diffeomorphism $\xi$,  this equality can be written as $\hat g^{\I\circ \xi}_{00}[\ell]=\hat g^\xi_{00}[\ell]$. Applying on both sides the diffeomorphism $\xi^{-1}$, we obtain $\hat  g^{\xi^{-1}\circ \I\circ \xi}_{00}[\ell]=\hat  g^{\xi^{-1}\circ\xi}_{00}[\ell]=\hat g_{00}[\ell]$. Using \Eq{transfo}, this means that
\be
{1\over [(\xi^{-1}\circ \I\circ \xi)']^2}\,  \hat g_{00}[\ell] = \hat g_{00}[\ell]\, .
\label{dife}
\ee
Since we are free to choose the form of the fiducial metric, 
let us make  for this proof the particular choice 
\be
\hat g_{00}[\ell](\tau)=\ell^2\quad  \mbox{defined for}\quad   \tau\in [0,1]\, .
\label{fid}
\ee
In that case, \Eq{dife} becomes $(\xi^{-1}\circ \I\circ \xi)'=\pm 1$, which leads to $(\xi^{-1}\circ \I\circ \xi)(\tau)=\pm \tau+\cst$.  Imposing  the sets $\{(\xi^{-1}\circ \I\circ \xi)(0),(\xi^{-1}\circ \I\circ \xi)(1)\}$ and $\{0,1\}$ to be equal for the domains of definition of both sides of \Eq{dife} to be the same, we are left with only two possibilities: $(\xi^{-1}\circ \I\circ \xi)(\tau)=\tau$ or $(\xi^{-1}\circ \I\circ \xi)(\tau)=1-\tau$. As a result we have $\I=\xi\circ \xi^{-1}=\Id$ or $\I=\xi\circ (\tau \to 1-\tau)\circ \xi^{-1}$. The isometry group is thus
\be
\Z_2=\big\{\Id, \cR\big\} \, , \quad ~~\where\quad~~ \cR=\xi\circ (\tau \to 1-\tau)\circ \xi^{-1}\, , \quad \cR\circ\cR=\Id\, .
\ee


\subsection{Modulus of the line segment}
\label{A3}

\begin{theorem}
The moduli space of a  line segment is of real dimension 1.
\end{theorem}

\noindent The moduli correspond to the deformations of the metric that are ``orthogonal'' to those obtained by diffeomorphisms. Let us consider any metric $g_{00}$  in class $\ell$ and defined on $[x^0_{\rm E i},x^0_{\rm E f}]$.  We look for an infinitesimal deformation, denoted $\delta' g_{00}$, that is orthogonal to all deformations $\delta_\xi g_{00}$, where $\xi\in\Diff[g_{00}]$. 

Using \Eq{xiloc} and the definition~(\ref{defin}), we have to solve
\begin{align}
0&=\big(\delta' g,\delta_\xi  g\big)_\ell=\big(\delta'  g,-2 \nabla \delta x_{\rm E} \big)_\ell\nonumber \\
&=\int_{ x^0_{\rm Ei}}^{ x^0_{\rm Ef}} \d  x^0_{\rm E}\, \sqrt{ g_{00}}\; \delta'  g_{00}\,(-2)  \nabla_0 \delta x^0_{\rm E} \;  g^{00} \nonumber \\
&= \Big[\sqrt{ g_{00}}\; \delta'  g_{00}\,(-2)\, \delta x^0_{\rm E }  \; g^{00}\Big]_{ x^0_{\rm Ei}}^{ x^0_{\rm Ef}} + \int_{ x^0_{\rm Ei}}^{ x^0_{\rm Ef}} \d  x^0_{\rm E}\, \sqrt{ g_{00}}\; 2 \nabla^0 \delta'  g_{00}\,\delta x^0_{\rm E} \nonumber \\
&= \Big[\sqrt{ g_{00}}\; \delta'  g_{00}\,(-2)\, \delta x^0_{\rm E }  \; g^{00}\Big]_{ x^0_{\rm Ei}}^{ x^0_{\rm Ef}} +\big(2 \nabla\delta'  g, \delta x_{\rm E} \big)_\ell\, .
\label{0=}
\end{align}
Let us restrict for a moment to the diffeomorphisms obeying 
\be
\delta x^0_{\rm E }(x^0_{\rm Ei})=0 \,, \qquad \delta x^0_{\rm E }(x^0_{\rm Ef})=0\, , 
\ee
so that  the boundary term in \Eq{0=} is absent. Since $\delta x^0_{\rm E}$ is still arbitrary on  $(x^0_{\rm Ei}, x^0_{\rm E{f}})$, we obtain  
\be
0= 2 \nabla^0\delta'  g_{00} = 2\,\partial_0\big( g^{00}\, \delta'  g_{00}\big)\, , 
\ee
whose solution can be written as
\be
 \delta'  g_{00}=2\, {\delta\ell\over \ell}\,  g_{00}\, , 
 \label{mode}
\ee
where $\delta\ell$ is any integration constant. 

Turning back to a generic diffeomorphism, \Eq{0=} becomes
\be
0= \Big[\sqrt{ g_{00}}\; 2\, {\delta \ell\over \ell}\,(-2)\, \delta x^0_{\rm E }  \Big]_{ x^0_{\rm Ei}}^{ x^0_{\rm Ef}} = -4\,\delta \ell\, \big[\delta \tau  \big]_{0}^{1}\, ,
\label{dl}
\ee
where we have used the fact that $\sqrt{ g_{00}}\, \delta x^0_{\rm E}$ is diffeomorphism invariant and, therefore, can be expressed in the coordinate system associated with the metric in class $\ell$ given in \Eq{fid}. However, we know that 
\be
\delta\tau(1)-\delta\tau(0)=0\, ,
\ee
since otherwise a diffeomorphism would change the proper length of a segment of constant metric. As a result, \Eq{dl} is trivial in the sense that it leaves $\delta\ell$ arbitrary.  

Integrating \Eq{mode}, one obtains
\be
g_{00}[\ell]=\ell^2 \, g_{00}[1]\, , 
\ee
where the classes are indicated in brackets. 
 Thus the only moduli deformation of a line segment is that associated with the variation of its length $\ell$.  
 
Notice that the above equation also shows that the diffeomorphisms acting on $g_{00}[\ell]$ are actually  acting only on $g_{00}[1]$. Therefore, the reparametrization group of the line segment is independent of its length. As a result, 
 \be
 \Diff[g^\xi_{00}[\ell]]= \Diff[g_{00}[\ell]]=\Diff[g_{00}[1]]\equiv \Diff\, , 
 \ee
 where the last equality means that the group depends only on the topology of the line segment.
 

\subsection{Independence on the fiducial metric}
\label{A4}


 In this Appendix, we check explicitly that the Faddeev--Popov Jacobian and  the contributions $Z_\epsilon(a_0)$ to the semi-classical expression of the wave function $\Psi(a_0)$  are independent of the choice of fiducial metric, \ie that that they are invariant under diffeomorphisms.

To this end, let us consider the change of coordinate $\tau(\hat y^{\hat 0}_{\rm E})$ that results in the change of lapse function
\be
\ell \, \d\tau = \sqrt{\hat g_{\hat 0 \hat 0}[\ell](\hat y^{\hat 0}_{\rm E})}\; \d \hat y^{\hat 0}_{\rm E}\, .
\ee
The commuting tensors and the ghosts in  \Sect{cfm} transform as 
\begin{align}
\beta^{\hat 0\hat 0}(\hat y^{\hat 0}_{\rm E}) &= \sum_{k\ge 0} \boldsymbol{\beta}_k\,  \chi^{\hat 0\hat 0}_k(\hat y^{\hat 0}_{\rm E}) \, , \qquad  \delta x_{\rm E\hat 0}(\hat y^{\hat 0}_{\rm E}) = \sum_{k\ge 1} \boldsymbol{\gamma}_k\,  \sigma_{\hat 0,k}(\hat y^{\hat 0}_{\rm E}) \,,\nonumber \\
b^{\hat 0\hat 0}(\hat y^{\hat 0}_{\rm E}) &= \sum_{k\ge 0} \b_k\,  \chi^{\hat 0\hat 0}_k(\hat y^{\hat 0}_{\rm E}) \,, \qquad \,~~~~\, c_{0}(\tau) = \sum_{k\ge 1} \c_k\,  \sigma_{\hat 0,k} (\hat y^{\hat 0}_{\rm E})\,,
\end{align} 
where the expansion modes are given by 
\be
\chi^{\hat 0\hat 0}(\hat y^{\hat 0}_{\rm E})={\ell^2\over\hat g_{\hat 0 \hat 0}[\ell](\hat y^{\hat 0}_{\rm E})}\, \chi^{00}(\tau(\hat y^{\hat 0}_{\rm E}))\, , \qquad \sigma_{\hat 0,k}(\hat y^{\hat 0}_{\rm E})={\sqrt{\hat g_{\hat 0 \hat 0}[\ell](\hat y^{\hat 0}_{\rm E})}\over \ell}\, \sigma_{0,k}(\tau(\hat y^{\hat 0}_{\rm E}))\, .
\ee
However, \Eqs{pieces},~(\ref{meacom}),~(\ref{meaanti}) remain unchanged, since  they  are derived from the reparametrization-invariant inner product given in  \Eq{defin}. Hence, the computation  of the Faddeev--Popov Jacobian displayed in \Eq{compu}  is valid in any gauge. 

 Moreover, denoting 
$\xi(\tau)\equiv \tau(\hat y^{\hat 0}_{\rm E})$, the scalar field $\delta a$ in \Sect{moexp} is transformed into $\delta a^\xi$, which can be expanded as
\begin{align}
\delta a^\xi(\hat y^{\hat 0}_{\rm E})&\equiv \delta a(\tau(\hat y^{\hat 0}_{\rm E}))=\sum_{k\ge 1}\deltaa_k\,  \phi^{\epsilon\xi}_k(\hat y^{\hat 0}_{\rm E})\, ,\nonumber \\
\where \quad \phi^{\epsilon\xi}_k(\hat y^{\hat 0}_{\rm E}) &\equiv  \phi^{\epsilon}_k(\tau(\hat y^{\hat 0}_{\rm E}))\, ,\quad ~~(\phi^{\epsilon\xi}_k,\phi^{\epsilon\xi}_{k'})_{\bell_\epsilon}=\delta_{kk'}\, .
\end{align}
The modes $\phi_k^{\epsilon \xi}$ are eigenvectors of the operator $\S_\epsilon^\xi$ 
in the new coordinate system, with eigenvalues $\nu^\epsilon_k$, since 
\be
\S^\xi\phi_k^{\epsilon\xi}(\hat y^{\hat 0}_{\rm E})\equiv (\S_\epsilon\phi^\epsilon_k)^\xi(\hat y^{\hat 0}_{\rm E})=(\S_\epsilon \phi^\epsilon_k)(\tau(\hat y^{\hat 0}_{\rm E}))=\nu_k^\epsilon \phi^\epsilon_k(\tau(\hat y^{\hat 0}_{\rm E})) = \nu_k^\epsilon \phi_k^{\epsilon\xi}(\hat y^{\hat 0}_{\rm E})\, .
\ee 
As a result, the argument of the exponential in the definition of $Z_\epsilon(a_0)$ in \Eq{psii}, along with the norm of the scalar field and its path-integral measure are invariant, 
\begin{align}
(\delta a^{\xi},\S^\xi_\epsilon \delta a^{\xi})_{\bell_{\epsilon}}&=(\delta a,\S_\epsilon \delta a)_{\bell_{\epsilon}}=\nu^\epsilon_k\,(\deltaa_k)^2\, , \nonumber\\
||\delta a^\xi||^2_{\bell_\epsilon}&= (\delta a^\xi,\delta a^\xi)_{\bell_\epsilon}= (\delta a,\delta a)_{\bell_\epsilon}=||\delta a||^2_{\bell_\epsilon}~~\Longrightarrow~~ \D\delta a^\xi = \D\delta a= \bigwedge_{k\ge 1}\d \deltaa_k\, .
\end{align}
 This shows the gauge invariance of $Z_\epsilon(a_0)$. 


\subsection{Hermiticity of the Hamiltonian}
\label{A5}

In this subsection, we retrieve the results of \Sect{hqh} using refined arguments that provide insight.  The following proofs include a generalization of  some treatment given in \Refe{DeWitt} in order to take into account a positive measure $\mu_A(a_0)$ to be determined and an unknown function $\rho_A(a_0)$ in the definition of the operator $H_A/N$. 

Let us consider the integral
\be
\J = \int_0^{+\infty} \d a_0\, \mu_A\, \Xi\, \Big\{\Big({H_A\over N}\,\Psi_{A1}\Big)^*\, \Psi_{A2} - \Psi_{A1}^*\, {H_A\over N}\,\Psi_{A2}\Big\}\, , 
\label{debut}
\ee
where $\Psi_{A1}$, $\Psi_{A2}$ are arbitrary complex functions. Moreover,  $\Xi$, which is also complex, is an arbitrary test function. This means that it is identically vanishing outside some closed interval $[a_{\rm min},a_{\rm max}]\subset \R_+^*$. Using the notation of \Eq{pr1}, we can write the first term as follows
\be
\int_0^{+\infty} \d a_0\, \mu_A\, \Xi\, \Big({H_A\over N}\,\Psi_{A1}\Big)^*\, \Psi_{A2} = \big\langle {H_A\over N}\,\Psi_{A1},\Xi\, \Psi_{A2}\big\rangle = \big\langle \Psi_{A1},{H_A\over N}\,(\Xi\, \Psi_{A2})\big\rangle\, , 
\ee
where we have assumed that \Eq{hh} holds. Indeed, the last equality  is obtained by integrating by parts and by noticing that no boundary term arises thanks to presence of the test function which vanishes identically in neighborhoods of $a_0=0$ and $+\infty$. Hence, we obtain 
\begin{align}
\J &=\int_0^{+\infty} \d a_0\, \mu_A\, \Psi_{A1}^*\, \Big\{{H_A\over N}\,(\Xi\Psi_{A2}) - \Xi \, {H_A\over N}\,\Psi_{A2}\Big\}\\\nonumber
& = {\hbar^2\over 12v_3}\int_0^{+\infty}\d a_0\, \mu_A\, \Psi_{A1}^*\, {1\over a_0\rho_A}\,\Big\{{\d\over \d a_0}\Big({\d\Xi\over \d a_0}\, \rho_A\, \Psi_{A2}\Big)+{\d\Xi\over \d a_0}\, \rho_A\,{\d\Psi_{A2}\over \d a_0}\Big\}\nonumber \esp\\
& = -{\hbar^2\over 12v_3}\int_0^{+\infty}\d a_0\, \Xi\, {\d\over \d a_0}\Big\{\rho_A \Big({\mu_A\over a_0\rho_A}\, \Psi^*_{A1} \, {\d \Psi_{A2}\over \d a_0}-{\d\over \d a_0}\Big({\mu_A\over a_0\rho_A}\, \Psi^*_{A1}\Big)\Psi_{A2}\Big)\Big\}\, ,\esp
\end{align}
where the second line follows from the definition of $H_A/N$ given in \Eq{WDWtil2}, while the last one is obtained by integrating by parts. Again, no boundary term arises thanks to the test function. Comparing the last expression with \Eq{debut} and remembering that the test function is arbitrary,\footnote{One can choose Dirac distributions centered at an arbitrary point in $\R_+^*$.} we conclude that the integrands are identical, namely
\begin{align}
\mu_A\, \bigg\{\Big({H_A\over N}\,\Psi_{A1}\Big)^*\, &\Psi_{A2} - \Psi_{A1}^*\, {H_A\over N}\,\Psi_{A2}\bigg\}\nonumber \\
&= -{\hbar^2\over 12v_3}\, {\d\over \d a_0}\left\{\rho_A\!\left({\mu_A\over a_0\rho_A}\, \Psi^*_{A1} \, {\d \Psi_{A2}\over \d a_0}-{\d\over \d a_0}\Big({\mu_A\over a_0\rho_A}\, \Psi^*_{A1}\Big)\Psi_{A2}\right)\right\} ,\esp
\end{align}
which is the local version of \Eq{con}. 

Since $\Psi_{A1}$, $\Psi_{A2}$ are arbitrary complex functions, we can specialize to $\Psi_{A1}=\Psi_{A2}$ real, which yields
\be
0 = {\d\over \d a_0}\Big\{ \rho_A \Psi_{A1}^2\, {\d\over \d a_0}\Big({\mu_A\over a_0\rho_A}\Big)\Big\}\, .
\ee
The choice $\Psi_{A1}\equiv 1/\sqrt{\rho_A}$ leads to 
\be
{\mu_A\over a_0\rho_A} = \kappa+\kappa_1 a_0\, , 
\ee
where $\kappa$, $\kappa_1$ are constants. Taking instead $\Psi_{A1}$ not identically equal to $1/\sqrt{\rho_A}$ yields finally  $\kappa_1=0$, \ie $\mu_A=\kappa a_0\rho_A$. Hence we obtain \Eq{local}. 

\end{appendices} 
 



\begin{thebibliography}{99}

\bibitem{HH} 
  J.~B.~Hartle and S.~W.~Hawking,
  ``Wave function of the universe,''
  Phys.\ Rev.\ D {\bf 28} (1983) 2960
   [Adv.\ Ser.\ Astrophys.\ Cosmol.\  {\bf 3} (1987) 174].

\bibitem{Vilenkin1}
  A.~Vilenkin,
  ``Creation of universes from nothing,''
  Phys.\ Lett.\ B {\bf 117} (1982) 25.
  
\bibitem{Vilenkin2}
  A.~Vilenkin,
  ``The birth of inflationary universes,''
  Phys.\ Rev.\ D {\bf 27} (1983) 2848.
  
\bibitem{Vilenkin3} 
  A.~Vilenkin,
  ``Quantum creation of universes,''
Phys. Rev. D \textbf{30} (1984), 509.
    
\bibitem{Vilenkin4}
A.~Vilenkin,
``Boundary conditions in quantum cosmology,''
Phys. Rev. D \textbf{33} (1986), 3560.

\bibitem{Halli}
J.~J.~Halliwell,
``Derivation of the Wheeler-De Witt equation from a path integral for minisuperspace models,''
Phys. Rev. D \textbf{38} (1988), 2468.

\bibitem{Coleman}
S.~Coleman,
``Aspects of symmetry: Selected Erice lectures,''
Cambridge University Press (2010).
 
\bibitem{DeWitt}
B.~S.~DeWitt,
``Quantum theory of gravity. I. The canonical theory,''
Phys. Rev. \textbf{160} (1967), 1113-1148.

\bibitem{Linde} 
  A.~D.~Linde,
  ``Quantum creation of an inflationary universe,''
  Sov.\ Phys.\ JETP {\bf 60}, 211 (1984)
  [Zh.\ Eksp.\ Teor.\ Fiz.\  {\bf 87}, 369 (1984)]; A.~D.~Linde,
  ``Quantum creation of the inflationary universe,''
  Lett.\ Nuovo Cim.\  {\bf 39}, 401 (1984).

\bibitem{HawkingL=0}
S.~W.~Hawking,
``The cosmological constant is probably zero,''
Phys. Lett. B \textbf{134} (1984), 403.

\bibitem{HalliHawk}
J.~J.~Halliwell and S.~W.~Hawking,
``The origin of structure in the universe,''
Phys. Rev. D \textbf{31} (1985), 1777.

\bibitem{Schleich:1986db}
K.~Schleich,
``Semiclassical wave function of the universe at small three geometries,''
Phys. Rev. D \textbf{32} (1985), 1889-1898.

\bibitem{Vilenkin5}
A.~Vilenkin,
``Predictions from quantum cosmology,''
NATO Sci. Ser. C \textbf{476} (1996), 345-367
[arXiv:gr-qc/9507018 [gr-qc]]. 

\bibitem{Halli2}
J.~J.~Halliwell and J.~Louko,
``Steepest descent contours in the path integral approach to quantum cosmology. 1. The de Sitter minisuperspace model,''
Phys. Rev. D \textbf{39} (1989), 2206.

\bibitem{Turok1}
J.~Feldbrugge, J.~L.~Lehners and N.~Turok,
``Lorentzian quantum cosmology,''
Phys. Rev. D \textbf{95} (2017) no.10, 103508
[arXiv:1703.02076 [hep-th]].

\bibitem{Halli-Hartle}
J.~Diaz Dorronsoro, J.~J.~Halliwell, J.~B.~Hartle, T.~Hertog and O.~Janssen,
``The real no-boundary wave function in Lorentzian quantum cosmology,''
Phys. Rev. D \textbf{96} (2017) no.4, 043505
[arXiv:1705.05340 [gr-qc]].

\bibitem{Quevedo}
S.~Cespedes, S.~P.~de Alwis, F.~Muia and F.~Quevedo,
``Lorentzian vacuum transitions: Open or closed universes?,''
[arXiv:2011.13936 [hep-th]].

\bibitem{Davidson1}
A.~Davidson, D.~Karasik and Y.~Lederer,
``Wavefunction of a brane-like universe,''
Class. Quant. Grav. \textbf{16} (1999), 1349-1356
[arXiv:gr-qc/9901003 [gr-qc]].

\bibitem{Davidson2}
A.~Davidson and B.~Yellin,
``Quantum black hole wave packet: Average area entropy and temperature dependent width,''
Phys. Lett. B \textbf{736} (2014), 267-271
[arXiv:1404.5729 [gr-qc]].

\bibitem{LindeBook}
A.~D.~Linde,
``Particle physics and inflationary cosmology,''
Contemp. Concepts Phys.~\textbf{5} (1990), 1-362
[arXiv:hep-th/0503203 [hep-th]].

\bibitem{Pol1}
J.~Polchinski,
``String theory. Vol. 1: An introduction to the bosonic string,''
Cambridge University Press (1998).

\bibitem{Lihui} Lihui Liu, private communication.

\bibitem{pearls}
J.~J.~Halliwell and R.~C.~Myers,
``Multiple sphere configurations in the path integral representation of the wave function of the universe,''
Phys. Rev. D \textbf{40} (1989), 4011.

\bibitem{Callan}
C.~G.~Callan, Jr. and S.~R.~Coleman,
``The fate of the false vacuum. II. First quantum corrections,''
Phys. Rev. D \textbf{16} (1977), 1762-1768.

\bibitem{He:2020wzj}
D.~He and Q.~y.~Cai,
``Wheeler-DeWitt equation rejects quantum effects of grown-up universes as a candidate for dark energy,''
Phys. Lett. B \textbf{809} (2020), 135747 
[arXiv:2009.05187 [gr-qc]].

\bibitem{wkb}
See \eg D.~J.~Griffiths,
``Introduction to quantum mechanics,'' Cambridge University Press (2016). 

\bibitem{He:2015wla}
D.~He, D.~Gao and Q.~y.~Cai,
``Dynamical interpretation of the wavefunction of the universe,''
Phys. Lett. B \textbf{748} (2015), 361-365 
[arXiv:1507.06727 [gr-qc]].

\bibitem{Nelson:2008vz}
W.~Nelson and M.~Sakellariadou,
``Unique factor ordering in the continuum limit of LQC,''
Phys. Rev. D \textbf{78} (2008), 024006 
[arXiv:0806.0595 [gr-qc]].

  \bibitem{Peskin}
See \eg  M.~E.~Peskin and D.~V.~Schroeder,
``An Introduction to quantum field theory,'' Perseus Book Publishing (1995).

\bibitem{Gibbons}
G.~W.~Gibbons, S.~W.~Hawking and J.~M.~Stewart,
``A natural measure on the set of all universes,''
Nucl. Phys. B \textbf{281} (1987), 736.


\bibitem{HHH1}
J.~B.~Hartle, S.~W.~Hawking and T.~Hertog,
``Quantum probabilities for inflation from holography,''
JCAP \textbf{01} (2014), 015
[arXiv:1207.6653 [hep-th]].

\bibitem{HHH2}
J.~Hartle, S.~W.~Hawking and T.~Hertog,
``Local observation in eternal inflation,''
Phys. Rev. Lett. \textbf{106} (2011), 141302
[arXiv:1009.2525 [hep-th]].

\bibitem{HHH3}
J.~B.~Hartle, S.~W.~Hawking and T.~Hertog,
``No-boundary measure of the universe,''
Phys. Rev. Lett. \textbf{100} (2008), 201301
[arXiv:0711.4630 [hep-th]].

\end{thebibliography}
\end{document}